\pdfoutput=1
\newcommand{\mode}{1} 
\newcommand{\movie}{1} 

\ifnum\mode=0
\documentclass[iop,numberedappendix,twocolumn,tighten]{aastex62}
\usepackage{etoolbox}
\submitjournal{The Astrophysical Journal}
\newtoggle{AAS}
\toggletrue{AAS}

\else
\documentclass[iop,twocolappendix,numberedappendix]{emulateapj}
\usepackage{etoolbox}
\newtoggle{AAS}
\togglefalse{AAS}
\usepackage{fancyhdr}
\fancypagestyle{plain}{ %
  \fancyhf{} 
  
}

\submitted{The Astrophysical Journal}

\makeatletter
\newcommand{\Mc}{}
\DeclareRobustCommand{\Mc}{%
  M%
  \raisebox{\dimexpr\fontcharht\font`M-\height}{%
    \check@mathfonts\fontsize{\sf@size}{0}\selectfont
    \underline{c}%
  }%
}
\makeatother

\fi

\received{September 11, 2018}
\revised{\today}

\begingroup\lccode`\~=`\*\lowercase{\endgroup\def~}{\mspace{1.5mu}}
\AtBeginDocument{\mathcode`\*="8000 }

\iftoggle{AAS}{}{
\usepackage[backref,breaklinks,colorlinks,citecolor=blue,linkcolor=magenta,hyperfigures]{hyperref}

}

\newcommand{\refemp}[1]{{#1}}

\usepackage{graphicx}
\graphicspath{{../_Figures/}}
\usepackage{media9}
\usepackage{amssymb}
\usepackage{amsmath}
\usepackage{upgreek}
\usepackage{fancyhdr}
\usepackage{float}
\usepackage{tabu}
\usepackage[nointegrals]{wasysym}
\usepackage[inline,shortlabels]{enumitem}
\usepackage{dcolumn}
\usepackage[nice,tight]{units}
\usepackage{savesym}
\savesymbol{tablenum}
\usepackage[separate-uncertainty=true]{siunitx}
\restoresymbol{SIX}{tablenum}
\usepackage[normalem]{ulem}
\usepackage{backref}

\usepackage[nameinlink,capitalize]{cleveref}
\usepackage[all]{hypcap} 
\makeatletter
\usepackage{etoolbox}
\patchcmd\H@refstepcounter{\protected@edef}{\protected@xdef}{}{}
\makeatother

\def\dbar{{\mathchar'26\mkern-12mu \mathrm{d}}}
\def\Lya{\textrm{Ly}\mkern-2mu \upalpha}
\def\LymanA{Lyman-$\mkern-0.5mu \upalpha$}

\makeatletter

\patchcmd{\NAT@citex}
  {\@citea\NAT@hyper@{%
     \NAT@nmfmt{\NAT@nm}%
     \hyper@natlinkbreak{\NAT@aysep\NAT@spacechar}{\@citeb\@extra@b@citeb}%
     \NAT@date}}
  {\@citea\NAT@nmfmt{\NAT@nm}%
   \NAT@aysep\NAT@spacechar\NAT@hyper@{\NAT@date}}{}{}

\patchcmd{\NAT@citex}
  {\@citea\NAT@hyper@{%
     \NAT@nmfmt{\NAT@nm}%
     \hyper@natlinkbreak{\NAT@spacechar\NAT@@open\if*#1*\else#1\NAT@spacechar\fi}%
       {\@citeb\@extra@b@citeb}%
     \NAT@date}}
  {\@citea\NAT@nmfmt{\NAT@nm}%
   \NAT@spacechar\if*#1*\else#1\NAT@spacechar\fi\NAT@hyper@{\NAT@date}}
  {}{}

\def\footnoterule{
  \kern -2pt
  \hrule width .58\columnwidth
  \kern 2pt} 

\makeatother

\ifxetex
  \usepackage{letltxmacro}
  \LetLtxMacro\SavedIncludeGraphics\includegraphics
  \def\includegraphics#1#{
    \IncludeGraphicsAux{#1}%
  }%
  \newcommand*{\IncludeGraphicsAux}[2]{%
    \XeTeXLinkBox{%
      \SavedIncludeGraphics#1{#2}%
    }%
  }%
\fi

\usepackage{filecontents}

\backcite {Fulton17}{{1}{1}{section.1}}
\backcite {Owen13}{{1}{1}{section.1}}
\backcite {Lopez13}{{1}{1}{section.1}}
\backcite {Vidal03}{{1}{1}{section.1}}
\backcite {Ben07}{{1}{1}{section.1}}
\backcite {Vidal08}{{1}{1}{section.1}}
\backcite {Ehrenreich08}{{1}{1}{section.1}}
\backcite {Lecavelier10}{{1}{1}{section.1}}
\backcite {Bourrier13}{{1}{1}{section.1}}
\backcite {Ehrenreich12}{{1}{1}{section.1}}
\backcite {Kulow14}{{1}{1}{section.1}}
\backcite {Ehrenreich15}{{1}{1}{section.1}}
\backcite {Bourrier16}{{1}{1}{section.1}}
\backcite {Lavie17}{{1}{1}{section.1}}
\backcite {Vidal04}{{1}{1}{section.1}}
\backcite {Ben10}{{1}{1}{section.1}}
\backcite {Vidal13}{{1}{1}{section.1}}
\backcite {Linsky10}{{1}{1}{section.1}}
\backcite {Loyd17}{{1}{1}{section.1}}
\backcite {Jensen12}{{1}{1}{section.1}}
\backcite {Barnes16}{{1}{1}{section.1}}
\backcite {Ehrenreich15}{{1}{3}{section.1-footnote.3}}
\backcite {Spake18}{{2}{1}{section.1}}
\backcite {Seager00}{{2}{1}{section.1}}
\backcite {Oklopcic18}{{2}{1}{section.1}}
\backcite {Parker58}{{2}{1}{section.1}}
\backcite {Yelle04}{{2}{1}{section.1}}
\backcite {Garcia07}{{2}{1}{section.1}}
\backcite {Murray09}{{2}{1}{section.1}}
\backcite {Owen14}{{2}{1}{section.1}}
\backcite {Tripathi15}{{2}{1}{section.1}}
\backcite {Debrecht18}{{2}{1}{section.1}}
\backcite {Schneiter16}{{2}{1}{section.1}}
\backcite {Stone09}{{2}{1}{section.1}}
\backcite {Tremblin13}{{2}{1}{section.1}}
\backcite {Bisikalo13}{{2}{1}{section.1}}
\backcite {Trammell14}{{2}{1}{section.1}}
\backcite {Owen14}{{2}{1}{section.1}}
\backcite {Cohen11}{{2}{1}{section.1}}
\backcite {Matsakos15}{{2}{1}{section.1}}
\backcite {Carroll17}{{2}{1}{section.1}}
\backcite {Trammell14}{{2}{1}{section.1}}
\backcite {Stone09}{{2}{1}{section.1}}
\backcite {Holmstrom08}{{2}{1}{section.1}}
\backcite {Tremblin13}{{2}{1}{section.1}}
\backcite {Tripathi15}{{2}{1}{section.1}}
\backcite {Tripathi15}{{2}{1}{section.1}}
\backcite {Tripathi15}{{2}{1}{section.1}}
\backcite {Debrecht18}{{2}{1}{section.1}}
\backcite {Owen14}{{2}{1}{section.1}}
\backcite {Draine11}{{3}{2.1.1}{equation.2.2}}
\backcite {Verner96}{{3}{2.1.1}{equation.2.3}}
\backcite {Osterbrock89}{{3}{2.1.1}{equation.2.7}}
\backcite {Murray09}{{3}{2.1.1}{equation.2.7}}
\backcite {Osterbrock89}{{3}{2.1.1}{equation.2.8}}
\backcite {Black81}{{3}{2.1.1}{equation.2.9}}
\backcite {Lamers99}{{4}{2.2}{subsection.2.2}}
\backcite {Pierre10}{{5}{2.3.2}{subsubsection.2.3.2}}
\backcite {Murray09}{{5}{2.3.2}{equation.2.23}}
\backcite {Murray09}{{5}{2.3.2}{equation.2.23}}
\backcite {Stone09}{{6}{2.3.4}{subsubsection.2.3.4}}
\backcite {Tremblin13}{{6}{2.3.4}{subsubsection.2.3.4}}
\backcite {Carroll17}{{6}{2.3.4}{subsubsection.2.3.4}}
\backcite {Mikic99}{{6}{2.3.4}{subsubsection.2.3.4}}
\backcite {Lamers99}{{6}{2.3.4}{subsubsection.2.3.4}}
\backcite {Vidotto18}{{6}{2.3.4}{subsubsection.2.3.4}}
\backcite {Gombosi18}{{6}{2.3.4}{equation.2.25}}
\backcite {Murray09}{{7}{2.4}{subsection.2.4}}
\backcite {Stone08}{{7}{3.1}{subsection.3.1}}
\backcite {Krumholz07}{{7}{3.1}{subsection.3.1}}
\backcite {Tripathi15}{{7}{3.1}{subsection.3.1}}
\backcite {Stone10}{{7}{3.1}{subsection.3.1}}
\backcite {Tripathi15}{{7}{3.1}{subsection.3.1}}
\backcite {Beckwith11}{{7}{3.1.1}{subsubsection.3.1.1}}
\backcite {Murray09}{{7}{12}{equation.2.28-footnote.12}}
\backcite {Krumholz07}{{8}{3.1.1}{subsubsection.3.1.1}}
\backcite {Murray09}{{8}{3.2.2}{subsubsection.3.2.2}}
\backcite {Murray09}{{8}{3.2.2}{subsubsection.3.2.2}}
\backcite {Vidotto18}{{9}{3.2.2}{equation.3.30}}
\backcite {Woods98}{{10}{3.3}{table.1}}
\backcite {Tripathi15}{{10}{3.3}{table.1}}
\backcite {Vidotto17}{{10}{3.3}{figure.2}}
\backcite {Vidotto18}{{10}{3.3}{figure.2}}
\backcite {Debrecht18}{{11}{4.1}{figure.3}}
\backcite {Stone09}{{11}{4.1}{figure.3}}
\backcite {Carroll17}{{11}{4.1}{figure.3}}
\backcite {Stone09}{{11}{4.1}{figure.3}}
\backcite {Carroll17}{{11}{4.1}{figure.3}}
\backcite {Garcia07}{{11}{4.1}{figure.3}}
\backcite {Murray09}{{11}{4.1}{figure.3}}
\backcite {Tripathi15}{{12}{4.2}{subsection.4.2}}
\backcite {Debrecht18}{{13}{4.3}{subsection.4.3}}
\backcite {Tripathi15}{{19}{4.7}{table.2}}
\backcite {Murray09}{{19}{4.7}{table.2}}
\backcite {Debrecht18}{{19}{4.7}{table.2}}
\backcite {Murray99}{{20}{5.1}{subsection.5.1}}
\backcite {Carroll17}{{20}{14}{figure.14}}
\backcite {Stone09}{{23}{5.2.3}{figure.20}}
\backcite {Stone09}{{24}{20}{figure.20-footnote.20}}
\backcite {Stone09}{{25}{5.2.3}{figure.20}}
\backcite {Murray09}{{25}{5.3}{subsection.5.3}}
\backcite {Murray09}{{26}{5.4}{subsection.5.4}}
\backcite {Vidal04}{{26}{5.4}{subsection.5.4}}
\backcite {Vidal03}{{26}{5.4}{subsection.5.4}}
\backcite {Ben07}{{26}{5.4}{subsection.5.4}}
\backcite {Bourrier13}{{26}{5.4}{subsection.5.4}}
\backcite {Vidal08}{{26}{24}{subsection.5.4-footnote.24}}
\backcite {Vidotto18}{{26}{5.4}{subsection.5.4}}
\backcite {Ehrenreich15}{{26}{5.4}{subsection.5.4}}
\backcite {Bourrier16}{{26}{5.4}{subsection.5.4}}
\backcite {Bourrier16}{{26}{5.4}{subsection.5.4}}
\backcite {Bourrier16}{{27}{5.4}{subsection.5.4}}
\backcite {Bourrier16}{{27}{5.4}{subsection.5.4}}
\backcite {Ehrenreich12}{{27}{5.4}{subsection.5.4}}
\backcite {Turk11}{{28}{6}{Item.6}}
\backcite {Watson81}{{29}{A.1}{subsection.A.1}}
\backcite {Parker58}{{29}{A.1}{equation.A.5}}
\backcite {Chamberlain61}{{29}{A.1}{equation.A.5}}
\backcite {Chamberlain61}{{29}{A.1}{equation.A.6}}
\backcite {Lamers99}{{30}{A.3}{equation.A.15}}
\backcite {Lamers99}{{31}{A.3}{equation.A.18}}
\backcite {Tripathi15}{{33}{C}{equation.C.9}}

\begin{document}
\title{Morphology of Hydrodynamic Winds:\\A Study of Planetary Winds in Stellar Environments%
\ifnum\movie=1
\footnotemark[1]
\fi}
\ifnum\movie=1
\ifnum\mode=0
\footnotetext[1]{This pdf contains animated figures %
(\hyperref[fig:noWinds]{Figures \ref{fig:noWinds}}, %
\hyperref[tidal_burp]{\ref{tidal_burp}}, %
\hyperref[fig:fullSnap]{\ref{fig:fullSnap}}, %
\hyperref[fig:full_burp]{\ref{fig:full_burp}},\\%
and \hyperref[fig:wwind]{\ref{fig:wwind}}) %
viewable by doubling clicking in Adobe Acrobat Pro.\\Otherwise available at \href{https://gitlab.com/athena_ae/athena_ae}{https://gitlab.com/athena\_ae/athena\_ae}.}
\else
\footnotetext[1]{This pdf contains animated figures %
(\hyperref[fig:noWinds]{Figures \ref{fig:noWinds}}, %
\hyperref[tidal_burp]{\ref{tidal_burp}}, %
\hyperref[fig:fullSnap]{\ref{fig:fullSnap}}, %
\hyperref[fig:full_burp]{\ref{fig:full_burp}}, %
\hyperref[fig:wwind]{\ref{fig:wwind}}, %
\hyperref[fig:wwind]{\ref{fig:iwind}}, %
and \hyperref[fig:wwind]{\ref{fig:swind}}) %
viewable by doubling clicking in Adobe Acrobat Pro. Otherwise available at \href{https://gitlab.com/athena_ae/athena_ae}{https://gitlab.com/athena\_ae/athena\_ae}.}
\fi
\fi
\iftoggle{AAS}{
\author{John McCann}
\affiliation{Department of Physics, University of California, Santa Barbara, USA}
\affiliation{Department of Astronomy and Astrophysics, University of California, Santa Cruz, USA}

\author{Ruth A. Murray-Clay}
\affiliation{Department of Astronomy and Astrophysics, University of California, Santa Cruz, USA}

\author{Kaitlin Kratter}
\affiliation{Department of Astronomy and Steward Observatory, Univ. of Arizona, Tucson, USA}

\author{Mark R. Krumholz}
\affiliation{Research School of Astronomy and Astrophysics, Australian National University, Canberra, Australia}
}{
\author{John R. McCann\altaffilmark{1,2}\href{https://orcid.org/0000-0002-5155-6645}{\includegraphics[height=6pt]{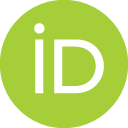}}, %
Ruth A. Murray-Clay\altaffilmark{2}\href{https://orcid.org/0000-0001-5061-0462}{\includegraphics[height=6pt]{orcid_128x128.png}}, %
Kaitlin Kratter\altaffilmark{3}\href{https://orcid.org/0000-0001-5253-1338}{\includegraphics[height=6pt]{orcid_128x128.png}}, %
and Mark R. Krumholz\altaffilmark{4}\href{https://orcid.org/0000-0003-3893-854X}{\includegraphics[height=6pt]{orcid_128x128.png}}}
\affil{
$\sp{1}$Department of Physics, University of California, Santa Barbara, USA; \href{mailto:mccann@ucsb.edu}{mccann@ucsb.edu}\\
$\sp{2}$Department of Astronomy and Astrophysics, University of California, Santa Cruz, USA\\
$\sp{3}$Department of Astronomy and Steward Observatory, University of Arizona, Tucson, USA \\
$\sp{4}$Research School of Astronomy and Astrophysics, Australian National University, Canberra, Australia\\
}
}
\begin{abstract}
Bathed in intense ionizing radiation, close-in gaseous planets undergo hydrodynamic atmospheric escape, which ejects the upper extent of their atmospheres into the interplanetary medium. Ultraviolet detections of escaping gas around transiting planets corroborate such a framework. Exposed to the stellar environment, the outflow is shaped by its interaction with the stellar wind and by the planet's orbit. We model these effects using \texttt{Athena} to perform 3-D radiative-hydrodynamic simulations of tidally-locked hydrogen atmospheres receiving large amounts of ionizing extreme-ultraviolet flux in various stellar environments for the low-magnetic-field case. Through a step-by-step exploration of orbital and stellar wind effects on the planetary outflow, we find three structurally distinct stellar wind regimes: weak, intermediate, and strong. We perform synthetic \LymanA{} observations and find unique observational signatures for each regime. A weak stellar wind\textemdash which cannot confine the planetary outflow, leading to a torus of material around the star\textemdash has a pre-transit, red-shifted dayside arm and a slightly redward-skewed spectrum during transit. The intermediate regime truncates the dayside outflow at large distances from the planet and causes periodic disruptions of the outflow, producing observational signatures that mimic a double transit. The first of these dips is blue-shifted and precedes the optical transit. Finally, strong stellar winds completely confine the outflow into a cometary tail and accelerate the outflow outwards, producing large blue-shifted signals post-transit. Across all three regimes, large signals occur far outside of transit, offering motivation to continue ultraviolet observations outside of direct transit.
\end{abstract}


\section{Introduction}
\label{sec:intro}

Atmospheric escape plays a key role in the evolution of planetary bodies. At their most extreme, processes that drive escape from the upper atmosphere may substantially transform the atmospheric composition of a body over its lifetime. The importance of this mechanism is underscored by the recent confirmation that for short-period exoplanets the radius distribution has a gap near \SI{1.8}{R_\oplus} (Fulton \citeyear{Fulton17}). The gap's existence was predicted prior to its discovery as a consequence of the complete erosion of lower-mass planets' atmospheres by photoionization-driven atmospheric escape (Owen and Wu \citeyear{Owen13}; Lopez and Fortney \citeyear{Lopez13}). While evidence of atmospheric evolution thus appears imprinted on planet demographics, observations of contemporary atmospheric escape in the regimes governing these populations are limited, making model validation difficult.  Loss from highly-irradiated planets typically occurs through hydrodynamic outflows, rather than via the kinetic loss mechanisms that currently dominate for Solar System planets.  Here, we explore the most approachable systems for which improved observational constraints on hydrodynamic escape can be obtained\textemdash hot Jupiters.

Ultraviolet observations of the hot Jupiter HD 209458 b have found up to a \SI{15}{\percent} occultation in the wings of hydrogen \LymanA, with effective Doppler shifts of up to \SI{\pm 150}{km.s^{-1}}, significantly larger than the \SI{5}{\percent} occultation observed at optical (Vidal-Madjar {et al.} \citeyear{Vidal03}; Ben-Jaffel \citeyear{Ben07}; Vidal-Madjar {et al.} \citeyear{Vidal08}; Ehrenreich {et al.} \citeyear{Ehrenreich08}). The high occultation and large velocity are indicative of a fast and extended component of the atmosphere, interpreted as an escaping planetary wind. Similar outflows have been reported for another hot and one warm Jupiter: HD 189733 b (Lecavelier des Etangs {et al.} \citeyear{Lecavelier10}; Bourrier {et al.} \citeyear{Bourrier13}), and 55 Cnc b (Ehrenreich {et al.} \citeyear{Ehrenreich12}\footnote{Along with a non-detection for a super Earth, 55 Cnc e, placing an upper limit on its mass loss.}); and for one hot Neptunian planet, GJ 435 b (Kulow {et al.} \citeyear{Kulow14}; Ehrenreich {et al.} \citeyear{Ehrenreich15}; Bourrier {et al.} \citeyear{Bourrier16}; Lavie {et al.} \citeyear{Lavie17}).  Interestingly, the outflow from GJ 435 b is asymmetric both temporally and spectrally,\footnote{Redshifted occultation of \SI{0.7\pm3.6}{\percent} pre-transit and \SI{8.0\pm3.1}{\percent} post-transit. Blueshifted occultation of \SI{17.6\pm5.2}{\percent} pre-transit and \SI{47.2\pm4.1}{\percent} post-transit (Ehrenreich et al. \citeyear{Ehrenreich15}).} suggesting a cometary-tail-like outflow moving rapidly away from the star. Tentative detections indicate that metals may be present in these escaping winds, including oxygen (Vidal-Madjar {et al.} \citeyear{Vidal04}; Ben-Jaffel and Hosseini \citeyear{Ben10}), magnesium (Vidal-Madjar {et al.} \citeyear{Vidal13}), and carbon and silicon (Linsky et al. \citeyear{Linsky10}; Lyod {et al.} \citeyear{Loyd17}). Additionally, hydrogen H$\upalpha$ absorption has been seen in HD 189733 b's transmission spectra (Jensen {et al.} \citeyear{Jensen12}), but its relation to hydrodynamic escape is still uncertain (Barnes {et al.} \citeyear{Barnes16}). Recently, the outflow from Wasp-107 b was detected in the \SI{1083}{nm} line of excited neutral helium (Spake et al. \citeyear{Spake18}). This line, predicted for exoplanet atmospheres by Seager and Sasselov (\citeyear{Seager00}), and in their outflows by Oklop{\v c}i{\'c} and Hirata (\citeyear{Oklopcic18}), provides an opportunity for ground-based observations.

Current observations have only detected atmospheric escape for exoplanets with orbital periods less than 20 days. As these close-in planets receive large amounts of external heating, they are believed to be undergoing hydrodynamic escape with outflow structures similar to a Parker wind (Parker \citeyear{Parker58}). In an attempt to model the observations, numerous one-dimensional hydrodynamic escape simulations have been produced (e.g., Yelle \citeyear{Yelle04}; Garc\`ia Mu\~noz \citeyear{Garcia07}; Murray-Clay et al. \citeyear{Murray09}). Yet, outflow velocities in these studies generally reach only tens of kilometers per second, comparable to the sound speed of the outflowing gas. Note that the temperature of the gas is limited by radiative cooling to less than a few times \SI{e4}{K}. To resolve inherently asymmetrical processes that cannot be modeled in 1-D and to investigate interactions between the outflowing wind and its environment that may produce high-velocity-offset neutral atoms, multidimensional simulations are needed.  A number of authors have produced such simulations studying pieces of the problem, including dayside photoionization heating in 2-D (Owen and Adams \citeyear{Owen14}) and 3-D (Tripathi et al. \citeyear{Tripathi15}; Debrecht et al. \citeyear{Debrecht18}), photoionization starting at the Hill radius (Schneiter et al \citeyear{Schneiter16}), stellar wind confinement in 2-D (Stone and Proga \citeyear{Stone09}; Tremblin and Chiang \citeyear{Tremblin13}) and 3-D (Bisikalo et al \citeyear{Bisikalo13}), magnetic fields in 2-D (Trammell et al. \citeyear{Trammell14}; Owen and Adams \citeyear{Owen14}) and 3-D (Cohen et al. \citeyear{Cohen11}; Matsakos et al. \citeyear{Matsakos15}), and the possibility of developing a circumstellar torus from the planetary outflow in global 3-D simulations (Nellenback-Carroll et al. \citeyear{Carroll17}).

Through these simulations, several possibilities have been identified that have the potential to enhance occultation in the \LymanA{} wings. One option is that given sufficiently high densities, absorption in the naturally-broadened line wings may be substantial enough that a large velocity offset between the gas and the planet is not required. For example, Trammell et al. (\citeyear{Trammell14}) demonstrated that HD 209458 b only needed a \SI{50}{G} dipole field to explain the observations by producing a dense and extended equatorial dead zone\textemdash a region where the outflow's ram pressure is insufficient to overcome the confining magnetic pressure. Likewise, stellar wind confinement may increase the column density by spatially restructuring the outflowing wind, generating a dense column where significant absorption in the line wings can occur (Stone and Proga \citeyear{Stone09}).  Alternative options seek to increase occultation via Doppler broadening by generating a fast neutral population through additional physics. One such method is the interaction between a slow neutral planetary wind and a fast ionized stellar wind, which can produce energetic neutral atoms through charge exchange (Holmstr\"om et al. \citeyear{Holmstrom08}). Furthermore, at the stellar-planetary wind interface, a Kelvin-Helmholtz instability will lead to stirring that increases the efficiency of charge exchange (Tremblin and Chiang \citeyear{Tremblin13}).

Regardless of whether stellar wind confinement is the correct or entire explanation for these particular observations, the stellar environment affects the structure of atmospheric outflows, with observational consequences. To investigate these consequences in 3-D, we take a bottom-up approach by deconstructing the stellar environment into three of its individual components\textemdash the ionizing flux, tidal potential, and stellar wind. In doing so we can illuminate how each physical process translates into observable properties of the outflow, and how one would expect those signatures to vary for different conditions. Of the previous simulations, the most complete calculation of stellar heating in 3-D was Tripathi et al. (\citeyear{Tripathi15}), which self-consistently calculated ionization and heating of the gas. \refemp{This self-consistent calculation is required to resolve the ionization structure of the planetary outflow, which is critical for synthetic observations. Otherwise, one must rely on previous work that has already done so for identical parameters, or assume approximate solutions for new parameters.} 

Yet, while the simulations of Tripathi et al. (\citeyear{Tripathi15}) did include tidal gravity, this study neglected the Coriolis force, a stellar wind, and magnetic fields. Expanding upon Tripathi et al. (\citeyear{Tripathi15}), we now seek to include stellar winds in a full-rotating frame. A concurrent study by Debrecht et al. (\citeyear{Debrecht18}) includes Coriolis force (but no stellar wind or magnetic fields) and focuses on varying planet mass and stellar flux. We defer exploration of magnetic fields, which should play a significant role in the outflow structure for fields above \SI{1}{G} (Owen and Adams \citeyear{Owen14}), to future work.  Though we focus on hot Jupiters, the concepts explored here should be applicable to a larger demographic, e.g., hot and warm gaseous planets.

An overview of the goals, model, and setup in this work are presented in \autoref{sec:model}. We provide a detailed description of the numerical methods and setup in \autoref{sec:nummethods}. Results are given in \autoref{sec:results}, followed by a discussion of observational consequences in \autoref{sec:obscon}. Future work and conclusions are discussed in \autoref{sec:conclusion}.

\section{Model}
\label{sec:model}

We aim to study the interaction between an escaping planetary atmosphere and its host star, in particular the star's stellar wind, gravitational force, and ionizing radiation. We will not consider atmospheres undergoing Roche lobe overflow or ablation by the stellar wind. Rather, our planetary winds will be self-consistently driven by the energy deposited, via photoionization heating, in the planet's upper atmosphere. To this end, we use radiative hydrodynamics to model the evolution of an atmospheric outflow.

To accurately track the evolution with hydrodynamics, it is necessary that the outflow remains in the collisional limit. We confirm this in post-processing with the evaluation of the Knudsen number

\begin{equation}
\textrm{Kn} = \frac{\lambda}{L},
\end{equation}

\noindent where $\lambda = (n \sigma_{\textrm{col}})^{-1}$ is the mean free path of  gas particles with volumetric number density $n$, $L = (\nabla \log{P})^{-1}$ is the characteristic length scale of the flow, and $P$ is the gas pressure. The collisional cross section, $\sigma_{\textrm{col}}$ is obtained by assuming the responsible collision mechanism for an individual population: Coulomb scattering for ion-ion interactions, and hardbody collisions for neutral-neutral or neutral-ion interactions.\footnote{Neutral-ion interactions are better modeled by induced dipole scattering and charge exchange, but we ignore those for simplicity and receive an upper bound for Kn.}

If one is only interested in the planetary evolution, e.g., mass-loss rate, it is sufficient for the outflow to remain collisional only to the sonic surface, where a Mach number of one is achieved.\footnote{As, neglecting any magnetic effects, conditions in a sonic region cannot propagate information to the subsonic region.} However, we additionally seek to model the large-scale interaction with the stellar environment, which is always governed by the ambient conditions\textemdash modeled here as a stellar wind. Moreover, there exist ambient conditions which impede the formation of a sonic surface, in which case the planetary evolution is also regulated by the ambient conditions. Therefore, we require the entire outflow to be sufficiently collisional to be properly modeled by hydrodynamics. Indeed, we find that within a few sonic radii, both the neutrals and the ions are well within the collisional limit. Farther out as the density decreases the neutrals become only marginally collisional, but are not dynamically or observationally significant precisely because they reach such low densities.

In \autoref{ssec:physics} we briefly describe all the physics taken into account in our models. For \autoref{ssec:bern} we discuss the usefulness of Bernoulli's constant for analyzing these winds. In \autoref{ssec:setup} we discuss the physical setup of the problem, and in \autoref{ssec:lscales} we give intuition for the relevant length scales in our flows.

\subsection{Physical processes}
\label{ssec:physics}

\subsubsection{Radiative transfer for radiative hydrodynamics}
\label{sssec:radhyro}

Our planetary winds are launched by photoionization heating from the host star. The frequency ($\nu$) dependent optical depth to ionizing photons along a given ray, parametrized by $s$, is given as

\begin{equation}
\tau_\nu(s) = \int_{s_0}^s \alpha_\nu(s') \, \textrm{d}s'.
\end{equation}

\noindent Here by definition $\tau_\nu(s_0) = 0$. In the work presented here, the opacities only come from neutral hydrogen absorption, i.e., $\alpha_\nu(s) = n_{\textrm{HI}}(s) \, \sigma_{\nu,\textrm{HI}}$, where $n_{\textrm{HI}}$ is the number density of neutral hydrogen. The near-ionization frequency-dependent cross section for photon absorption in neutral hydrogen is approximately (e.g., Draine \citeyear{Draine11}, \S\!~13.1)

\begin{equation}
\label{csaprx}
\sigma_{\nu,\textrm{HI}} \approx 6.3 \times 10^{-18} \left(\frac{\nu}{\nu_0}\right)^{-3} \si{cm^{2}}.
\end{equation}

\noindent For our simulations we implement the Verner et al. (\citeyear{Verner96}) analytic fits to get more accurate cross sections.

In our study, we will only consider \refemp{ionizing} monochromatic light, \refemp{without any radiation pressure}. Therefore, we will now drop any $\nu$ subscripts. Taking the optical depth from the star to the edge of our simulation to be negligible, we equate the incoming photon number flux of the simulation, $F_{0}$, with the photon number flux of the host star. Then the flux as a function of optical depth $\tau$ is

\begin{equation}
F = F_{0} * \textrm{e}^{-\tau}.
\end{equation}

\noindent The ionization rate is then given by

\begin{equation}
\mathcal{I} = \sigma_{\textrm{HI}} * n_{\textrm{HI}} * F_0 * \textrm{e}^{-\tau}.
\end{equation}

\noindent The rate of photoionization heating is the ionization rate times the energy of the photoelectron released per ionization

\begin{equation}
\mathcal{G} = \mathcal{I} * E_{\textrm{pe}}.
\end{equation}

\noindent Here $E_{\textrm{pe}} = h\nu - I_\textrm{H}$, where $h$ is Planck's constant and $I_\textrm{H} = \SI{13.6}{eV}$ is the ionization threshold energy of hydrogen.

Within a comoving fluid parcel neutrals are repopulated only via recombination. Since we only follow the direct stellar ionizing radiation field, and not the diffuse field generated by recombinations, we will adopt case-B recombination. This is the appropriate case when optically thick to ionization, as is the case where the wind is launched. Farther out in the flow, where the gas is almost completely ionized, this may be inappropriate. In spite of this, we will ubiquitously adopt case-B, making the recombination rate

\begin{equation}
\mathcal{R} = \alpha_\textrm{B} * n_{\textrm{HII}} * n_{e}.
\end{equation}

\noindent Here $n_{\textrm{HII}}$ is the number density of ionized hydrogen, $n_{e}$ is the number density of electrons, and we use an approximate form for $\alpha_\textrm{B} = 2.59 \times 10^{-13} * (T/\SI{e4}{K})^{-0.7} \si{cm^{3}.s^{-1}}$ (Osterbrock \citeyear{Osterbrock89}). Since the atmosphere is solely composed of hydrogen, we take the gas to be electrically neutral so that $n_{e} = n_{\textrm{HII}}$. We leave the question of metal cooling and entrainment in the winds for future work.

While our flows near the planet are certainly optically thick in \LymanA, it is argued in Appendix C of Murray-Clay et al. (\citeyear{Murray09}) that similar flows are sufficiently ``thin'' that \LymanA{} emission is scattered into the line wings and escapes before being thermalized back into the fluid via collisions. Therefore, in our cooling rates we consider both \LymanA{} emission from collisionally excited neutral hydrogen and radiative recombination emission

\begin{equation}
\mathcal{L} = \mathcal{L}_{\textrm{rr}} + \mathcal{L}_{\Lya}.
\end{equation}

The rate of energy loss from recombination is given by (Osterbrock \citeyear{Osterbrock89})

\begin{equation}
\mathcal{L}_{\textrm{rr}} \approx \Lambda_{\textrm{rr}} * k_{\textrm{B}} * T^{0.11} * n_\textrm{HII} * n_{e}.
\end{equation}

\noindent The constant $\Lambda_{\textrm{rr}} = \SI{6.11e-10}{K^{0.89}.cm^{3}.s^{-1}}$, and $k_\textrm{B}$ is the Stefan-Boltzmann constant. The rate of energy loss from Lyman-$\alpha$ is given by (Black \citeyear{Black81})

\begin{equation}
\mathcal{L}_{\Lya} = \Lambda_{\Lya} * n_{\textrm{HI}} * n_{e} * \textrm{e}^{\unit[-118348]{K}/T}.
\end{equation}

\noindent Here $\Lambda_{\Lya} = \SI{7.5e-19}{erg.cm^{3}.s^{-1}}$. Both the \LymanA{} and recombination cooling are temperature-dependent, where the temperature is calculated as

\begin{equation}
T = \frac{P}{n * k_\textrm{B}}. 
\end{equation}

\noindent The variable $n$ is the total number density of all species, $n = \sum_s n_s = n_{\textrm{HI}} + n_{\textrm{HII}} + n_{e}$.

\subsubsection{Fluid equations}
\label{sssec:fluid}

In our simulations we solve the conservative form of the fluid equations:

\begin{equation}
\label{masscont}
\frac{\partial \rho}{\partial t} + \vec{\nabla} \cdot \left(\rho \vec{u} \right) = 0,
\end{equation}

\begin{equation}
\label{momcont}
\frac{\partial \rho \vec{u}}{\partial t} + \vec{\nabla} \cdot \left(\rho \vec{u} \otimes \vec{u} + \overline{\overline{P}} \right) = - \rho \vec{\nabla}\phi - 2 * \rho \left(\vec{\Omega} \times \vec{u}\right),
\end{equation}

\begin{equation}
\label{Econt}
\frac{\partial E}{\partial t} + \vec{\nabla} \cdot \left( \left( E + P \right) \vec{u} \right) = - \rho \vec{u} \cdot \vec{\nabla}\phi + \mathcal{G} - \mathcal{L},
\end{equation}

\begin{equation}
\label{HIcont}
\frac{\partial n_{\textrm{HI}}}{\partial t} + \vec{\nabla} \cdot \left(n_{\textrm{HI}} \vec{u} \right) = \mathcal{R} - \mathcal{I},
\end{equation}

\begin{equation}
\label{eos}
E = \frac{1}{2} \rho * \vec{u} \cdot \vec{u} + \frac{P}{\gamma -1}.
\end{equation}

\noindent Here, $\rho$ is the mass density of the gas, $\vec{u}$ is the bulk velocity, $\overline{\overline{P}}$ is the isotropic pressure tensor with scalar value $P$, $\phi$ is the mechanical potential, $\vec{\Omega}$ is the frame rotation vector, and $\gamma$ is the adiabatic index, for which we will adopt $\gamma = 5/3$. Recall that $\vec{u} \otimes \vec{u}$ is the outer product of the velocity with itself, sometimes written as the dyadic product $\vec{u} \vec{u}$.

\Crefrange{masscont}{Econt} have the familiar conservative forms on the left-hand side, with the relevant source terms for our problem on the right-hand side. For mass continuity there are no sources, momentum sources are external forces arising from the potential and the Coriolis force, and for energy there is the change from advecting through a potential field along with the energy gained and lost from radiative processes. Recall that the Coriolis force can do no work. Note that since the centrifugal force can be expressed as the gradient of a scalar potential, we place it in our mechanical potential, $\phi$, as discussed in \autoref{sssec:frame}. The continuity equation for neutrals, \cref{HIcont}, is similar to that for total density with recombination as a source and ionization as a sink. Note that in \cref{eos} we have used an ideal equation of state for a perfect gas.

\subsection{Bernoulli constant}
\label{ssec:bern}

Bernoulli's constant is a useful tool for analyzing our simulations, picking a stellar wind, and setting up the initial conditions of the atmosphere. Lamers and Cassinelli (\citeyear{Lamers99}, \S\! 4.1.1) provide a derivation of the Bernoulli constant for a spherically-symmetric wind; in \autoref{apnss:revbern} we provide a generalized derivation along any streamline. For a reversible ideal gas the Bernoulli constant is

\begin{equation}
\label{BernHeat}
b = \frac{1}{2}u^2 + h + \phi - \Delta q,
\end{equation}

\noindent where $u$ is the bulk velocity, $h$ is the enthalpy, $\phi$ is the mechanical potential and $\Delta q$ is the heat added to the fluid (see \autoref{apn:bern}). The heat flow along the path is given by the total local heating rate, $\mathcal{G}-\mathcal{L}$,  and therefore $\Delta q$ can be expressed as

\begin{equation}
\Delta q = \int_{C(s)} \frac{\mathcal{G}-\mathcal{L}}{\rho * u} * \textrm{d} s,
\end{equation}

\noindent where $\rho$ is the density and $C(s)$ is the streamline, parameterized by $s$.

Something that immediately becomes apparent from the Bernoulli constant is which parts of the domain are energetically forbidden. Consider a system in which there exists a surface defined by $\phi_{z} \equiv b + \Delta q$. At this surface, the kinetic energy and enthalpy have gone to zero (hence the subscript ``z'' in $\phi_{z}$), and all the energy is necessarily in the potential energy. Thus, the fluid is bounded by this surface\textemdash the absolute-zero-velocity surface. Therefore, a condition for an unbounded flow is that $(b + \Delta q) > \phi$ at all points along its streamline. We use this criterion to help determine our initial conditions, picking only bounded planetary atmospheres and unbounded stellar winds.

When a flow is unbounded it is meaningful to talk about its asymptotic velocity. For frames where both $\phi \rightarrow 0$ and $h \rightarrow 0$ as $r \rightarrow \infty$, the asymptotic velocity can be calculated as

\begin{equation}
\label{termv}
u_{\infty} = \sqrt{2 \left(b + \Delta q_\infty \right)}.
\end{equation}

\noindent For an ideal gas, $h \rightarrow 0$ is equivalent to $T \rightarrow 0$.  In practice, as long as the flow is supersonic as $r \rightarrow \infty$, \cref{termv} is approximately correct as $u_\infty^2 \gg h_\infty \approx c_{\textrm{s},\infty}^2$, where $c_{\textrm{s}, \infty}$ is the sound speed as $r \rightarrow \infty$. Here, the notation $\Delta q_\infty$ reminds us that it includes all the energy injected into the wind out to infinity, where we have assumed that the differential heat flow is zero.

\subsection{Physical setup}
\label{ssec:setup}

\subsubsection{Reference frame}
\label{sssec:frame}

We place our planet on a circular orbit and adopt a rotating reference frame in which the planet and star are at fixed locations. This frame centered on the barycenter, has the rotation vector given by the Third law of Kepler 

\begin{equation}
\label{omega}
\vec{\Omega} = \sqrt{\frac{G*(M_\textrm{p}+M_\star)}{a^3}} \, \hat{z}.
\end{equation}

\noindent Here $G$ is the Newtonian gravitational constant, $M_\star$ is the primary mass, $M_\textrm{p}$ is the secondary mass, $a$ is the semi-major axis of the secondary's orbit around the primary, and the $z$\textendash axis is the axis of rotation. In such a frame the static potential is given by

\begin{equation}
\label{mechpot}
\phi = - \frac{G*M_\textrm{p}}{r} - \frac{G*M_\star}{r_\star} - \frac{1}{2} \Omega^2 * r_\perp^2,
\end{equation}

\noindent where, for a given point, $r$ is the distance to the secondary, $r_\star$ is the distance to the primary, and $r_\perp$ is the distance to the barycenter projected into the orbital plane.

\subsubsection{Initial Planetary atmosphere setup}
\label{sssec:patm}

Within this reference frame we place a point mass to simulate the planet's core, on top of which sits an atmosphere. As the wind is launched in the upper atmosphere of a planet, the gas is relatively dilute and warm enough to be well described as an ideal gas. We ignore any viscous dissipation, and take all processes to be reversible. Absent any external energy input, the most stable solution of an atmosphere will be an adiabat. We thus construct an isentropic atmosphere that satisfies the polytropic relationship, $P = K \rho^\Gamma$, with an polytropic exponent, $\Gamma$, equal to the heat capacity ratio, $\gamma$.

In reality the bolometric flux from the star drives the upper atmosphere of a planet to be approximately isothermal at the skin temperature (See \S\! 3.6 of Pierrehumbert \citeyear{Pierre10}). Due to a lack of incorporating the bolometric flux in our simulation, and since $\Gamma \approx 1$ is not an appropriate approximation throughout our simulation domain, e.g, in the ionized outflow, we use an isentropic atmosphere with the temperature at the base of the wind equal to the planet's skin temperature.

\begin{figure}
\label{fig:setup}
\includegraphics[width=\columnwidth]{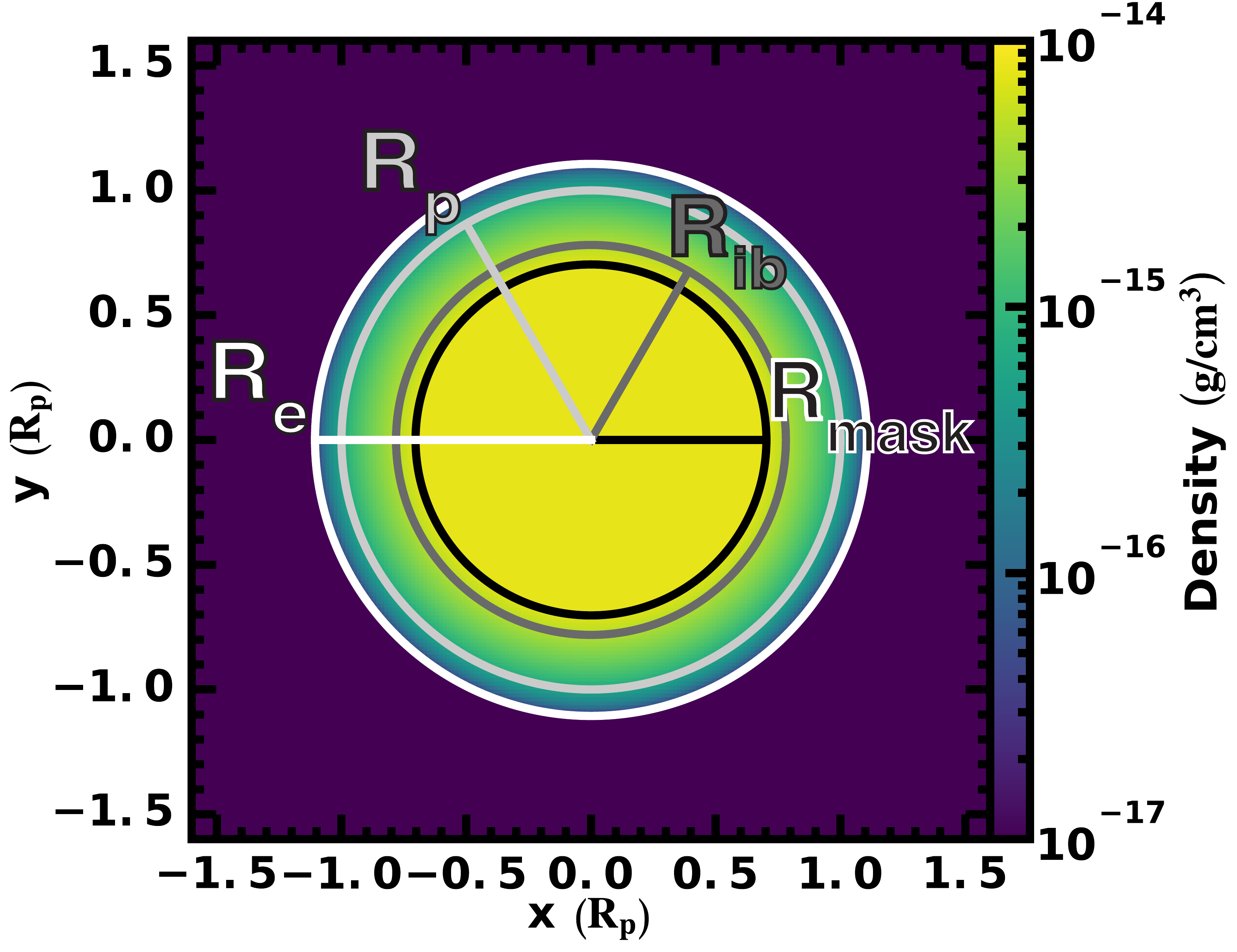}
\caption{The initial density profile of a hydrostatic atmosphere. Within the innermost circle, $r < R_{\textrm{mask}}$, the density is held to a constant value, $\rho_{\textrm{mask}}$. Between $R_{\textrm{ib}}$ and $R_{\textrm{mask}}$ the density is held fixed at every time step to its original analytic hydrostatic solution. The radius of the planet, $R_{\textrm{p}}$, is where $\tau=1$ initially. The outermost circle, $R_{\textrm{e}}$, is the edge of the atmosphere where the density tends to zero, past which is a low density, pressure matched ambient medium.}
\end{figure}

Our atmosphere is contained well within the planet's Hill sphere, so for now we can ignore stellar gravity when calculating its analytical initial profile. \refemp{Note that in practice we will use the full potential so that the atmosphere is not technically spherically symmetric, but for simplicity of discussion we will assume such symmetry.\footnote{See \autoref{fig:setup}, which shows the solution for the full non-spherical potential. Yet, by visual inspection the atmosphere is virtual spherically symmetric.}} Furthermore, we consider planets that are tidally locked to their host stars, and will ignore any rotational effects tending to make the atmosphere oblate. Under these conditions we require the atmosphere to be in hydrostatic equilibrium. We ignore the gravitational effect of the gas itself, both analytically and numerically in our simulations.\footnote{\label{fn:polytrope}Thus while the atmosphere has a polytropic equation of state, it is not a polytrope as it is not a solution to the Lane-Edmen equation, \textit{i.e.} no self-gravity, and there are an infinite number of analytic solutions.} From %
\crefmultiformat{equation}{#2Eqs.~(#1)#3}{, #2(#1)#3}{, #2(#1)#3}{, #2(#1)#3,}%
\cref{BernHeat,eq:idealEnthalpy},%
\crefmultiformat{equation}{#2Eqs.~(#1)#3}{ and~#2(#1)#3}{, #2(#1)#3}{, and~#2(#1)#3} %
and the adiabatic equation of state, an adiabatic atmosphere that is everywhere static has a density profile

\begin{equation}
\label{isenatm}
\rho(r) = \rho_\textrm{p} \left[1 +  \frac{\left( \phi_\textrm{p} - \phi(r) \right)}{h_\textrm{p}}  \right]^{1/(\gamma-1)}.
\end{equation}

\noindent Here variables with subscript ``p'' denote their value at the planetary surface, $R_\textrm{p}$, which we define to be the radius where $\tau = 1$. We emphasis that, $\tau = 1$ at $R_{\textrm{p}}$ is only the initial condition at time step zero in our simulations of our atmosphere. The optical depth to ionizing photons at $R_{\textrm{p}}$ is not fixed during the simulation, as the wind self-consistently picks its base. We pick the atmosphere's Bernoulli constant such that the atmosphere is bound, which is found by solving the equation $\phi = \phi_\textrm{p} + h_\textrm{p}$.\footnote{Thus the relevant parameters are the radius, mass and skin temperature of the planet. For our simulations we pick parameters of numerically convenient hot Jupiters, see \autoref{sssec:initconds} for more details.} Since our atmosphere is (nearly) spherically symmetric we shall call this the zero radius, $R_\textrm{z}$, as all variables go to zero at this surface. This is often called the homentropic atmosphere since it is isentropic, with constant entropy along rays, and is spherically symmetric\textemdash thus, of ``the same entropy'' everywhere throughout. For an ideal gas in a point mass potential this is equivalent to

\begin{equation}
\label{dProfVars}
\rho(r) = \rho_\textrm{p} \left[1 + \frac{(\gamma-1)*G*M_\textrm{p}}{\gamma * c_{\textrm{s,p}}^2} \left(\frac{1}{r} - \frac{1}{R_\textrm{p}} \right) \right]^{1/(\gamma-1)}.
\end{equation}

\noindent Here $c_{\textrm{s,p}}$ is the isothermal speed of sound at $R_\textrm{p}$.\footnote{Notice that from this equation, and the limit definition of $e$, the $\lim_{\gamma \rightarrow 1} \rho(r) = \rho_{\textrm{p}} \exp(*G*M_\textrm{p}/c_{\textrm{s,p}}^2*(1/r-1/R_\textrm{p}))$, which is the profile of an isothermal atmosphere.}

Numerically fixed inner-boundary conditions, which are extrapolated from \cref{isenatm} with the initial conditions at $R_\textrm{p}$, are set at an inner boundary radius. The inner boundary, $R_{\textrm{ib}}$, is deep enough within the atmosphere that conditions there do not affect the outflow. \refemp{While we setup an initial atmosphere with predetermined conditions, the stellar flux boundary condition self-consistently evolves the atmosphere above $\tau = 1$. Therefore, below $\tau = 1$ our atmosphere is relatively static, and our model should not necessarily be considered valid there\textemdash even though it is far outside our inner boundary. Justification for the insensitivity of the solution above $\tau = 1$ to what lies below is given in Appendix A of Murray-Clay et al. (\citeyear{Murray09}).}

See \autoref{fig:setup} for the initial atmosphere with labeled radii. We note that our choice of $R_\textrm{p}$ as the radius at which $\tau = 1$ is not the same as the radius of the optical surface of the planet (as probed, for example, in optical transit measurements).  We make this choice for $R_\textrm{p}$ because our model is only valid beyond this radius.  For typical hot Jupiters, the difference between the radii of the optical surface and the $\tau = 1$ surface to ionizing photons is of order \SI{10}{\percent} (Murray-Clay et al. \citeyear{Murray09}). 

\subsubsection{Ambient medium setup}
\label{sssec:amb}

Ideally, conditions outside of the atmosphere would be those of a typical interplanetary medium, which for most star systems takes the form of a stellar wind. However, as we wish to break the stellar environment down into its components, in some simulations we do not include a stellar wind.  When the stellar wind is absent, we instead use an ambient medium that is as low-density as numerically feasible and that is pressure matched to the outer edge of the planetary atmosphere. For numerical reasons we truncate the planetary atmosphere just prior to the zero radius \refemp{(defined in \autoref{sssec:patm})} at a radius we call the atmospheric edge, $R_\textrm{e}$. 

Initially it is important that the ambient medium is both pressure matched to the numerical edge of our atmosphere, and that the ambient medium is itself pressure supported so it does not collapse onto the planet. This is accomplished with another hydrostatic atmosphere similar to \cref{isenatm}

\begin{equation}
\label{amb1}
\rho(r) = \rho_{\textrm{a}} \left(1 + \frac{\phi_{\textrm{a}} - \phi(r)}{h_{\textrm{a}}} \right)^{1/(\gamma-1)}, \qquad r > R_{\textrm{e}}.
\end{equation}

\noindent Here the subscripts ``a'' denotes an ambient medium reference variable, where the reference location is the ambient medium/planetary atmosphere interface ($R_{\textrm{e}}$).

By design the ambient medium's reference density $\rho_{\textrm{a}} < \rho_{\textrm{e}}$\textemdash the density of the atmosphere at $R_{\textrm{e}}$. Thus by also requiring pressuring matching at the interface ($\phi_{\textrm{a}} = \phi_{\textrm{e}}$), $P_{\textrm{a}} = P_{\textrm{e}}$, so that $h_{\textrm{a}} > h_{\textrm{e}}$. Since $\phi_{z,a} = \phi_\textrm{a} + h_\textrm{a}$ determines the ambient absolute-zero-velocity surface, then the zero radius of the ambient medium is larger than the planetary atmosphere's bounding surface.\footnote{Since $\phi_{z,\textrm{a}} = \phi_{\textrm{a}} + h_{\textrm{a}}$ and $\phi_{z,\textrm{p}} = \phi_{\textrm{p}} + h_{\textrm{p}} = \phi_{\textrm{e}} + h_{\textrm{e}}$, but $h_{\textrm{a}} > h_{\textrm{e}}$ and $\phi_{\textrm{a}} = \phi_{\textrm{e}}$, then $\phi_{z,\textrm{a}} > \phi_{z,\textrm{p}}$ or $R_{z,\textrm{a}} > R_{z,\textrm{p}}$.\label{fn:azvspz}} Due to numerical considerations, a balancing act between too high or too low of values for $h_{\textrm{a}}$ (or relatedly $\rho_{\textrm{a}}$) occurs and is discussed in \autoref{apn:ambient}.

\subsubsection{Stellar wind setup}
\label{sssec:stellar}

When we seek to include the interplanetary medium in the form of a stellar wind, we \refemp{set a stellar wind inflow boundary condition} and initialize our domain with that of a steady-state stellar wind instead of an ambient medium. One challenging aspect of our simulation is the injection of a realistic stellar wind. As our energy deposition into the planetary atmosphere is detailed, we use an adiabatic index of $\Gamma = \gamma = 5/3$ throughout the entire simulation. This differs from numerous previous simulations \refemp{only} in that we have no ``hidden'' energy injection or ad hoc redistribution built into our adiabatic index, such as an isothermal Parker Wind with $\Gamma \approx 1$ (\refemp{e.g., Stone \& Proga \citeyear{Stone09}; Tremblin \& Chaing \citeyear{Tremblin13}; Caroll-Nellenback et al. \citeyear{Carroll17}}). This means our fluid behaves isentropically in the absence of ionizing radiation. 

\refemp{Typically, stellar wind models that do not resolve the heating sources use such ``polytropes'' ($\Gamma \approx 1$).} Unfortunately for our simulation, as discussed in \autoref{apnss:polybern}, there is no transonic wind solution for a polytropic fluid with polytropic index $\Gamma = 5/3$. \refemp{Note that, stellar wind models which do resolve heating terms also use $\Gamma = \gamma = 5/3$, but include things such as: heat conduction, Alfv\'en wave dissipation, resistive and viscous dissipation, or coronal heating sources (Miki\'c et al. \citeyear{Mikic99}).} As our only energy injection is from ionization, we cannot generate an optically thin transonic stellar wind without another energy injection method, e.g., via magnetic fields as seen in Alfv\'en-driven winds (Lamers \& Cassinelli \citeyear{Lamers99}, Ch.\! 10). Fortunately, by using the Bernoulli constant we can set the boundary conditions to induce a stellar wind that within the domain of our simulation mimics a transonic wind locally. The only catch is that the stellar wind is always sub- or supersonic \refemp{(see \autoref{sssec:initconds} for more details)}.

Therefore, our stellar wind is modeled as a spherical isentropic wind,\footnote{\refemp{Necessarily due to the fact that there is no energy injection mechanism for the optically-thin stellar wind, and that the entire work done in our hydrodynamic simulation is reversible.}} \refemp{i.e., a ``polytrope'' with index $\Gamma = \gamma$ (see \cref{fn:polytrope} and \autoref{apnss:polybern})}, for which the velocity profile is derived in \autoref{apnss:polybern}. \refemp{Note that spherical symmetry may not be realistic, even at close distances, but we will make this assumption regardless (Vidotto et al. \citeyear{Vidotto18}).} Therefore, we use \cref{analvel} with $\Gamma = \gamma$ to evaluate the stellar wind's velocity structure. Notice that since $u(r)$ is independent of $\rho$. Therefore, given a $u_{\star,0}$ and $T_{\star,0}$ at $r_{\star,0}$ (a reference radius not necessarily equal to the stellar radius), we can adjust the total pressure of the stellar wind by scaling the stellar proton number density, $n_{\star,0}$, while leaving all other profiles unaltered

\begin{equation}
\label{pressureScaling}
\begin{split}
P_{\star,\textrm{total}}(r) &= \rho_\star u_\star^2 + P_\star \\
&=  n_{\star,0} \Bigg( (m_{\textrm{HII}}+m_e)* \left( \frac{u_{\star,0} \, r_{\star,0}^2}{u(r) \, r^2}\right) \, u(r)^2 \\
& \qquad\qquad\qquad + 2*k_B T_{\star,0} \left( \frac{u_{\star,0} \, r_{\star,0}^2}{u(r) \, r^2}\right)^\gamma \Bigg).
\end{split}
\end{equation}

\noindent Here $m_e$ is the mass of an electron and $m_{\textrm{HII}}$ is the mass of an ionized hydrogen. Note that there is a factor of two on the thermal pressure to account for both the protons and electrons. Thus, we only need to alter $n_{\star,0}$ to tune the stellar wind strength. Controlling $n_{\star,0}$ grants us a handle on where the bow-shock interface of the planetary and stellar winds occurs, and on whether the planetary outflow is a wind or breeze. Moreover, this allows us to use the same velocity and temperature profile, and therefore Mach profile, across all of our stellar winds.

\refemp{We note that realistic stellar winds do not merely differ between one another by a density scaling, and that we are only probing a small area of the possible phase space of stellar winds. In this study we focus on the effects of a confining stellar wind, and therefore only require a handle on the total wind pressure to make comparisons between wind strengths. Studies probing charge exchange, or stellar magnetic fields, will require more realistic modeling of stellar winds. For a review of realistic stellar-wind numerical modeling see Gombosi et al. (\citeyear{Gombosi18}).} 

\subsection{Length scales}
\label{ssec:lscales}

To determine an appropriately sized domain for any simulation, it is important to understand the length scales of the problem to ensure that everything can be captured within the domain. Additionally, length scales offer insight into the resulting structures and their origins. For our problem, the important scales for the outflow are: the outflow's scale height (the importance of which has already been discussed in \autoref{sec:model}), the optical depth one surface for ionizing photons, the sonic point where the planetary outflow transitions from subsonic to supersonic, the planet's Hill radius, the Coriolis length of the planetary outflow, and the bow shock radius where the planetary and stellar winds collide.

\textit{Optical depth one to ionizing photons}: While $\tau = 1$ along the substellar ray was initially set at $R_\textrm{p}$, we allow the simulation to self-consistently choose where $\tau = 1$ as the flow evolves. This will depend on the incoming ionizing flux, recombination rate, and advection of neutrals in the planetary outflow. However, one can still chose $n_{\textrm{p}}$ such that $\tau = 1$ ends up near $R_{\textrm{p}}$. This choice depends on the strength of the ionizing flux, and is briefly discussed in \autoref{apn:sheight}.

\textit{Sonic point}: The sonic point is where the outflow goes transonic ($|\vec{u}| = c_{\textrm{s}}$ and $\vec{u} \cdot \vec{\nabla} |\vec{u}| > 0$). However, for a self-consistently launched outflow with ionization heating its location is difficult to know a priori. For reference, the sonic point radius for an isothermal Parker wind is $r_\textrm{s,iso} = G*M_\textrm{p}/c_{\textrm{s}}^2$. This expression can be used to approximate the location in a self-consistently launched outflow by evaluating $c_{\textrm{s}}$ using the temperature is at the base of the wind, $T \sim \SI{e4}{K}$. For a fuller understanding of how the sonic point depends on ionization heating see the discussion in \S\! 2.2.3 of Murray-Clay et al. (\citeyear{Murray09}).

\textit{Effective Hill radius}: The Hill radius, where stellar tidal gravity balances the planet's gravity, is well approximated as

\begin{equation}
\label{eq:hill}
r_\textrm{H} = a  \sqrt[\leftroot{-1}\uproot{2}\scriptstyle 3]{\frac{M_\textrm{p}}{3M_\star}}.
\end{equation}

\noindent However, the Hill radius is derived for particles that do not experience a pressure force. Just as the pressure force causes gas in a protoplanetary disk to orbit at sub-Keplerian velocities by effectively reducing the central mass, here the pressure force reduces the Hill radius by effectively reducing the planet's mass. We call this radius where the planet, tidal, and pressure forces balance the ``effective Hill radius,'' which will be interior to \cref{eq:hill} for monotonically decreasing pressure profiles. 

\textit{Coriolis length}: Now imagine a ballistic particle only experiencing the Coriolis force, $\vec{a} = -2\vec{\Omega} \times \vec{v}$. The particles moves in a circle with a period of $t = \pi/\Omega$, since the frequency of the acceleration is twice the frame rotation, $2\Omega$, and a full revolution is $2 \pi$. Recall that the Coriolis force does no work, $ \left(-2\vec{\Omega} \times \vec{v}\right) \cdot \vec{v} = 0$, and only serves to transfer momentum between coordinate axes. Therefore, we arbitrarily define when a particle has been significantly affected by the Coriolis force as when its initial momentum has been deflected by one radian, i.e., it has traversed one radius\textemdash the length scale of the circle. Thus the time scale on which particles are significantly affected by the Coriolis force is $(2 \Omega)^{-1}$. Given the average velocity over that trajectory, $\langle v \rangle$, the Coriolis length scale is

\begin{equation}
\label{eq:CoriL}
L_{\Omega} = \frac{\langle v\rangle}{2 \Omega}.
\end{equation}

\noindent One could similarly define the Coriolis length to be the scale on which the outflow reaches a Rossby number of one, $\textrm{Ro} \equiv v/(2 \Omega L_{\Omega})$. For an outflow with its sonic point interior to the Hill sphere, $\langle v \rangle$ is on the order of $u_{\infty}$ from \cref{termv}.

\textit{Bow shock}: The bow shock radius, $R_{\textrm{bow}}$ occurs where there is a pressure match between the stellar wind and the planetary outflow

\begin{equation}
\label{eq:bowshock}
\left[ P_\star + \rho_\star u_{\star,\perp}^2 = P + \rho u_{\perp}^2 \right]\rvert_{R_{\textrm{bow}}}.
\end{equation}

\noindent Here the ``$\star$'' subscripts denote the stellar wind, which is not constant throughout the domain. Subscript ``$\perp$'' denotes the normal component of the velocity to the bow shock interface. As it is the stellar wind that \refemp{shrouds} the planet, the bow shock is roughly spherical with respect to the planet's origin between the planet and star, but asymptotes to a radial line far past the planet.\footnote{See Figure 10 of Murray-Clay et al. (\citeyear{Murray09}) for a cartoon of the geometry.} Since we can analytically solve for the stellar wind structure, we can numerically solve for the location of the standoff radius given the numerical structure of the planetary wind in the absence of a stellar wind. We note that because the shocked stellar wind does not radiate effectively, the width of the shocked stellar wind region can be substantial, and care must be taken to choose a box size large enough to enclose the shock.

\section{Numerical Methods}
\label{sec:nummethods}

\subsection{\textnormal{\texttt{Athena}}}
\label{ssec:ATHENA}

To solve the model described in \autoref{sec:model} we use the publicly available magnetohydrodynamics code \texttt{Athena v4.2} (Stone et al. \citeyear{Stone08}). This Eulerian code has been rigorously tested and highly parallelized, making an ideal starting point to solve our radiative-hydrodynamic model. Two additional packages are utilized to add physics beyond the ideal fluid equations. The first is the ionization package from Krumholz et al. (\citeyear{Krumholz07}), updated to incorporate static mesh refinement (SMR) for plane-parallel ionization (Tripathi et al. \citeyear{Tripathi15}). The radiative transfer in this package is operationally split from the hydrodynamic update and is done by radiative sub-cycling between hydro time steps. Secondly, \texttt{Athena}'s shearing-box physics package (Stone et al. \citeyear{Stone10}) is used to implement the Coriolis force, as described in \autoref{sssec:Amods}.

Standard fluid algorithmic choices are as follows. We use the piecewise-parabolic method, a third-order spatial reconstruction method, for reconstructing the fluid variables at cell interfaces. For the Riemann problem of the interface fluxes, Roe's linearized solver is used. Our integrator is the 3-D directionally unspilt corner transport upwind (CTU) scheme. Static mesh refinement is used around the planet to ensure scale heights within the fluids are well resolved. As discussed in Tripathi et al. (\citeyear{Tripathi15}), we use the H\textendash correction algorithm to avoid the carbuncle instability from the wind's convergences on the nightside.

\subsubsection{Modifications and use of \textnormal{\texttt{Athena}}'s features}
\label{sssec:Amods}

In order to run the simulations successfully and to improve their accuracy, we modify the default \texttt{Athena} code and implement a few non-standard features. First, we added first-order flux correction to the corner transport upwind (CTU) integrator, using the same method as already implemented in \texttt{Athena}'s van Leer integrator (Beckwith and Stone \citeyear{Beckwith11}). This scheme detects when calculation of the flux at higher order leads to a negative density or pressure, and self-consistently redoes the flux calculation at the boundary of each affected cell using more diffusive, and hence more stable, first-order fluxes. Resorting to first-order fluxes turns out to only be necessary initially while the wind as launching, for similar reasons as discussed below regarding our prolongation slope limiter changes. 

Second, we modified \texttt{Athena}'s shearing box physics feature so that the Coriolis force without a centrifugal term can be used with non-periodic boundary conditions. To do so without adding additional fictitious forces one can set the shear parameter to zero, $q = -\partial{\log \Omega}/\partial{\log r} = 0$. A shear parameter of zero suggests solid body rotation, for which central forces are balanced everywhere and the only force felt is the Coriolis force. Thus the only fictitious force the shearing-box module adds is the Coriolis force, while the tidal forces are taken care of in the static potential without any of the usual shearing-box approximations. Additional steps need to be taken as the shearing-box approximation assumes periodic boundary conditions, with \texttt{Athena}'s shearing-box feature hardcoded to remap boundaries without consideration to the user-input boundaries.

As some simulations are in a rotating frame, issues can arise when a flow is bending near a boundary. To prevent unphysical inflow from being extrapolated from the boundary conditions, we use what are sometimes referred to as dipole boundary conditions. That is to say, we use the standard outflow boundary conditions when the bulk velocity normal to the boundary is outwards, but restrict the mass inflow to the initial ambient medium when the normal velocity is inwards. While this sufficiently reduces undesired inflow, there are still reflections at the boundary that are not damped and can lead to oscillations. While these oscillations are present in our simulations, they occur where the density and pressure are orders of magnitude smaller than those around the planet in the domain of interest. Note that these are not standard \texttt{Athena} boundary conditions, and are implemented as user defined boundary conditions.

Next, we also found it necessary to add an additional slope limiter to \texttt{Athena}'s prolongation operator as applied at SMR boundaries. Since radiative cooling in our simulations is rapid, when the ionization front in the planetary atmosphere first expands it is characterized by a very large density jump. When this structure passes over SMR boundaries, the prolonged density or pressure can be negative unless it is appropriately limited. We therefore limit the slope in each direction to be less than 4/3,\footnote{This limit comes from the cell-centered distances between child cells and their parents, $1/4$, and the number of dimensions, 3.} which is sufficient to ensure that the prolonged quantities remain positive definite. This limit only modifies the simulation during the early phase of evolution when the outflow is expanding, and does not apply in the steady-state configuration reached at later times as this structure has passed over all of our SMR level boundaries.

Finally, we made minor improvements to Krumholz et al.'s (\citeyear{Krumholz07}) ionization package by modifying the temperature calculation to produce higher accuracy. Our modified version of \texttt{Athena}, other tools used in the production of this paper, a version of this paper with embedded movies, and more are freely available for download and use at our GitLab repository.\footnote{\href{https://gitlab.com/athena_ae/athena_ae}{https://gitlab.com/athena\_ae/athena\_ae}}

\subsection{Domain setup}
\label{ssec:dsetup}

\subsubsection{Reference frame}
\label{sssec:reframe}

All simulations have the planet at the origin, and depending on the simulation, we adopt either an inertial or a non-inertial rotating reference frame. By using a frame centered on the secondary we can use a smaller domain, which keeps both total computational expense and error from using plane-parallel radiation to a minimum. For rotating frames, let the primary be located at $\vec{r}_\star = -a \, \hat{x}$ and the barycenter at $\vec{r}_\textrm{b} = -a * M_\star/(M_\star+M_\textrm{p}) \, \hat{x}$, so that the potential in \cref{mechpot} is recast as

\begin{equation}
\label{rotpot}
\phi(\vec{r}) = - \frac{G*M_\textrm{p}}{|\vec{r}|} - \frac{G*M_\star}{|\vec{r}-\vec{r}_\star|} - \frac{1}{2} \Omega^2 * |\vec{r}_\perp-\vec{r}_\textrm{b}|^2.
\end{equation}

The first term is included in all of our simulations as the planet is always present, while the second and third term are only included in simulations where tidal potential is considered. In \texttt{Athena} this is implemented by setting the static potential function to \cref{rotpot}. This allows us to examine the tidal forces aspect of a rotating frame without considering the Coriolis force, which can be included with the shearing-box package described in \autoref{sssec:Amods}.

\subsubsection{Fluid initial conditions}
\label{sssec:initconds}

Within the simulation is a planetary atmosphere surrounded by an ambient medium. The parameters that characterize the planet and its atmosphere are the planet's mass, $M_{\textrm{p}}$, and radius, $R_{\textrm{p}}$, and the temperature, $T_{\textrm{p}}$, at $R_{\textrm{p}}$. While the mass and radius can be treated as free parameters used to model any desired planet, to launch a wind from ionization heating the planet's temperature must be well below that of the launched wind (Appendix A of Murray-Clay et al. \citeyear{Murray09}). Thus the temperature sets a physical scale within the problem ($T_{\textrm{p}} \ll \SI{e4}{K}$ for hot Jupiters), making planets with larger escape speeds or smaller wind temperatures, which have a smaller scale height to planetary radius ratio, more challenging to simulate.

Other atmospheric quantities can be calculated from these primary parameters and certain assumptions. One such assumption\textemdash accurate for ionizing fluxes studied here\textemdash is that planetary escape occurs in the energy-limited regime rather than the recombination-limited regime, as described in Murray-Clay et al. (\citeyear{Murray09}). Therefore, the surface of optical depth unity to ionizing photons does not move appreciably between the start of the simulation and when the wind has reached steady state. As the initial isentropic atmosphere's density and pressure scale heights are independent of the particle number density, we can always set the location of the optical depth unity surface to be at $R_\textrm{p}$ by scaling the number density. We do so by setting

\begin{equation}
\label{eq:planetn}
n_\textrm{p} = \frac{1}{\sigma_{\textrm{HI}} * \widetilde{H}(R_\textrm{p})},
\end{equation}

\noindent where $\widetilde{H}$ is a scale height defined in \autoref{apn:sheight}. Otherwise, in the recombination-limited regime one could balance recombination with ionization, to get a proper number density such that detailed balance is achieved near $R_\textrm{p}$.

Setting aside the temperature, a planet's mass and radius will determine the scale height and in turn the simulation resolution required. For an isentropic atmosphere, the scale heights become infinitesimally small near the edge of the atmosphere and at the core. This does not matter near the atmospheric edge as we truncate before reaching the zero radius, and once the simulation begins to run the edge will be replaced by a wind in any event. However, the singularity near the core requires that we adopt an inner boundary in our simulation to avoid numerical difficulties.

To set the inner boundary, $R_\textrm{ib}$, we require a few scale heights between $\tau = 1$ and $R_{\textrm{ib}}$, so that the base of the wind will be self-consistently found without interference from the inner boundary. Here we define a scale height to be one e\textendash folding in density (see \cref{Fdef}), and choose $R_\textrm{ib}$ to be two scale heights below the initial atmosphere's $\tau = 1$. We then set $R_\textrm{mask}$ five cells below $R_\textrm{ib}$ (number of ghost cells plus one), within which we hold the all variables to the same fixed value. By resetting all the cells between $R_\textrm{ib}$ and $R_\textrm{mask}$ at every time step to their hydrostatic solution, we create an internal boundary condition, as no information within $R_\textrm{mask}$ can propagate out by construction. By using a masking radius instead of a softening radius (i.e., a Plummer radius) in our potential we avoiding introducing artificial errors in our potential. 

Past the numerical edge of the planetary atmosphere is the ambient medium. When the ambient medium is not supposed to represent a stellar wind, our goal is to minimize its impact on the planetary wind. The detailed structure of our ambient medium is described in \autoref{apn:ambient}, but it suffices to say that by using a low density, initially pressure-matched, ambient medium, we prevent infall onto the atmosphere and stop the planetary wind from entering a snowplow phase. When we do include a stellar wind, we initialize the ambient medium with the velocity structure given by \cref{analvel}. With the velocity structure and a stellar mass-loss rate we can then calculate the density and temperature profiles of the wind with \cref{eq:Amcont,eq:ATprof} respectively. To indefinitely sustain the wind, the same formulation used to refresh the planet's hydrostatic lower atmosphere with a fixed inner boundary is used. That means within a fixed inner boundary radius, $r_{\star,\textrm{ib}} = \SI{4e11}{cm}$, centered on the star's origin, stellar wind conditions are held constant throughout time. Lastly, the domain's dipole boundary conditions (\autoref{sssec:Amods}) are then modified to respect these conditions.

\refemp{As mentioned in \autoref{sssec:stellar}, the stellar wind is either always sonic or sub-sonic due to the stellar wind's polytropic index of $\Gamma = 5/3$. Therefore, we chose a sonic stellar wind, such that it will shock on the planetary outflow. While the wind is technically isentropic, it has been carefully chosen to mimic a more realistic stellar wind within the domain of the simulation. Specifically, the stellar wind starts at $r_{\star,0}$ with $v = \SI{200}{km.s^{-1}}$, but by the time it reaches the planet it has accelerated to $v = \SI{290}{km.s^{-1}}$ (similar stellar outflow velocities are found, e.g., in a stellar wind model for HD 219134 by Vidotto et al. (\citeyear{Vidotto18})). It continues to accelerate past the planet with a maximum of nearly $\SI{300}{km.s^{-1}}$. We note that outside of our simulation box, the isentropic stellar wind is not a good model for the stellar wind profiles.}

\subsubsection{Stellar radiation}
\label{sssec:srad}

A planetary wind is launched by irradiating our computational domain with monochromatic plane-parallel radiation. Ionizing flux enters from the negative $x$\textendash axis, altering the ionization state and depositing energy into the fluid. To prevent transients from impacting the early simulation, the flux is ramped up so that the wind is gently launched into the ambient medium. Let the true physical stellar flux be equal to our flux at time equal infinity, $F(\infty) = F_\infty$. Our initial flux will start at a factor $f_0$ of the physical flux, such that $F(0) = f_0 F_\infty$. We gradually increase the flux using a Gaussian rate of change ramp function with standard deviation $\sigma$, so that after time $t$ the flux is

\begin{equation}
F(t) = \frac{F_\infty}{2} \, \left(1 + \textrm{erf}\left(\textrm{erf}^{-1}\left(1-2 f_0\right) \left(\frac{t}{t_{1/2}} - 1\right)\right)\right).
\end{equation}

\noindent Here erf$()$ is the error function and erf$^{-1}()$ is the inverse error function. The halfway point of the flux ramping occurs at $t_{1/2} = \sqrt{2} \sigma \, \textrm{erf}^{-1}\left(1-2 f_0\right)$, such that $F(t_{1/2}) = \frac{1}{2} F(\infty)$. In practice, once $F(t) = 0.999 F_\infty$ we set $F(t) = F_\infty$ for the rest of the simulation duration.

\subsection{\textit{Parameters used}}
\label{ssec:params}

\capstartfalse
\begin{deluxetable}{l D{?}{\times}{0.5}} 
\tablewidth{\columnwidth} 
\tablecaption{Simulation parameters \label{tab:params}} 
\tablehead{Parameter & \colhead{Value}}
\startdata 
Planet\\
\hline
\rule{0pt}{1.\normalbaselineskip}Mass, $M_{\textrm{p}}$ (g) & 5.0?10^{29}\\
Radius, $R_{\textrm{p}}$ (cm) & 1.5?10^{10}\\
Temperature at $R_{\textrm{p}}$, $T_{\textrm{p}}$ (K) & 1.1?10^{3}\\
Density at $R_{\textrm{p}}$,$^{\texttt{\#}}$ $\rho_{\textrm{p}}$ (\si{g.cm^{-3}}) & 6.64?10^{-16}\\
\\
\hline
\rule{0pt}{1.\normalbaselineskip}Orbital Parmeters\\
\hline
\rule{0pt}{1.\normalbaselineskip}Semi-major axis, $a$ (cm) & 1.0?10^{12}\\
Orbital period,$^{\texttt{\#}}$ P (s) & 5.5?10^{5}\\
\\
\hline
\rule{0pt}{1.\normalbaselineskip}Star\\
\hline
\rule{0pt}{1.\normalbaselineskip}Mass, $M_\star$ (g)& 1.989?10^{33}\\
Radius, $R_\star$ (cm) & 6.957?10^{10}\\
Ionizing flux, $F_0$ (\si{cm^{-2}.s^{-1}}) & 2.0?10^{13}\\
Photon energy, $h\nu$ (eV) & 1.6?10^{1}\\
\\
\hline
\rule{0pt}{1.\normalbaselineskip}Stellar wind\\
\hline
\rule{0pt}{1.\normalbaselineskip}Reference radius, $r_{\star,0}$ (cm) & 4.0?10^{11}\\
Temperature at $r_{\star,0}$, $T_{\star,0}$ (K) & 1.35?10^{6}\\
Velocity at $r_{\star,0}$, $v_{\star,0}$ (\si{cm.s^{-1}}) & 2.0?10^{7}\\
Proton number density at $r_{\star,0}$, $n_{\star,0}$ (\si{cm^{-3}}) & (1.5, 15, 70)?10^{3}\\
Mass-loss rate,$^{\texttt{\#}}$ $\dot{M}_{\star}$ (\si{M_\odot.yr^{-1}}) & (1.3, 13, 59)?10^{-16}\\
Pressure at planet,$^{\texttt{\#}}$ $P_{\star,\textrm{total}}(a)$ (\si{dyn.cm^{-2}}) & (1.3, 13, 61)?10^{-7}
\enddata 
\tablecomments{Parameters used in simulations. Parameters that are not free parameters, but are determined by other variables are denoted with a $\texttt{\#}$ superscript. Tuple values in parentheses correspond to the weak, intermediate, and strong stellar winds accordingly.}
\end{deluxetable}
 \capstarttrue

To efficiently model atmospheric escape around hot Jupiters, we pick parameters that enable good resolution at reasonable cost while still being physically motivated. For the planet we use ${M_\textrm{p} =  \SI{5.0e29}{g} \approx M_{\saturn}}$, ${R_\textrm{p} = \SI{1.5e10}{cm} \approx 2 R_{\jupiter}}$, and ${T_\textrm{p} = \SI{1100}{K}}$ with a semi-major axis ${a = \SI{e12}{cm} \approx \SI{0.07}{au}}$. Recall that $\saturn$ is the astronomical symbol for Saturn and $\jupiter$ the symbol for Jupiter. Our choices correspond to a large planetary radius, which helps increase the scale height by decreasing the local surface gravity $g$. For a full understanding of how these parameters affect the scale height, see \autoref{apn:sheight}. Our parameters are close to those of Wasp-17 b, one of the more extreme exoplanets. We run all of our simulations for \SI{2e6}{s}, which is over $3.5$ orbital periods. After which, all simulations appear to have reached a steady or quasi-steady state.

The domain varies depending on which simulation we run. As explained in \autoref{sec:results}, we run three classes of simulation depending on which physical processes we include; we denote these runs \textit{Rogue}, \textit{Tidal}, and \textit{Rotating}. Our \textit{Rogue} simulations are carried out in a $(50 * R_\textrm{p})^3$ box centered on the planet. The \textit{Tidal} simulations use a similar domain, but extended an additional $25 * R_\textrm{p}$ along the negative $x$\textendash axis towards the star (($75\times 50 \times 50) * R_\textrm{p}$). Lastly the \textit{Rotating} simulations take the \textit{Tidal} domain and extend an additional $25 * R_\textrm{p}$ along both the negative and positive $y$\textendash axes (($75\times 100 \times 50) * R_\textrm{p}$). For the planet's inner boundary conditions, we use $R_{\textrm{mask}} = 44/64 * R_\textrm{p}$ and $R_{\textrm{ib}} = 49/64 * R_\textrm{p}$. The edge of the planetary atmosphere is at $R_\textrm{e,0} = 89/80 * R_\textrm{p}$, and the first ambient medium's edge is at $R_\textrm{e,1}  = 2 \sqrt{3} * R_\textrm{p}$ (defined in \autoref{apn:ambient}).

\refemp{Computational expense varies greatly across our several different simulations. We used 115 Intel Xeon E5-2650 CPUs in parallel for all simulations. Wall times ranged from roughly 2 days (\textit{Rogue}) to a month (\textit{Rotating} with a strong stellar wind). However, most simulations ran in under 10 days, with the strong stellar wind being the outlier (for the \textit{Rotating} domain, the intermediate stellar wind took 13.5 days and the weak stellar wind took 10 days). The two main reasons for the range of cost were the varying domain sizes, and the inclusion of a stellar wind. The most prohibiting factor, a strong-stellar-wind, remains present on the finest level of resolution throughout the duration of the simulation. As the stellar wind is both hot and fast, the Courant condition leads to smaller time-steps for the stellar wind than the planetary outflow at the same spatial resolution. Ideally in the future, adaptive mesh refinement or better static mesh refinement will be chosen to avoid excessively high spatial resolution in the stellar wind.}

Resolution for the simulation is set such that within a cell there is at most a scale height. While we do not know the structure of the wind in steady state a priori, the hydrostatic isentropic atmosphere contains some of the smallest scale heights throughout the duration of the simulation. The number of scale heights between two points in our hydrostatic isentropic atmosphere is given by \cref{nheight}. Since the hydrostatic atmosphere and base of the wind require the most resolution, we use a statically-refined mesh. The highest resolution region, which is a box of size $(2.5 * R_\textrm{p})^3$ centered on the planet with $\delta x = R_\textrm{p}/64$, encapsulates both the wind base and hydrostatic atmosphere. We then include four coarser levels of refinement around that, leading to a minimum resolution $\delta x = R_\textrm{p}/4$ in the region outside a box of size $(4 * R_\textrm{p})^3$ centered on the planet.

The star is modeled after the Sun, with $M_\star = \SI{1.989e33}{g} = M_\odot$ and ${R_\star = \SI{6.957e10}{cm} = R_\odot}$. The ionizing flux is $F_\infty = \SI{2e13}{cm^{-2}.s^{-1}}$, which is comparable to our sun's ``moderate to low solar activity'' extreme-ultraviolet flux\footnote{Photon energies of \SIrange{13.6}{40}{eV} (\SIrange{91}{30}{nm}).} scaled to \SI{0.05}{au} (Woods et al. \citeyear{Woods98}). Note our ionizing flux is scaled to the center of our box as we use plane-parallel radiation. To ramp the flux to avoid transients we use $f_0 = 10^{-2}$, and $t_{1/2} = \SI{5e4}{s}$.  We use a monochromatic ionizing flux of $h\nu = \SI{16}{eV}$, which is in line with previous studies and reasonable for a hot Jupiter around a quiet solar analog (Tripathi et al. \citeyear{Tripathi15}). 

\begin{figure*}
\centering
\label{fig:noWinds}
\ifnum\movie=0
\includegraphics[width=\textwidth]{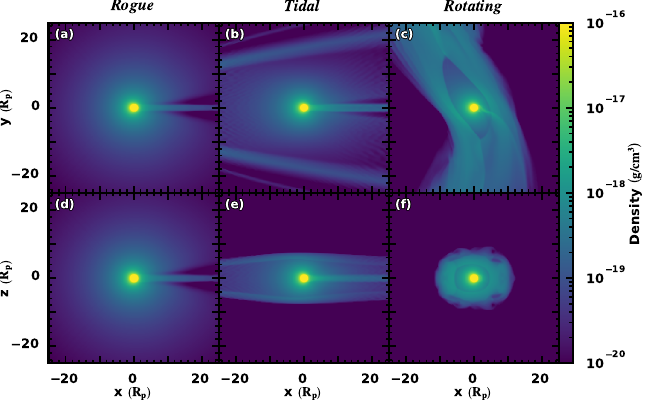}
\else
\includemedia[width=\textwidth,
playbutton=none,
addresource=movie02.mp4,
flashvars={source=movie02.mp4}
]{\includegraphics[width=\textwidth]{Figure02.pdf}}{VPlayer.swf}
\fi
\caption{The fully-launched density structure of a hydrodynamic escaping planetary outflow without a stellar wind. Ionizing radiation enters from the left side of each box. The first row shows slices in the orbital plane, $z=0$, while the second row shows vertical slices perpendicular to the planet-star axis, $y=0$. \textit{Left}: Our \textit{Rogue} planet, which only feel the effects of an external ionizing source and the planet's gravity. \textit{Middle}: Our \textit{Tidal} simulation, which includes tidal gravity but no Coriolis term. \textit{Right}: A true rotating frame with the Coriolis force, called our \textit{Rotating} simulation. This figure is available online as an animation, showing the launching of the outflow from $t = \SI{0}{s}$ until $t = \SI{2e6}{s}$.}
\end{figure*}

For simulations that include a stellar wind, we consider a range of wind strengths, as parameterized by $n_{\star,0}$, \cref{pressureScaling}. Our choices for $n_{\star,0}$ are \SIlist{1.5e3;1.5e4;7.0e4}{cm^{-3}} at reference radius ${r_{\star,0} = \SI{4e11}{cm} \approx \SI{0.03}{au}}$. Note that the proton number density of the Sun's stellar wind at this distance is roughly ${n_{\star,0} = \SI{8.4e3}{cm^{-3}}}$, which falls directly between our low and intermediate value. \refemp{Vidotto \& Bourrier's (\citeyear{Vidotto17}) model of GJ 436 estimate a ram pressure of $P_{\textrm{ram}} = \SI{1.4e6}{dyn.cm^{-2}}$ at the location of GJ 436 b, which nearly corresponds to the pressure of our intermediate stellar wind. Vidotto et al. (\citeyear{Vidotto18}) modeled HD 219134, and found pressures near HD 219314 b similar to our strong stellar wind.} The other parameters of the stellar wind are $T_\star = \SI{1.35e6}{K}$ and an initial velocity of $v_\star = \SI{200}{km.s^{-1}}$, corresponding to a Mach number of $\textrm{M} = 1.04$. \refemp{The stellar mass-loss rates corresponding to our three wind strengths are \SIlist{1.3e-16;1.3e-15;5.9e-15}{M_\odot.yr^{-1}}.}

\section{Results}
\label{sec:results}

We now explore the effects of the tidal gravity, the Coriolis force, and the stellar wind on the planetary outflow piece-by-piece. Our base case is an atmosphere receiving ionizing flux in the planet's potential with no external stellar wind (\textit{Rogue} run, \autoref{ssec:rogue}). The impact of non-inertial forces from the planet's orbit around a star is first examined without the Coriolis force (\textit{Tidal} run, \autoref{ssec:tidal}) then with (\textit{Rotating} run, \autoref{ssec:rotating}). The morphology of the outflows until this point is recapped with a more quantitative analysis of their velocity structures in \autoref{ssec:orbital}. Next the effects of a stellar wind are demonstrated by contrasting the results of varying stellar wind strengths in both the \textit{Rogue} and \textit{Tidal} runs (\autoref{ssec:wind}). Finally the full suite of stellar environment physics is considered in the \textit{Rotating} run with a stellar wind (\autoref{ssec:fullenv}). A summary of quantitative variables, particularly the mass-loss rate, across all simulations is given in \autoref{ssec:massloss}, and observational consequences are delayed until \autoref{sec:obscon}. Summarized in \autoref{tab:params} are the important simulation parameters used across all runs.

\subsection{Rogue simulation: effects of ionizing radiation}
\label{ssec:rogue}

We begin with a planet receiving plane-parallel ionizing radiation in the absence of tidal gravity or a stellar wind, shown in \hyperref[fig:noWinds]{Figures \ref{fig:noWinds}(a)} and \hyperref[fig:noWinds]{\ref{fig:noWinds}(d)}. The closest physical analog would be an ejected rogue planet heated by high-energy photons from a nearby high-mass star.  Such a situation would be more likely to occur in extreme environments, for example the young star-forming regions near the Galactic Center. However, the motivation for this study is as a base case for comparison to more complex planetary conditions, i.e., those with a stellar host.

\begin{figure}
\label{fig:uniTemp}
\includegraphics[width=\columnwidth]{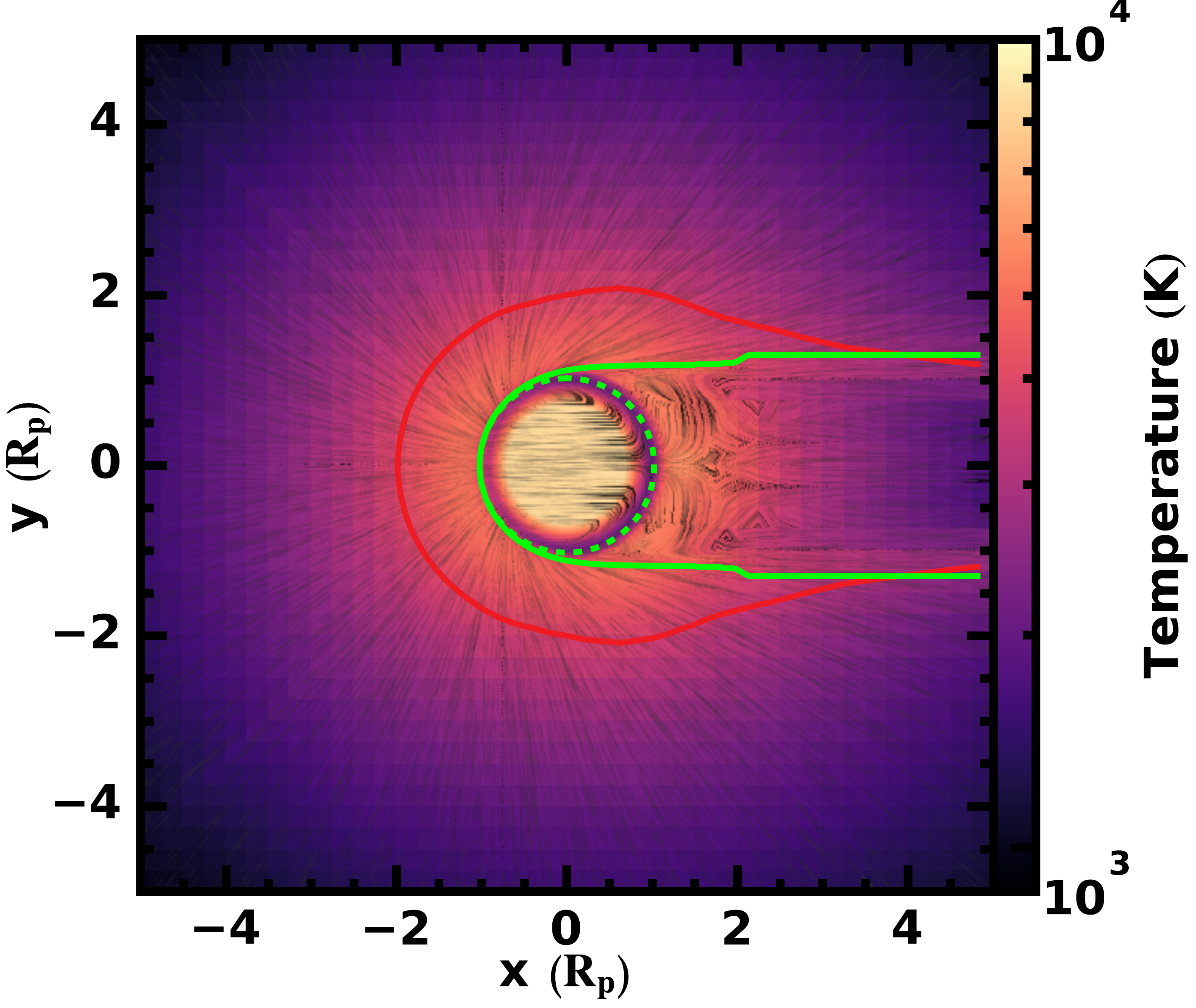}
\caption{The temperature of the gas for our \textit{Rogue} planet. The green contour is the surface where $\tau = 1$ for ionizing radiation entering from the left side of the plot. The corresponding dashed green circle is of constant radius equal to the substellar $\tau=1$ radius. While the nightside is cold just at and below the dashed circle, the temperature just above is practically uniform around the entire planet. The pencil shadings are the line integral convolution to assist in visualizing the flow. We can see a stagnation point in the shadow near $4 * R_\textrm{p}$, which leads to the flaring around the neutral shadow exhibited in \hyperref[fig:noWinds]{Figures \ref{fig:noWinds}(a)} and \hyperref[fig:noWinds]{\ref{fig:noWinds}(d)}. The sonic surface is marked in red and does not form a closed surface.}
\end{figure}

Despite inherent asymmetric heating, the outflow at large scales is strikingly spherically symmetric. The symmetry arises from azimuthal pressure gradients freely redistributing material along equipotential surfaces, the efficiency of which can be seen in the steady-state solution of the temperature distribution, \autoref{fig:uniTemp}. At larger distances material streams radially outwards, having subdued the azimuthal pressure gradients. It is only on the nightside, originating from the planet's shadow, that a planetary tail and the surrounding dearth of material break the symmetry. \refemp{Note that the ``planetary tail'' is not the nightside or downstream arm of the outflow, but the noticeably neutral-enriched density enhancement that forms in the shadow of the planet.}

The planetary tail is a new prediction in multi-dimensional simulations \refemp{(also seen in Debrecht et al. \citeyear{Debrecht18})}, as previous models had not resolved the stellar heating. Instead these models used an adiabatic index near one, and fixed temperature boundary conditions that varied as the angular distance from the substellar point, to emulate an anisotropic, isothermal Parker wind (Stone and Proga \citeyear{Stone09}; Carroll-Nellenback et al. \citeyear{Carroll17}). In place of a planetary tail, these prior studies show a sustained nightside compression shock, which arises as hot dayside material flows around the planet and enters the planet's cold nightside atmosphere (See Figures 1 and 4 in Stone and Proga \citeyear{Stone09}; Figure 3 ``ANISO'' panel in Carroll-Nellenback et al. \citeyear{Carroll17}). We should note that this phenomenon is also seen in our simulations, but only initially as our boundary conditions are set deeper than the base of the wind. Thus, when inflowing dayside material advects heat to the nightside, it cannot remain cold as it does in the fixed anisotropic Parker wind.

Setting boundary conditions interior to the wind base enables the nightside to readjust, preventing material from reentering the atmosphere. Material advected to the nightside, unable to receive ionizing radiation in the planet's shadow, begins to form the planetary tail. The tail then serves to lessen azimuthal pressure gradients and shuts down the compression shock, explaining why we only see the shock as an initial transient in our simulations. Even with typical ISM cooling rates at solar metallicity, we find that the timescale for the material to flow from the day to nightside would be too fast for the tail to sufficiently cool and collapse. We note that both the previous models and ours lack consideration of conductive heating and cooling. While conduction was found not to be dynamically important on the substellar ray due to outflow timescales (Garc\`ia Mu\~noz \citeyear{Garcia07}; Murray-Clay et al. \citeyear{Murray09}), if hot stagnant gas persists on the nightside, evident in \autoref{fig:uniTemp}, heat conduction may be dynamically important\textemdash something future multi-dimensional simulations need to take into consideration.

The formation of the tail is also responsible for the surrounding dearth of material. The tail grows by accruing material via the tip, thus new material must travel farther, spending more time in ionizing radiation, before it can enter the shadow. Eventually there comes a point at which this material has received enough ionizing energy to become unbound, and it flows radially outwards as this is now the path of least resistance. Thus, drawing a radial line from the origin to this point at the edge of the shadow demarcates the dearth boundary. With this description the length, $\ell$, from the planet where the flaring outflow begins to diverge from the tail can be estimated. We equate the potential energy to the ionizing heating rate multiplied by the time spent getting there, i.e., the distance divided by the flow velocity, ${\phi = \mathcal{G} * \ell/v}$. For a point-mass potential this yields  

\begin{equation}
\label{stagpoint}
\ell = \sqrt{\frac{G * M_\textrm{p} * \mu * v}{E_{\textrm{pe}} * \sigma_{\textrm{HI}} * F_\infty * \textrm{e}^{-\tau}}}.
\end{equation}

With a flow velocity of order $v = \SI{10}{km.s^{-1}}$, and marginally bound gas near the $\tau = 1$ surface, we find $\ell = 3.6 * R_\textrm{p}$\textemdash close to the simulation's $4 * R_\textrm{p}$. Naively one might expect the tail to truncate at this distance as it can no longer receive material at the tip. However, the tail is hot and adiabatically expands outwards, allowing new material to enter at points prior to the truncation, sustaining its growth.

Lastly, we comment that irrespective of the nightside steady state, a cosine temperature function does not match the dayside temperature profile of the self-consistently simulated hot Jupiter. At the altitude of the wind base, the temperature is uniformly $\SI{\sim7000}{K}$, shown in \autoref{fig:uniTemp}. This directly calls into question whether simple cosine anisotropic heating functions are valid in the regime of hot Jupiters. We believe not, as the cosine function inherently assumes the temperature is set by the flux received. Instead, we find that for the dayside of hot Jupiters the temperature is set by the cooling equilibrium temperature, as the entire face is bathed in ample flux to achieve such temperatures. We therefore suggest a cosine anisotropic heating function would only be suitable for planets whose dayside temperature is set by the heating rate and not the cooling rate, such as a planet similar to Earth.

\begin{figure}
\label{fig:tball}
\includegraphics[width=\columnwidth]{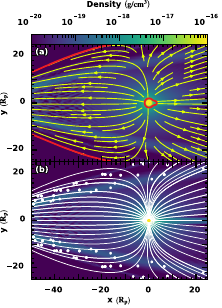}
\caption{\textit{Panel (a)}: Streamlines (chartreuse yellow) overlaid on gas density from the steady-state \textit{Tidal} run (\hyperref[fig:noWinds]{Figure \ref{fig:noWinds}(b)}). The sonic surface is demarcated in red, with the outer sonic surface coinciding with the interface between the planetary outflow and the virtually static ambient medium. Note that while supersonic material does shock within the flow, the shock is oblique and thus there is no corresponding sonic surface. \textit{Panel (b)}: Ballistic trajectories (white) of particles launched radially from the surface of the planet with velocity $v_{\textrm{p}}$ (see text). Ballistic particles launched near the planet's terminator ($x=0$, $y=\pm R_\textrm{p}$) achieve the furthest $y$\textendash extent for gas infalling onto the star. These trajectories bound the rest of the flow and are in good agreement with the outer sonic surface that demarcates the interface with the ambient medium. Notice no significant trajectory crossing seems to occur in the orbital plane. Shocking results instead from trajectories crossing the orbital plane from above and below, having been accelerated to do so by stellar gravity\textemdash white dots demarcate where a few such trajectories, i.e., those originating with $z \neq 0$, pass through the orbital plane. Trajectories launched near the poles ($z = \pm R_{\textrm{p}})$ are excluded, as they shock gas far out of the orbital plane.}
\end{figure}

\begin{figure}
\label{fig:ballistic}
\centering
\includegraphics[width=\columnwidth]{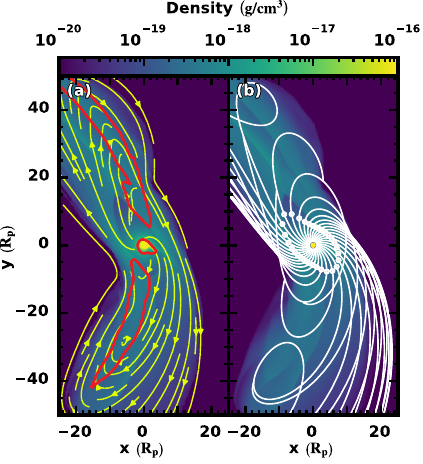}
\caption{Gas streamlines and ballistic particles launched from the planet surface overlain on the steady-state density map from our \textit{Rotating} simulation (\hyperref[fig:noWinds]{Figure \ref{fig:noWinds}(c)}). The ballistic trajectories are launched with initial velocity $v_{\textrm{p}} = \sqrt{v_{\textrm{esc}}^2+v_\infty^2}$ at $R_{\textrm{p}}$, which gives the particles a total energy equal to that of the gas at infinity for the \textit{Rogue} case. Given this energy, the ballistic trajectories show the farthest possible extent of the gas. The white dots along the ballistic trajectories occur where the velocities have been deflected by one radian from their original direction. This is what we call our Coriolis length, which roughly matches the shock we see in the simulation.}
\end{figure}

\subsection{Tidal simulation: effects of tidal gravity and ionizing radiation}
\label{ssec:tidal}

Our \textit{Tidal} run is shown in \hyperref[fig:noWinds]{Figures \ref{fig:noWinds}(b)} and  \hyperref[fig:noWinds]{\ref{fig:noWinds}(e)}. In addition to ionizing radiation, it includes tidal forces, with the star located at $x = - 66.\bar{6} * R_\textrm{p}$. Note that this is not a true rotating frame as we neglect the Coriolis force, but this frame is akin to that used in Tripathi et al. (\citeyear{Tripathi15}) and numerous 1-D models that included tidal gravity.

A clear difference between the \textit{Tidal} and \textit{Rogue} runs is the funneling of material towards the star. Escaping planetary gas is energetically restricted to a cone originating at the barycenter by the star's gravitational pull and the frame's centrifugal push (\autoref{fig:tball}). This can be further illuminated by considering ballistic particles. \hyperref[fig:noWinds]{Figure \ref{fig:tball}(b)} illustrates the ballistic trajectories of particles launched radially from the planet's surface at velocity $v_{\textrm{p}}$. For supersonic flow (external to the inner red contour in \hyperref[fig:noWinds]{Figure \ref{fig:tball}(a)})  the internal energy of the fluid is much less than the kinetic energy, and rough agreement between the streamlines and ballistic trajectories is expected. Given that the outflow quickly becomes supersonic, large-scale streamlines should match appropriately chosen ballistic trajectories. These trajectories are chosen such that the total energy at infinity, ignoring work done by tidal forces, is the same found in the \textit{Rogue} runs, $e_\infty = v_\infty^2/2$, so that $v_\textrm{p} = \sqrt{v_{\textrm{esc}}^2+v_\infty^2}$.

Along with the funneling comes a shock in the outflow with the same funnel geometry. Examination of ballistic particles in the orbital plane gives little hint of intersecting trajectories (\hyperref[fig:noWinds]{Figure \ref{fig:tball}(b)}). However, due to the lack of the centrifugal force in the $z$\textendash axis, intersections still readily occur. This can be seen by considering ballistic particles launched with the majority of their velocity parallel to the $z$\textendash axis (not shown). Trajectories able to escape the planet's gravitational pull nevertheless fall back into the orbital plane due to the stellar gravity. Trajectories originating both above and below the orbital plane oscillate through the orbital plane, intersecting many trajectories closer to the orbital plane which were not as strongly accelerated into the plane.

Notice that with tidal gravity the dearth of material around the tail appears less pronounced in the orbital plane and nonexistent in the $x$\textendash $z$ plane. The nonexistence in the $x$\textendash $z$ plane is due to the star's strong gravitational pull of all the gas back into the orbital plane. Eventually far away from the planet, the entire nightside stream will be compressed into a vertically pressure-supported plane due to the star's gravity. In the orbital plane the reduced flaring can be understood from the tidal forces stretching the outflow along the $x$\textendash axis, i.e., material unbound from the planet no longer moves purely radially. Unlike in the vertical plane, the dearth is not readily filled by the stellar gravitational pull towards the $x$\textendash axis, as the centrifugal force opposes the pull with a push.

\subsection{Rotating simulation: effects of tidal gravity and ionizing radiation in a rotating frame}
\label{ssec:rotating}

The \textit{Rotating} run is presented in \hyperref[fig:noWinds]{Figures \ref{fig:noWinds}(c)} and \hyperref[fig:noWinds]{\ref{fig:noWinds}(f)}, where we have now included the Coriolis force in addition to the tidal forces. A \textit{Rotating} frame without a stellar wind has also been explored with a separate code, \texttt{AstroBEAR}, finding qualitatively the same geometry and quantitatively the same mass-loss rates (Debrecht et al. \citeyear{Debrecht18}). Due to the Coriolis force, streamlines now wind in a clockwise fashion as our rotation vector points out of the page (\hyperref[fig:noWinds]{Figure \ref{fig:ballistic}(a)}). We can still understand the dayside and nightside arm dichotomy as owing to the tidal focusing previously described, but now deflected with the inclusion of the Coriolis force.

Including the Coriolis force in our ballistic trajectory analysis demonstrates that there now exist significant trajectory intersections, even in the orbital plane (\hyperref[fig:ballistic]{Figure \ref{fig:ballistic}(b)}). In \hyperref[fig:noWinds]{Figure \ref{fig:ballistic}(b)} the white dots along the ballistic trajectories denote where the velocity of each streamline has been deflected by one radian, which defines our Coriolis length scale. The surface created by these dots is similar to the teardrop-shaped shock, evident in the density map of \hyperref[fig:noWinds]{Figure \ref{fig:noWinds}(c)}, with an additional rotation of roughly $\pi/12$.  This additional rotation may arise because the shocked gas moves slower than the unshocked gas and therefore turns on a smaller scale. Material the outflow shocks on has rotated more than predicted from the ballistic analysis, modestly rotating the shock interface everywhere.

Note that the Coriolis length is not spherically symmetric for two reasons. First, the velocity profiles along each streamline may not be equal, owing to the dayside/nightside asymmetric heating. This effect seems small as our ballistic particles are all launched at the same velocity from $R_\textrm{p}$ and seem to give good agreement with the simulation. However, a detailed examination of the ballistic trajectories shows that the distance of the nightside shock from the planet is not as well matched as that on the dayside suggesting that the nightside would be better described by modestly lower ballistic velocities. The second and more significant asymmetry is due to the fact that particles launched from the planet in quadrant I (upper right) and III (lower left) are pulled by tidal forces in in the same direction that the Coriolis force bends them, towards the $x$\textendash axis ($y=0$) and away from the $y$\textendash axis ($x=0$). Conversely, quadrants II (upper left) and IV (lower right) are pulled in the opposite direction of their deflections. Thus, in a rotating frame tidal forces can either assist or hinder the turning of streamlines, and the shock occurs closer to the planet in quadrants I and III than in II and IV.

In \autoref{fig:ballistic} the ballistic trajectories completely bound the extent of the atmospheric escape. This is a statement about which regions of space are energetically forbidden for the escaping gas to reach. While the trajectories do a good job in all four quadrants of probing the maximal extent of the outflow, they dramatically overestimate the extent of the outflow in quadrant IV. This is due to the fact that the farthest reach of the flow to positive $x$ values in quadrant IV comes from gas that originated or passed through the shadow of the planet\textemdash readily seen by following the nightside tail in \hyperref[fig:noWinds]{Figure \ref{fig:noWinds}(c)}. Unable to receive ionization heating in the shadow, the gas is not fully accelerated, turns on smaller lengths, and thus reaches a lesser extent in quadrant IV than predicted by ballistics.

The planetary tail remains present, but is now bent due to the Coriolis force. Its turning length appears longer than the other streamlines due to the column being virtually stagnant until the gas' effective Hill radius.\footnote{The effective Hill radius being smaller than the standard ballistic Hill radius, see \autoref{ssec:lscales}} Thus the tail follows ballistic particles launched from the effective Hill sphere near $2 * R_\textrm{p}$ on the nightside, giving the appearance of a longer turning length.

\subsection{Comparison of velocity structures}
\label{ssec:orbital}

\begin{figure}[t]
\centering
\label{fig:vel_rot}
\includegraphics[width=\columnwidth]{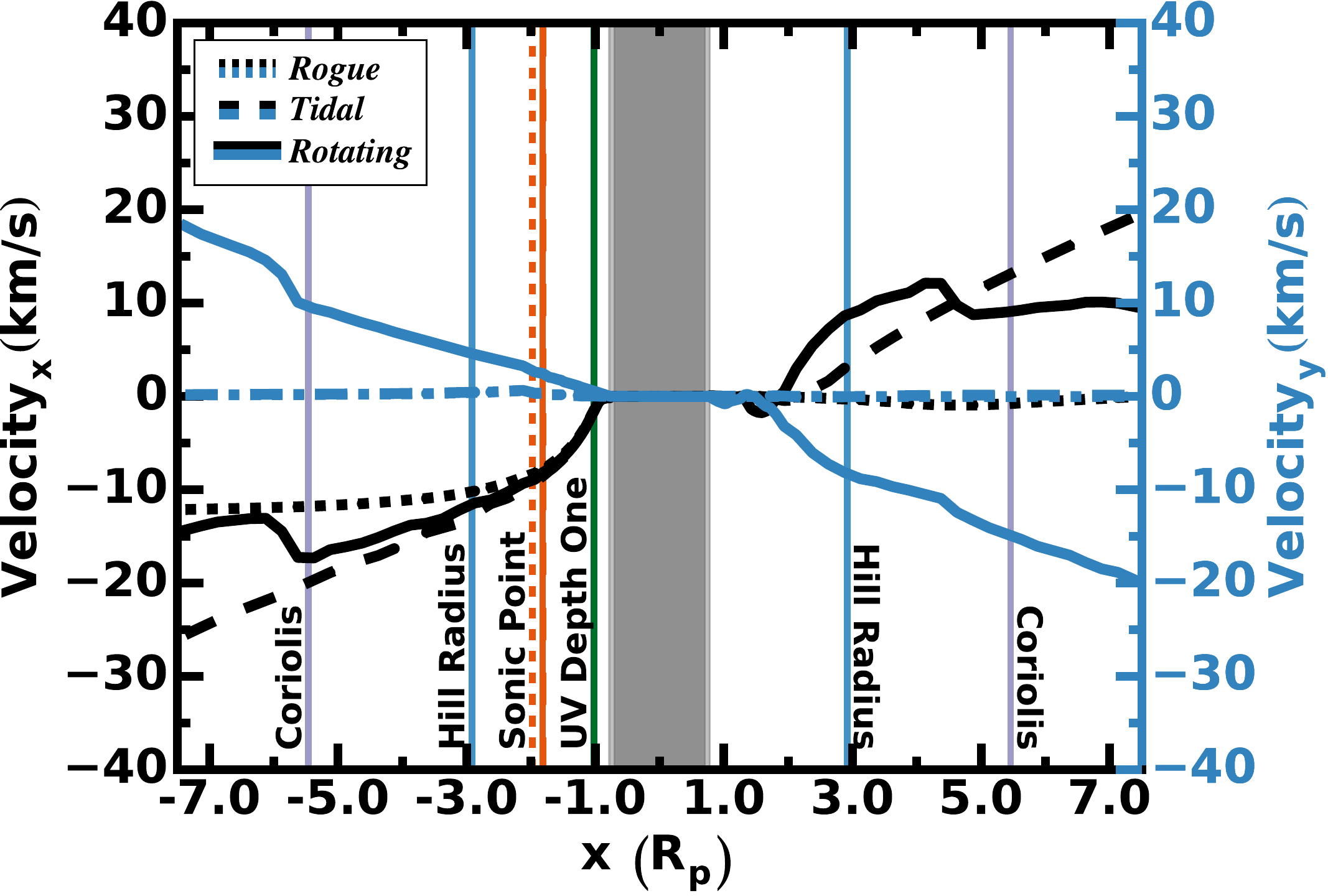}
\caption{Velocity along the $x$\textendash axis for three simulations: \textit{Rogue} (dotted), \textit{Tidal} (dashed) and \textit{Rotating} (solid). The black lines are the $x$ velocity and the blue lines are the $y$ velocity. In both the \textit{Tidal} and \textit{Rotating} runs there is significant acceleration outside the Hill radius from tidal forces. On the nightside the gas remains stagnant out to about $2 * R_\textrm{p}$,where tidal  forces begin to accelerate it away from the planet. The decrease in velocity in the \textit{Rotating} profiles at large separation is from a shock, which occurs near the Coriolis length on the dayside, and slightly closer to the planet on the nightside due to slower outflow velocities. The velocity on the nightside is faster in the \textit{Rotating} case compared to the \textit{Tidal} case because the $x$\textendash axis samples multiple streamlines in the \textit{Rotating} case. The verticals lines are clearly labeled, and overlap (or do not exist for some simulations, i.e., the Coriolis length), except for the sonic point. That is because due to tidal forces, \textit{Tidal} and \textit{Rotating} (overlapping sonic points) go transonic before the \textit{Rogue} simulation. The shaded regions denote regions that are fixed by boundary conditions (light grey for $R_{\textrm{ib}}$ and dark grey for $R_{\textrm{mask}}$), and should be ignored.}
\end{figure}

\begin{figure*}
\centering
\label{noCoriolis}
\includegraphics[width=\textwidth]{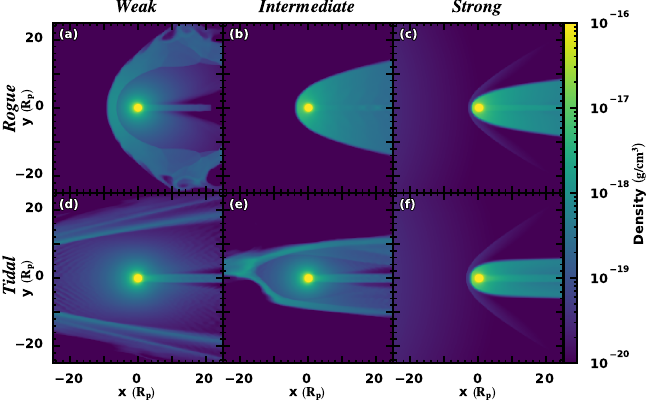}
\caption{Snapshots of the density structure of a fully launched hydrodynamic escaping planetary wind outflow with varying stellar wind strengths in the orbital plane, $z=0$. The first row shows our \textit{Rogue} simulations and the second row is our \textit{Tidal} runs. The columns are for the various wind strengths parameterized by: $n_{\star,0}= \SI{1.5e3}{cm^{-3}}$ (left), $n_{\star,0}= \SI{1.5e4}{cm^{-3}}$ (middle), and $n_{\star,0}= \SI{7.0e4}{cm^{-3}}$ (right). Panel (a) shows planetary gas forming a bow shock with the stellar wind. Panel (d) is unconfined as the stellar wind is not strong enough to overcome the planetary outflow, and panel (e) is not in a steady state. The rest show complete confinement and breezes.}
\end{figure*}

\autoref{fig:vel_rot} provides the $x$\textendash component (black) and the $y$\textendash component (blue) of the velocity profiles for the \textit{Rogue} (dotted), \textit{Tidal} (dashed) and \textit{Rotating} (solid) runs along the $x$\textendash axis.\footnote{\refemp{Along this cut, the $x$\textendash axis, the Cartesian variables directly correspond to the polar variables: $v_x$ = $v_r$ and $v_y = v_\phi$.}} The \textit{Rogue} planet has a velocity $x$\textendash component that asymptotes to a finite value on the dayside as expected from \cref{termv}. On the nightside, the column is roughly hydrostatic.\footnote{There is actually some infall coming from circulation in the shadow as material falls inwards along the $x$\textendash axis and moves outwards near the edge of the shadow. See the line integral convolution shading in \autoref{fig:uniTemp}.} For the \textit{Tidal} run, the velocity starts to significantly deviate from the \textit{Rogue} case outside the Hill radius (blue vertical). There is also a hydrostatic tail on the nightside out to about $2 * R_\textrm{p}$, after which tidal forces start significantly accelerating the gas. Recall that the effective Hill sphere for the gas is less than the Hill sphere in \cref{eq:hill} due to gas pressure forces (\autoref{ssec:lscales}), so that the effective Hill radius appears to be near $2 * R_\textrm{p}$. This is almost equivalent to reducing the planet's mass in  \cref{eq:hill} to $M_\textrm{p}/3$.

In a true rotating frame with the Coriolis force, streamlines are no longer radial and velocities along the $x$\textendash axis no longer probe a single streamline. Looking at the dayside, we can see the $x$\textendash component of velocity is less than that of the \textit{Tidal} run, as some of the velocity has been transferred into the $y$\textendash component. Note that even though the Coriolis force does no work, the velocity components will not add in quadrature to be equivalent to the \textit{Tidal} run, as the same $x$\textendash coordinate corresponds to different distances from the planet along a streamline in the two runs. This explains why the nightside velocity is greater in the \textit{Rotating} case than in the \textit{Tidal} case, as the streamlines sampled here are originating closer to the dayside and have had more time to accelerate. Note that eventually both on the dayside and nightside there is a shock even in the absence of a stellar wind. This occurs near the Coriolis length scale where streamlines have been significantly bent, and trajectories start to cross. On the nightside the shock is at smaller distances for reasons discussed in \autoref{ssec:rotating}.

\subsection{Stellar wind in Rogue and Tidal runs:\\effects of a stellar wind and ionizing radiation}
\label{ssec:wind}

In this section we add a stellar wind to both the \textit{Rogue} and \textit{Tidal} simulations. We study three different stellar wind strengths tuned by the stellar wind proton number density, $n_{\star,0}$: a weak ($n_{\star,0}=\SI{1.5e3}{cm^{-3}}$), an intermediate ($n_{\star,0}=\SI{1.5e4}{cm^{-3}}$), and a strong stellar wind ($n_{\star,0}=\SI{7.0e4}{cm^{-3}}$). Qualitatively, what separates the strong from the weak stellar wind is whether the planetary outflow solution is a wind or a breeze. The intermediate stellar wind is in between these two extremes, but is chosen to still be a wind but with a smaller bow shock radius. Shown in the top row of \autoref{noCoriolis} is the $x$\textendash $y$ orbital plane slice of density in simulations without tidal forces, i.e., \textit{Rogue} runs with stellar wind. The second row has tidal forces, i.e., \textit{Tidal} runs with a stellar wind. The first column is our weak stellar wind, the second our intermediate and last our strong stellar wind.

The weak stellar wind (\hyperref[noCoriolis]{Figures \ref{noCoriolis}(a)} and \hyperref[noCoriolis]{\ref{noCoriolis}(d)}), is able to confine the planetary wind when tidal forces are absent. In contrast when tidal gravity is included, the planetary wind is no longer confined and is able to exit our domain. The intermediate stellar wind (\hyperref[noCoriolis]{Figures \ref{noCoriolis}(b)} and \hyperref[noCoriolis]{\ref{noCoriolis}(e)}) is capable of confining the planetary wind for the most part. Without tidal forces it is actually confined so strongly that the planetary outflow is no longer a wind, but a breeze as seen by the absence of shocked planetary gas. With tidal forces the outflow is still capable of being a wind. However, we have provided a snapshot that hints that the stellar confinement is not complete. This is not due to picking a frame prior to steady state, but as we shall see is due to no steady state standoff shock being possible. Lastly our strong stellar wind (\hyperref[noCoriolis]{Figures \ref{noCoriolis}(c)} and \hyperref[noCoriolis]{\ref{noCoriolis}(f)}) is capable of confining both the \textit{Rogue} and \textit{Tidal} simulations to breezes. Note that the \textit{Tidal} simulation is more confined, as within roughly $10 * R_\textrm{p}$ on the nightside, the tidal forces are restorative to the $x$\textendash axis ($y=0$).

\begin{figure}
\centering
\label{tidal_burp}
\ifnum\movie=0
\includegraphics[width=\columnwidth]{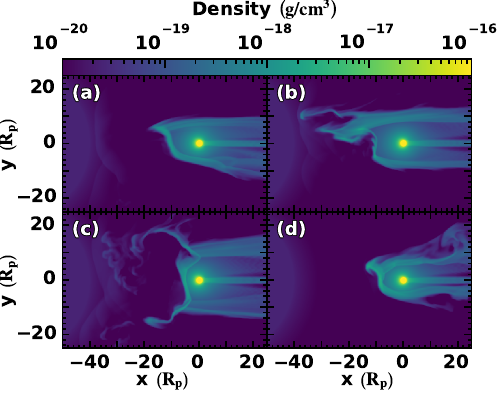}
\else
\includemedia[width=\columnwidth,
playbutton=none,
addresource=movie08.mp4,
flashvars={source=movie08.mp4&loop=true}
]{\includegraphics[width=\columnwidth]{Figure08.pdf}}{VPlayer.swf}
\fi
\caption{Quasi-steady state time series of the \textit{Tidal} simulations for the intermediate stellar wind case. The wind starts confined in panel (a) but continues to grow and extend outwards as seen in panel (b). Eventually the column grows out too far and can no longer maintain itself in the face of growing stellar wind strength. In panel (c) we see the disruption of the column, returning back to a confined state as seen in panel (d). The time lapsed between panel (a) and (d) is \SI{2.7e5}{s}. This process repeats continuously. This figure is available online as an animation, showing several periods of the quasi-steady state.}
\end{figure}

To explore the absence of a steady state in the intermediate \textit{Tidal} run we present a time series of simulation snapshots in \autoref{tidal_burp}. In \hyperref[tidal_burp]{Figure \ref{tidal_burp}(a)} the planetary wind is reasonably well confined by the stellar wind. However the confined region grows as the planetary wind continues to liberate more mass towards the star. Eventually in \hyperref[tidal_burp]{Figure \ref{tidal_burp}(b)} the dayside outflow grows too close to the star, where the stellar wind grows stronger, and is pushed back. Finding the path of least resistance the overextended gas is blown out and around the core-confined planetary wind (\hyperref[tidal_burp]{Figure \ref{tidal_burp}(c)}), reducing the planetary wind to a state similar to earlier, as seen in \hyperref[tidal_burp]{Figure \ref{tidal_burp}(d)}. Note that the time between \hyperref[tidal_burp]{Figure \ref{tidal_burp}(a)} and \hyperref[tidal_burp]{Figure \ref{tidal_burp}(d)} is $\SI{2.7e5}{s}$. This process repeats itself in perpetuity at roughly this timescale. A movie of several cycles is available online. Since this is not a truly physical frame, we leave a detailed explanation until \autoref{ssec:fullenv}\textemdash where a similar behavior is seen.

\begin{figure}
\label{fig:dLine}
\includegraphics[width=\columnwidth]{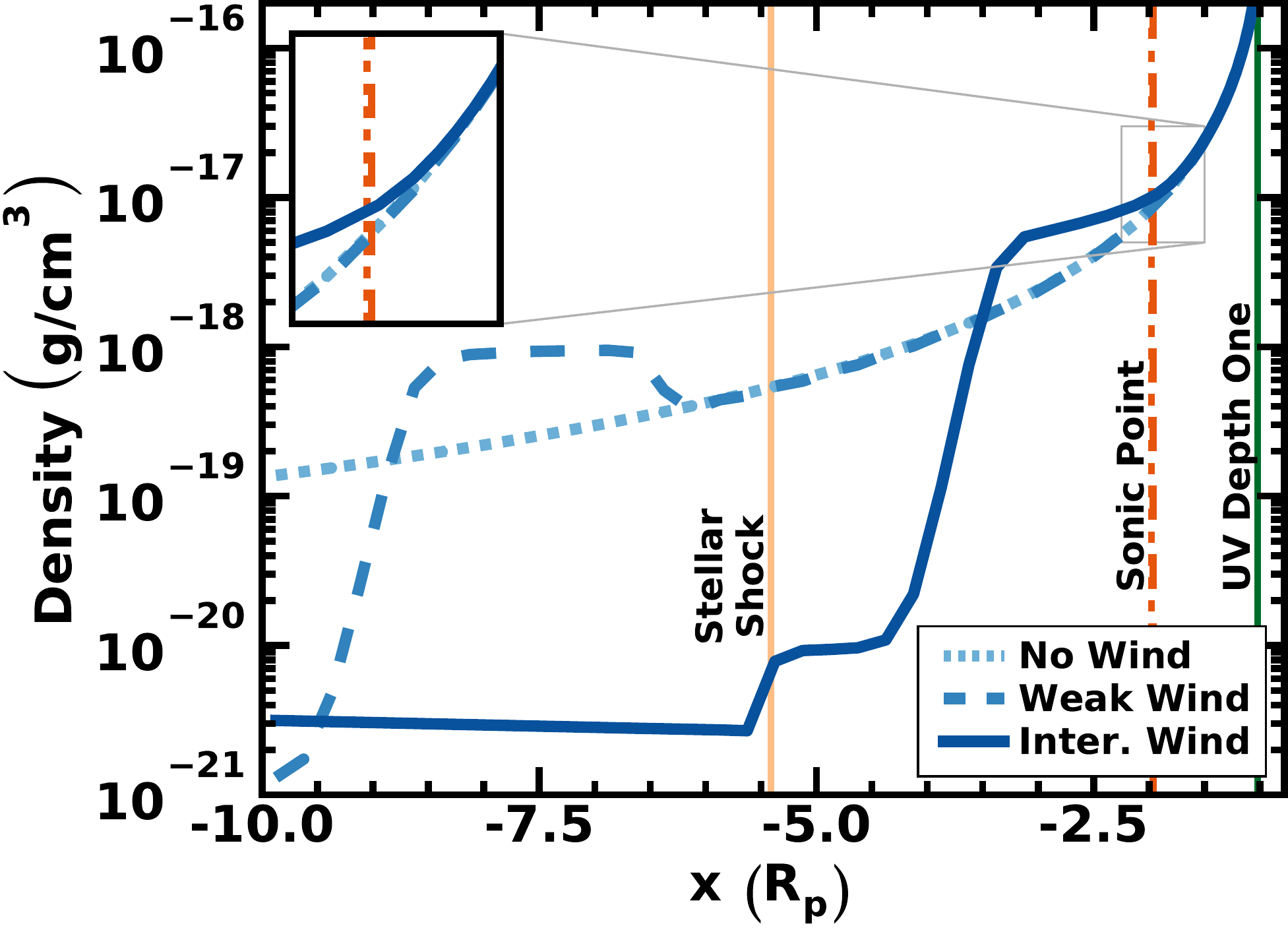}
\caption{Total densities along the substellar ray for the \textit{Rogue} simulations with no stellar wind (dotted), a weak stellar wind (dashed) and an intermediate stellar wind (solid). When a stellar wind is present, shocks and standoffs between the planetary and stellar wind occur. A several order of magnitude decrease in density occurs at the standoff location. Note that weak stellar winds do not affect the solution below the sonic point (red vertical), while a stronger stellar wind can alter the solution (see zoomed insert). In fact strong winds force a planetary breeze solution without a sonic point (lack of solid red vertical). The dark green vertical line is the UV depth one, and the orange vertical line shows the sonic surface of the stellar wind, corresponding to where it is shocked on the planetary wind (present only for the intermediate wind within the plotted region).}
\end{figure}

\begin{figure*}
\centering
\label{fig:fullSnap}
\ifnum\movie=0
\includegraphics[width=\textwidth]{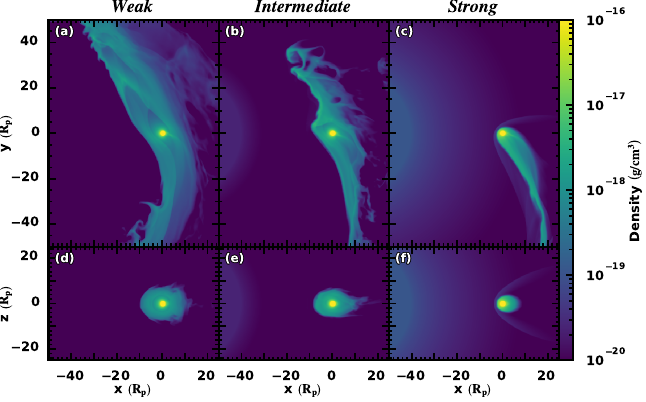}
\else
\includemedia[width=\textwidth,
playbutton=none,
addresource=movie10.mp4,
flashvars={source=movie10.mp4&loop=true}
]{\includegraphics[width=\textwidth]{Figure10.pdf}}{VPlayer.swf}
\fi
\caption{Snapshots of the density structure of a fully-launched hydrodynamic, escaping planetary outflow with varying stellar wind strengths in a full rotating frame. The first row shows slices in the orbital plane ($z=0$) and the second row shows vertical slices perpendicular to the planet\textendash star axis, $y=0$. The columns are for various wind strengths parameterized by: $n_{\star,0}=\SI{1.5e3}{cm^{-3}}$ (left), $n_{\star,0}=\SI{1.5e4}{cm^{-3}}$ (middle), and $n_{\star,0}=\SI{7.0e4}{cm^{-3}}$ (right). For the weakest wind, the arms are able to extend to the boundary of our domain, implying a significant torus-like structure may be possible. For the intermediate-stellar-wind case, the gas is still able to significantly extend outwards but is eventually confined. Lastly, the strongest wind is able to completely confine the outflow into a cometary-like structure. This figure is available online as an animation, showing the evolution of the outflow when a stellar wind is introduced at $t = \SI{5e5}{s}$ until $t = \SI{2e6}{s}$.}
\end{figure*}

To understand the differences between a breeze and a wind, we present a density comparison of the substellar ray for three stellar wind strengths from our \textit{Rogue} run in \autoref{fig:dLine}. Relative to when there is no stellar wind (dotted line), the weak stellar wind (dashed line) agrees well within \SI{\sim6}{\textit{R}_\textrm{p}}, i.e., prior to the planetary gas being shocked. The thickness of the planetary shock is \SI{\sim3}{\textit{R}_\textrm{p}}, extending to the contact discontinuity between the stellar and planetary winds at \SI{\sim 9}{\textit{R}_\textrm{p}}. By comparison the intermediate stellar wind case (solid line) has no shocked planetary wind, and the entire outflow is in causal contact with the stellar wind as the outflow never goes transonic and is a breeze. The zoomed insert demonstrates that interior to the sonic point, the weak-stellar-wind case is in excellent agreement with the no-stellar-wind case, whereas the intermediate stellar wind causes a planetary breeze that significantly diverges from the base case. Taken in conjunction with the fact that no information propagates upstream in a sonic flow, the stellar wind does not affect the planetary outflow interior to the sonic point unless it is strong enough to confine the planetary wind to a breeze.

\begin{figure*}
\centering
\label{fig:full_4var}
\includegraphics[width=\textwidth]{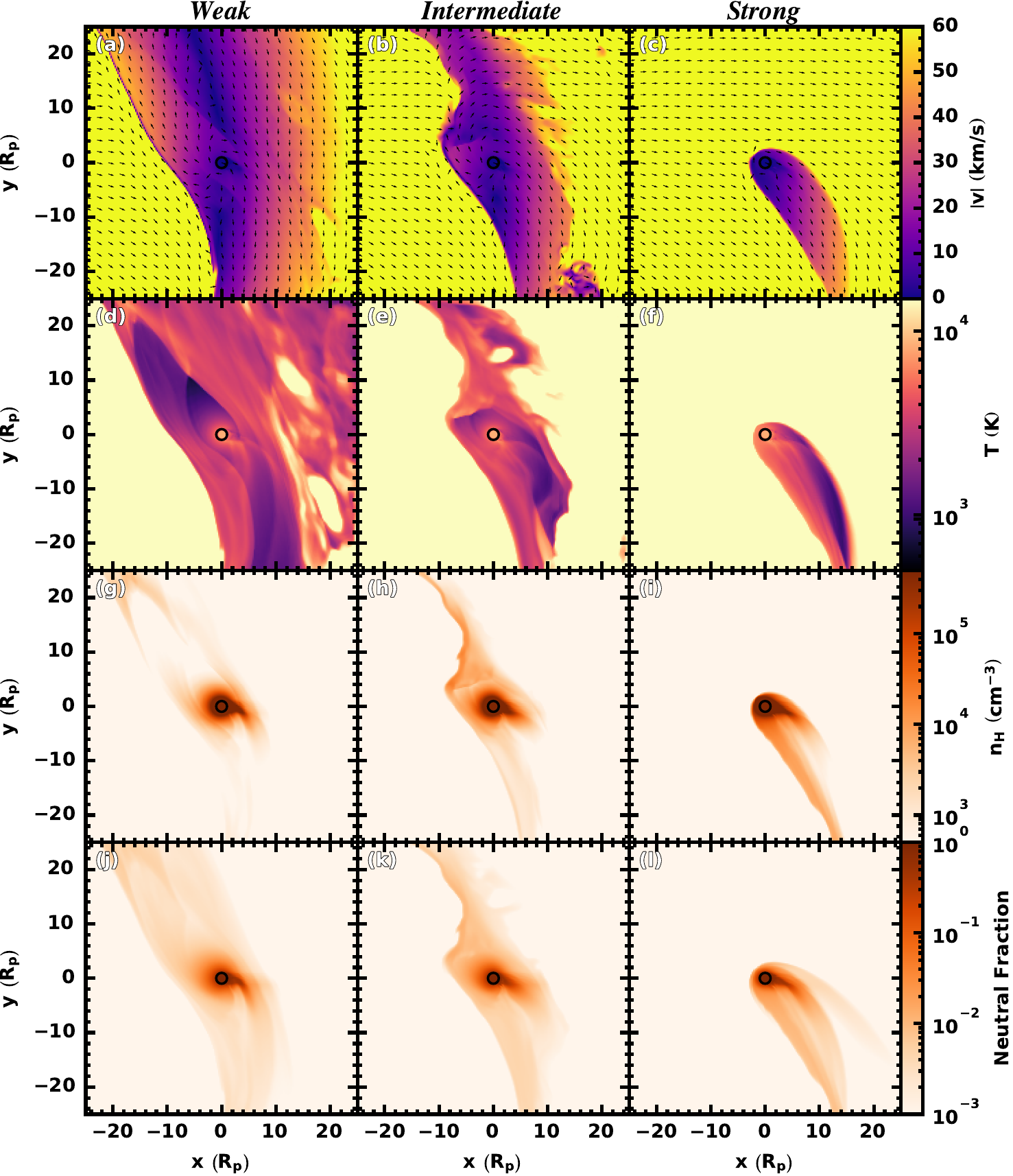}
\caption{The velocity magnitude (top), temperature (top middle), neutral density (bottom middle) and neutral fraction (bottom) for the \textit{Rotating} simulation for three stellar wind strengths: weak ($n_{\star,0}= \SI{1.5e3}{cm^{-3}}$, left), intermediate ($n_{\star,0}=\SI{1.5e4}{cm^{-3}}$, middle) and strong ($n_{\star,0}=\SI{7.0e4}{cm^{-3}}$, right). Quivers in top row indicate direction of velocity, and the black circles demarcate $R_{\textrm{p}}$.}
\end{figure*}

\subsection{Stellar wind in Rotating run: effects of a full stellar environment}
\label{ssec:fullenv}

Finally, we add our three stellar winds into our \textit{Rotating} simulation, shown in \autoref{fig:fullSnap}. The top row are the $x$\textendash $y$ orbital plane and the second rows the $x$\textendash $z$ plane density slices. The columns are from left to right: a weak, intermediate and strong stellar wind. Similar to the \textit{Tidal} simulations with a stellar wind, the three wind strengths exhibit three different regimes. The weak stellar wind (\hyperref[fig:fullSnap]{Figures \ref{fig:fullSnap}(a)} and \hyperref[fig:fullSnap]{\ref{fig:fullSnap}(d)}), is incapable of confining the planetary outflow within our domain. For the intermediate wind, we see a snapshot of the partially-confined planetary wind, which as in the \textit{Tidal} simulation (\autoref{ssec:wind}), undergoes periodic disruption events. Lastly, the strong stellar wind is capable of completely confining the planetary outflow and reducing it to a breeze. Zoomed-in plots of the velocity magnitude, temperature, neutral number density and neutral fraction for each simulation are shown in \autoref{fig:full_4var}. The columns from left to right again correspond to weak, intermediate and strong stellar winds.

\begin{figure}
\centering
\label{fig:full_burp}
\ifnum\movie=0
\includegraphics[width=\columnwidth]{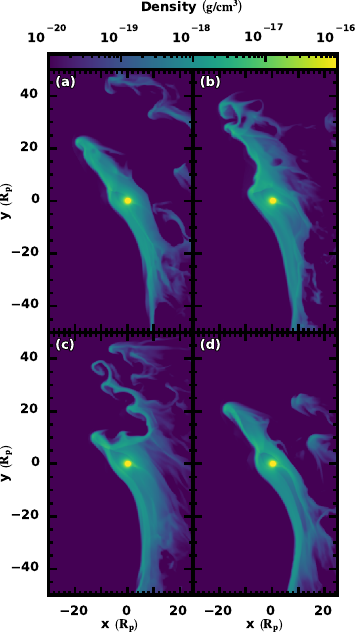}
\else
\includemedia[width=\columnwidth,
playbutton=none,
addresource=movie12.mp4,
flashvars={source=movie12.mp4&loop=true}
]{\includegraphics[width=\columnwidth]{Figure12.pdf}}{VPlayer.swf}
\fi
\caption{Time series of burping in a full-rotating frame for the intermediate wind for the \textit{Rotating} simulation. In panel (a), the outflow is confined within a dayside arm. As the arm continues to grow it get pushed outwards by the stellar wind as seen in panel (b). Eventually the arm grows out too far and is completely blown out from the system and detaches from the rest of the outflow as seen in panel (c). Afterwards in panel (d), we can see that the outflow is in a similar state to our initial panel and the process will repeat. The time between panels (a) and (d) is \SI{1.8e5}{s}. Available online as an animation, showing several periods of the quasi-steady state.}
\end{figure}

The intermediate wind cannot achieve a steady-state standoff shock. The time series for the \textit{Rotating} run with this behavior is displayed in \autoref{fig:full_burp}. Arranged from \hyperref[fig:full_burp]{Figures \ref{fig:full_burp}(a)} to \hyperref[fig:full_burp]{\ref{fig:full_burp}(d)} in chronological order, are snapshots of the $x$\textendash $y$ orbital plane gas density. Initially the planetary outflow is confined, but continues to grow as more gas is liberated from the planet. Eventually the arm grows too large and is blown out, resetting the confined planetary wind to something similar to an earlier state, after which the process repeats.

\refemp{To understand the timescale of the disruption event consider the following. The geometry of the outflow forms a torus around the star. such that the dayside arm is collimated into a cylinder, length $L$ and cross-sectional radius $s$. The mass enclosed in the cylinder grows proportional to $M(t) \propto \dot{M}*t/2$, where the factor of one-half accounts for the two arms of the outflow. Let the length of the cylinder grow at some average velocity $\langle u \rangle$, i.e., $L(t) \propto \langle u \rangle t$. Then a cylinder that has just been disrupted starts to grow according to, $L(t) = \langle u \rangle t$ and $M(t) = \dot{M} t /2$.}

\refemp{We imagine that the inflow ``nozzle'' growing the cylinder from the planetary outflow is sufficiently pressurized as to not be disrupted or displaced. Thus, we say that the cylinder is disrupted when the stellar wind has disconnected the cylinder from this inflow. Equivalently stated, the cylinder has been radially displaced a distance $\Delta d_{\textrm{col.}} = 2s$ away from the star. The stellar wind imparts a force on the column equal to the pressure times the cross sectional area, $A(t) = 2*s*L(t) = 2*s*\langle u \rangle t$, so that $F_{\textrm{wind}}(t) =  A(t) * P_{\star,\textrm{total}}(a) = 2*s * \langle u \rangle * t * P_{\star,\textrm{total}}(a) $. Thus, the acceleration of the cylinder from the stellar wind is $a_{\textrm{col.}}(t) = F_{\textrm{wind}}(t)/M(t) = 4 *s * \langle u \rangle P_{\star,\textrm{total}}(a)/\dot{M}$. Solving for the time of disruption from the equation of motion, $\Delta d_{\textrm{col.}} = a_{\textrm{col.}} t_{\textrm{disrupt}}^2/2 = 2*s$, yields}

\begin{equation}
t_{\textrm{disrupt}} = \sqrt{\frac{2*\dot{M}}{\langle u \rangle * P_{\star,\textrm{total}}(a)}}.
\end{equation}

\refemp{From \autoref{tab:massloss}, the intermediate regime has $\dot{M} = \SI{3.8e10}{g.s^{-1}}$. From \autoref{tab:params} $P_{\star,\textrm{total}}(a) = \SI{1.3e-6}{dyn.cm^{-2}}$. Approximating $\langle u \rangle = \SI{25}{km.s^{-1}}$ from \autoref{fig:full_4var}, we get $t_{\textrm{disrupt}} \approx \SI{1.5e5}{s}$. The time lapsed between \hyperref[fig:full_burp]{\autoref{fig:full_burp}(a)} and \hyperref[fig:full_burp]{\autoref{fig:full_burp}(d)} is $\SI{1.8e5}{s}$, close to our order of magnitude estimate. We can also plug the disruption time into the length at the time of disruption and get $L(t_{\textrm{disrupt}}) \approx 30 * R_{\textrm{p}}$, again in good agreement with what can be seen in \hyperref[fig:full_burp]{\autoref{fig:full_burp}(b)}.}

\begin{figure*}
\label{fig:full_profiles}
\includegraphics[width=.46\textwidth]{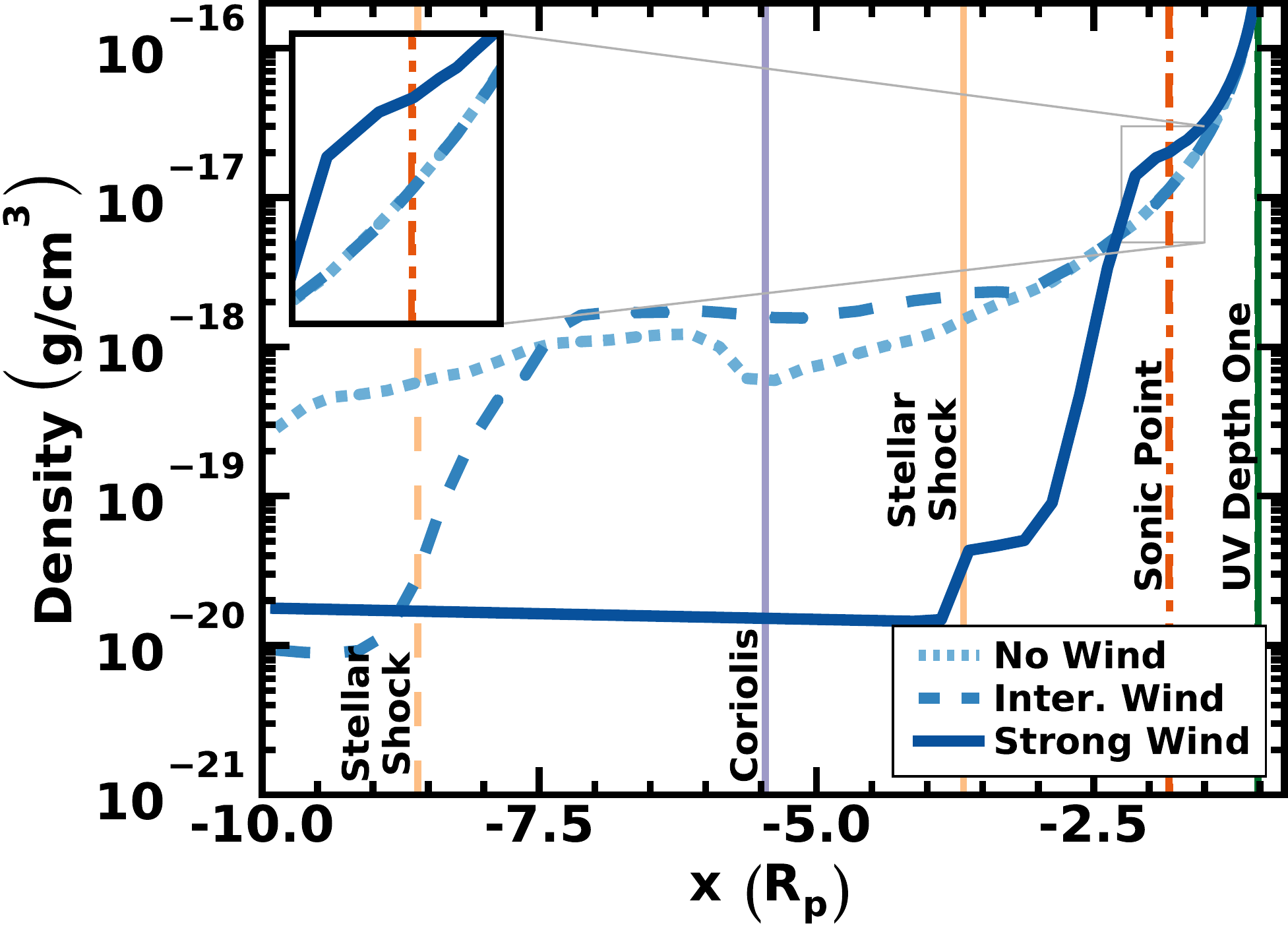}
\hspace{.04\textwidth}
\includegraphics[width=.50\textwidth]{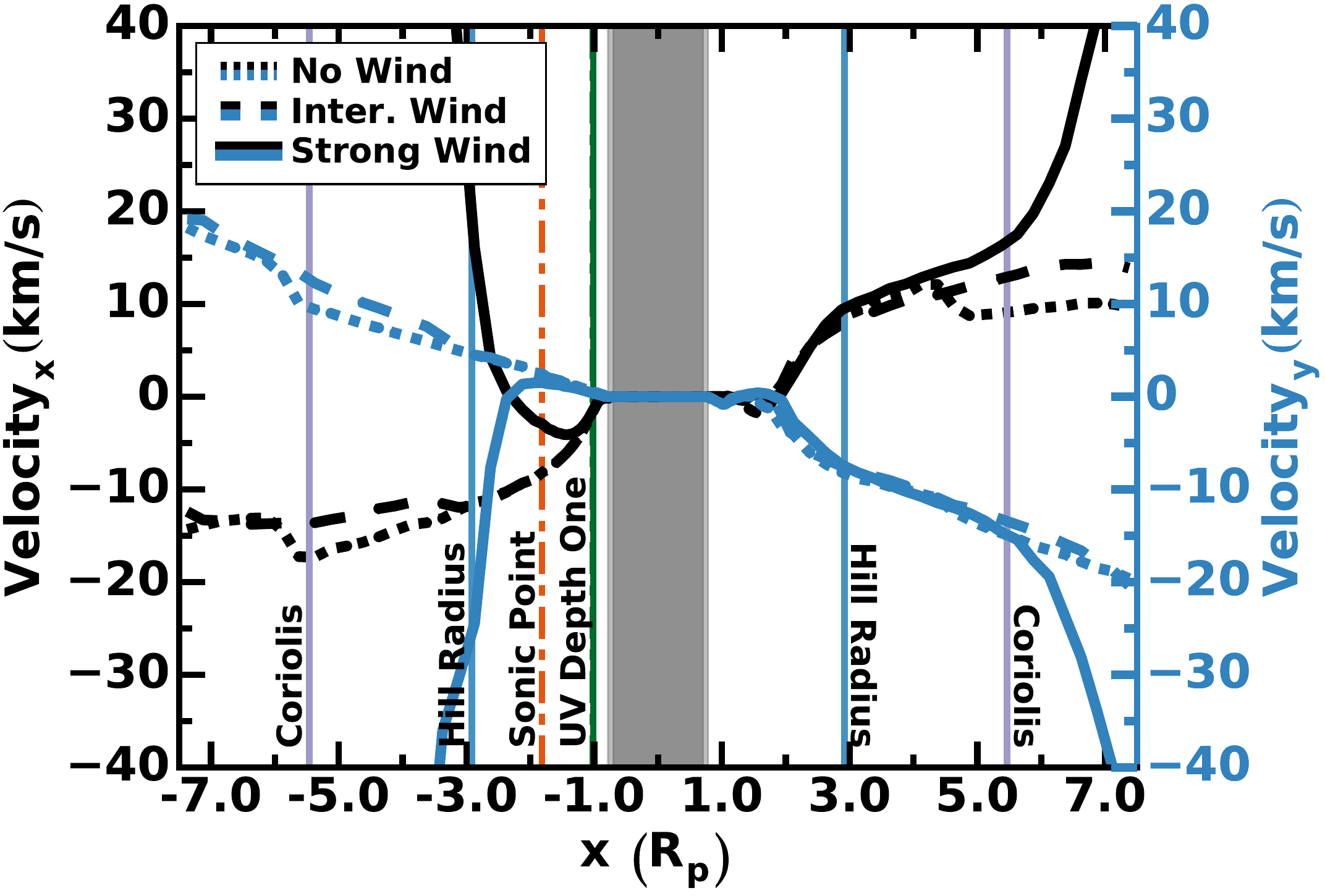}
\caption{\textit{Left}: Density profiles for the no-stellar-wind, intermediate-stellar-wind and strong-stellar-wind cases of the \textit{Rotating} simulation (\hyperref[fig:noWinds]{\autoref{fig:noWinds}(c)} and \hyperref[fig:fullSnap]{Figures~\ref{fig:fullSnap}(b)} and \hyperref[fig:fullSnap]{\ref{fig:fullSnap}(c)} respectively). From the zoomed inset we can see significant deviation in the density interior to the sonic point in the high wind case. Unlike for weaker winds, the high wind case does not have a sonic surface (notice the lack of a solid red line over plotted on the dashed and dotted lines), and has been confined to a breeze. \textit{Right}: Velocity profiles for the same three simulations. Where both the $x$\textendash\ and $y$\textendash velocities are zero is a standoff point between the planetary outflow and the stellar wind. Shocks in the velocity profile can be seen as sudden changes to the velocity profile, near the Coriolis length for the no-stellar-wind case, and purely by coincidence at the Hill Radius for the intermediate-stellar-wind case. The velocity profiles for the no-stellar-wind and intermediate-stellar-wind cases are in excellent agreement prior to shocking.}
\end{figure*}

Velocity and density profiles along the $x$\textendash axis are provided in \autoref{fig:full_profiles}. The zoomed inset again shows significant differences near the wind's sonic point between planetary outflows that are winds (no and intermediate stellar winds) and those that are breezes (strong stellar wind). In the velocity profiles, the locations where the $x$\textendash\ and $y$\textendash velocity profiles intersect at zero are the stagnation or standoff points of the outflow. The shocks can also be seen where the $x$\textendash velocity magnitudes suddenly decrease. Again all velocity profiles for planetary winds are identical until they are shocked on their stellar wind outside their sonic point, while the confined breeze solution differs substantially.

\subsection{Planetary mass-loss rates}
\label{ssec:massloss}

\capstartfalse
\begin{deluxetable}{D{?}{\times}{3.3} D{?}{\times}{0.3} D{?}{\times}{0.3} D{?}{\times}{0.3}}
\tabletypesize{\footnotesize} 
\tablecolumns{4} 
\tablewidth{\columnwidth} 
\tablecaption{Planetary mass-loss rates \label{tab:massloss}} 
\tablehead{\colhead{$n_{\star,0}$ (cm$^{-3}$)} & \colhead{\textit{Rogue} (\si{g.s^{-1}})} & \colhead{\textit{Tidal} (\si{g.s^{-1}})} & \colhead{\textit{Rotating} (\si{g.s^{-1}})}}
\startdata 
0.0             & 4.0 ? 10^{10} & 3.9 ? 10^{10} & 3.8 ? 10^{10} \\
1.5 ? 10^3 & 3.9 ? 10^{10} & 3.8 ? 10^{10} & 3.8 ? 10^{10} \\
1.5 ? 10^4 & 3.9 ? 10^{10} & 3.8 ? 10^{10} & 3.8 ? 10^{10} \\
7.0 ? 10^4 & 3.6 ? 10^{10} & 3.5 ? 10^{10} & 3.4 ? 10^{10} \\
\enddata 
\tablecomments{The time-averaged mass-loss rate of the sustained planetary outflow ($\SI{1e6}{s} \le t \le \SI{2e6}{s}$).}
\end{deluxetable}
\capstarttrue

For the four wind strengths studied here in the \textit{Rotating} case, \autoref{tab:massloss} list the mass-loss rates averaged over the last \SI{e6}{s} of our simulation. To determine the mass-loss rates we choose a sphere of radius larger than our reset radius and calculate the flux of material through the sphere. We use the marching cubes algorithm to generate a polygonal mesh of our data, the normal of which is given by the gradient. Then by projecting the interpolated velocity into this normal and multiplying by the interpolated density and area of a mesh element, we get the mass-loss rate through that surface. Summing the mass-loss rate over the entire triangulated sphere we then arrive at the total mass-loss rate of the planet. We perform the calculation at \SI{0.8}{\textit{R}_p},  \SI{2}{\textit{R}_p}, and \SI{15}{\textit{R}_p} and find excellent agreement. Presented here are the time-averaged results averaged over those three radii.

Notice that all rates for a given simulation are within \SI{5}{\percent} of each other across all wind strengths except for the strongest wind ($n_{\star,0} = \SI{7.0e4}{cm^{-3}}$). In this case, the outflow is a breeze and has a noticeably lower mass-loss rate, consistently about \SI{10}{\percent} lower than the average of the other three wind strengths. This is in agreement with the fact that the stellar wind is now in direct contact with the planetary wind base, which adds additional pressure that the planetary breeze must overcome.
 
These mass-loss rates are about an order of magnitude lower than those found in the fiducial case of Tripathi et al. \citeyear{Tripathi15}, who modeled an order of magnitude higher ionizing flux for a planet twice as massive in a simulation analogous to our \textit{Tidal} simulation with $n_{\star,0} = 0.0$. Their mass-loss rate of $\dot{M} = \SI{1.9e11}{g.s^{-1}}$, is in excellent agreement with ours when adjusted using the scaling from Murray-Clay et al. (\citeyear{Murray09}) for energy limited escape, $\dot{M} \propto M_\textrm{p}^{-1} F_0$. Furthermore, our mass-loss rate for the \textit{Rotating} simulation without a stellar wind finds good agreement with Debrecht et al. (\citeyear{Debrecht18}). Here they explored a similar planet and ionizing flux,\footnote{Same mass and flux; however, different initial conditions led to a different $\tau = 1$ surface.} finding an $\dot{M} = \SI{3.35e10}{g.s^{-1}}$.

For reference, in \autoref{tab:radius} we provide the time-averaged sonic radius and the radius where optical depth to ionizing photons is one for our simulations. These either did not change over the wind strengths, or were non-existent such as the sonic radius for the breezes. The only discernible difference is in the sonic radius between the \textit{Rogue} and \textit{Tidal} or \textit{Rotating} simulations. This difference results from the tidal forces helping to accelerate the gas to supersonic velocities earlier than occurs in the \textit{Rogue} simulation.

\capstartfalse
\begin{deluxetable}{c D{?}{.}{0.3} D{?}{.}{0.3} D{?}{.}{0.3}}
\tabletypesize{\footnotesize} 
\tablecolumns{4} 
\tablewidth{\columnwidth} 
\tablecaption{Length scales in outflows \label{tab:radius}} 
\tablehead{\colhead{} & \colhead{\textit{Rogue} ($R_{\textrm{p}}$)} & \colhead{\textit{Tidal} ($R_{\textrm{p}}$)} & \colhead{\textit{Rotating} ($R_{\textrm{p}}$)}}
\startdata 
$r_{\tau = 1}           $ & 1?02 & 1?02 & 1?02\\
$r_{\textrm{sonic}} $ & 1?97 & 1?81 & 1?81
\enddata 
\tablecomments{The substellar radii of various length scales.}
\end{deluxetable}
 \capstarttrue

\section{Observational Consequences:\\Lyman-$\mkern-0.5mu \upalpha$ transits and spectrum}
\label{sec:obscon}

To explore the observational consequences of the features found in \autoref{sec:results}, we perform synthetic observations of the \textit{Rotating} runs across three stellar wind strengths. Through these synthetic observations we are able to probe the spatial extent of escaping hydrogen via obscuration maps, transit light curves, and spectra. A number of lines have been seen in escaping planetary winds. We focus our synthetic observations around \LymanA{} as it is the feature with the greatest number of high-confidence detections. By comparing features seen in synthetic observations, we can search for unique signatures to observationally disentangle our different physical scenarios.

The synthetic observation procedure is discussed in \autoref{ssec:methobs}, and the results for our three scenarios are presented in \autoref{ssec:synobs}.  We comment on the conditions in which we expect observational signals to persist in \autoref{ssec:largeobs}. In \autoref{sec:realobs} we review current observations and tentatively suggest which regimes they may fall into. Note that our simulation is for a generic hot Jupiter, and we only seek to provide context for features that have already been observed and predict some we may yet to discover.

\subsection{Procedure}
\label{ssec:methobs}

\begin{figure}
\centering
\label{fig:raydomain}
\includegraphics[width=\columnwidth]{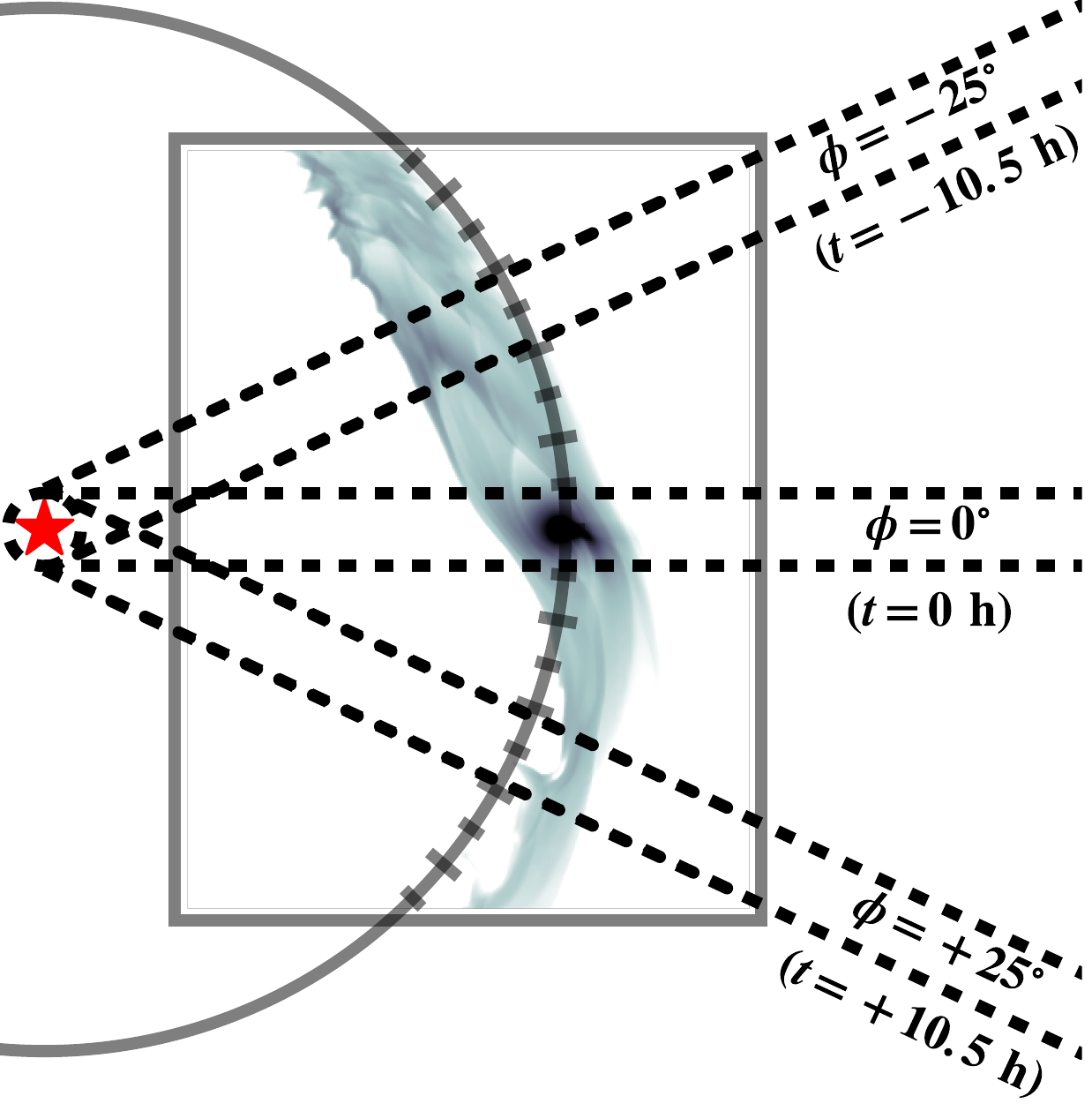}
\caption{Schematic of viewing angle through the simulation (bounded by the rectangle) in the orbital plane. The dashed lines show the extent of the outflow transiting the star at a given orbital phase. The half-circle centered on the star represents the orbit of the planet around the star, with major ticks every \SI{10}{\degree} and minor ticks every \SI{5}{\degree}. While the image plane is defined on this circle, rays extend through the entire domain by entering and exiting at the boundaries. We only take our observations out to $\phi = \SI{\pm 25}{\degree}$, as for some simulations, larger angles do not intersect any of the outflow. All distances and sizes are to scale. Refer to Figure 11 of Carroll-Nellenback et al. (\citeyear{Carroll17}) for a global schematic.}
\end{figure}

To perform synthetic observations we begin by constructing an image plane defined by $x=0$, which shares its origin with the planet and is perpendicular to the substellar ray between the planet and the star. We then solve the equation of radiative transfer along rays that emanate from the observer through each pixel in the image plane. The observer is arbitrarily chosen to be one-hundred parsecs away from the system's barycenter on the opposite side of the image plane from the star, such that rays within the domain are virtually parallel. This produces an obscuration map appropriate for the middle of the planet's transit across the star. We produce observations at various orbital phases by rotating the image plane around the barycenter, \autoref{fig:raydomain} (see \S\! 2.8 of Murray and Dermott (\citeyear{Murray99}) for an example of a rotation matrix in terms of Keplerian elements). As our simulation is local, the orbital phases which we can observe are limited by the box size. Yet more limiting is the extent of the outflow in our domain, which in practice conservatively limits the accessible orbital phases to $\phi = \SI{\pm 25}{\degree}$. 

We do not model the stellar Lyman-$\alpha$ spectrum or interstellar extinction in this work. We instead calculate the optical depth through the planetary outflow as a function of frequency by integrating the absorption coefficient along each ray, $\tau_\nu = \int \alpha_{\Lya}(\nu,s) \textrm{d} s$. The absorption coefficient for Lyman-$\alpha$ is given by

\begin{equation}
\alpha_{\Lya}(\nu, s) = n_\textrm{HI}(s) \frac{\pi e^2}{m_{e} c} f_{12} \left(1-\textrm{e}^{-h\nu/k_{\textrm{B}}T(s)}\right) \phi_{\textrm{Voigt}}(\nu,s).
\end{equation}

\noindent Here $e$ is the elementary charge, $m_e$ the mass of an electron, $c$ the speed of light, $f_{12} = 0.4164$ the absorption oscillator strength between $n=1$ and $n=2$ for hydrogen and $\phi_{\textrm{Voigt}}(\nu,s)$ is the Voigt profile. Note that the term in the parentheses is a correction to absorption for stimulated emission. Since our simulations are already in the star's rest frame, when calculating the Voigt line profile we only need to shift the Doppler broadening component by the projected bulk velocity of the fluid. We take for granted that observers account for systematic velocities, e.g., the relative velocity between the system and the observer and orbital motion of the star, and neglect these in our presentation.

From the optical depth, we calculate the stellar disc averaged obscuration fraction

\begin{equation}
\langle \mathcal{O}_\nu \rangle^{}_{\!*}  = \iint_{S} \left(\frac{I_0-I}{I_0}\right) \, \frac{\textrm{d} \Sigma}{A} = \iint_{S} \left(1 - \textrm{e}^{-\tau_\nu} \right) \, \frac{\textrm{d} \Sigma}{A},
\end{equation}

\noindent for comparisons to observational spectra. Here $A$ is the area of the stellar disc defined as the surface $S$, $\textrm{d} \Sigma$ is the infinitesimal area element, $I$ is the intensity observed and $I_0$ the intensity of light emitted from the star. Note we only consider absorption without emission, such that $\tau_\nu = \ln(I_0/I)$. We do not model the stellar line, nor do we consider spatial variation across the stellar surface, including limb effects. Typically obscuration will be strongest near line center, where observers cannot make accurate measurements due to geocoronal confusion and interstellar extinction, and weakest at large effective Doppler velocities. Thus, a quoted transit obscuration will depend on the frequency domain of the spectrograph and methodology for removing geocoronal confusion.

For our synthetic observations we chose to measure the obscuration between \SIrange{1215.26}{1216.08}{\angstrom} (equivalent Doppler velocities of \SIrange{-100}{100}{km.s^{-1}}), ignoring obscuration between \SIrange{1215.53}{1215.81}{\angstrom} (\SIrange{-35}{35}{km.s^{-1}}) to model geocoronal confusion and ISM absorption. Note that negative velocities probe gas moving away from the star with blueshifted absorption, and conversely positive velocities correspond to gas infalling towards the star with redshifted absorption. Within this spectral range we further subdivide our observations into four unique frequency domains to retrieve dimensionless equivalent widths for transit observations. The dimensionless equivalent widths are calculated as

\begin{equation}
W = \int_{\nu_a}^{\nu_b} \langle \mathcal{O}_\nu \rangle^{}_{\!*} \, \frac{\textrm{d} \nu}{\Delta \nu} = \int_{\nu_a}^{\nu_b} \langle 1 - \textrm{e}^{-\tau_\nu} \rangle^{}_{\!*} \, \frac{\textrm{d} \nu}{\Delta \nu}.
\end{equation}

\noindent Here $\nu_a$ and $\nu_b$ define the spectral range with a width of $\Delta \nu = \nu_b - \nu_a$. Observations integrated over the red range, \SIrange{1215.81}{1216.06}{\angstrom} (\SIrange{35}{100}{km.s^{-1}}), are referred to as ``Red'' and those over the blue range, \SIrange{1215.26}{1215.53}{\angstrom} (\SIrange{-100}{-35}{km.s^{-1}}), as ``Blue.'' The measurement called ``Both''  refers to an integration over the entire observational range (excluding the geocornal confusion region), which covers both the red and blue wings.

The last measurement called ``Full'' is integrated over the full observed spectrum as if there existed no geocoronal confusion or interstellar extinction. Typically obscuration is largest near line-center and the ``Full'' line displays the strongest obscuration. However, when the bulk velocity significantly shifts the Doppler broadening core away from line center, the dimensionless equivalent width will be larger in the respective wing as the wings are normalized over a smaller $\Delta \nu$. As the ``Full'' line probes the total column density, it is suggestive of other hydrogen lines, or metals with abundance proportional to hydrogen, that have no confusion at line center.

As light rays take approximately \SI{37.5}{s} to traverse the domain's width of $75 * R_{\textrm{p}}$, during which the fluid structure does not significantly evolve, we are well justified in using the fast-light approximation, and produce synthetic observations in post-processing. Less obviously justified is using a single static output for the entire orbit when performing our synthetic transits. Yet, there is only one regime of stellar winds, the intermediate regime, in which the simulation does not reach a steady-state. Using appropriately timed outputs from the hydrodynamic simulation would have no effect for systems in steady-state, so the only concern would be for the intermediate stellar wind regime. Yet, since the transit duration of the planetary disc takes \SI{\sim4}{\hour} and the non-steady-state intermediate regime has a periodicity of \SI{\sim 50}{\hour}, it is acceptable to use a static frame for the duration of a single transit observation. As it is arbitrary when a transit might occur during the periodic disruptions, since the disruption is not related to the orbital period but rather the stellar ionizing flux and wind strength, we perform synthetic observations at the three different phases shown in \hyperref[fig:full_burp]{Figures \ref{fig:full_burp}(a)}, \hyperref[fig:full_burp]{\ref{fig:full_burp}(b)}, and \hyperref[fig:full_burp]{\ref{fig:full_burp}(c)}.

\subsection{Synthetic observations of planetary escape in stellar environments}
\label{ssec:synobs}

\subsubsection{Weak-stellar-wind regime}
\label{sssec:weak}

\begin{figure}
\label{fig:wwind}
\ifnum\movie=0
\includegraphics[width=\columnwidth]{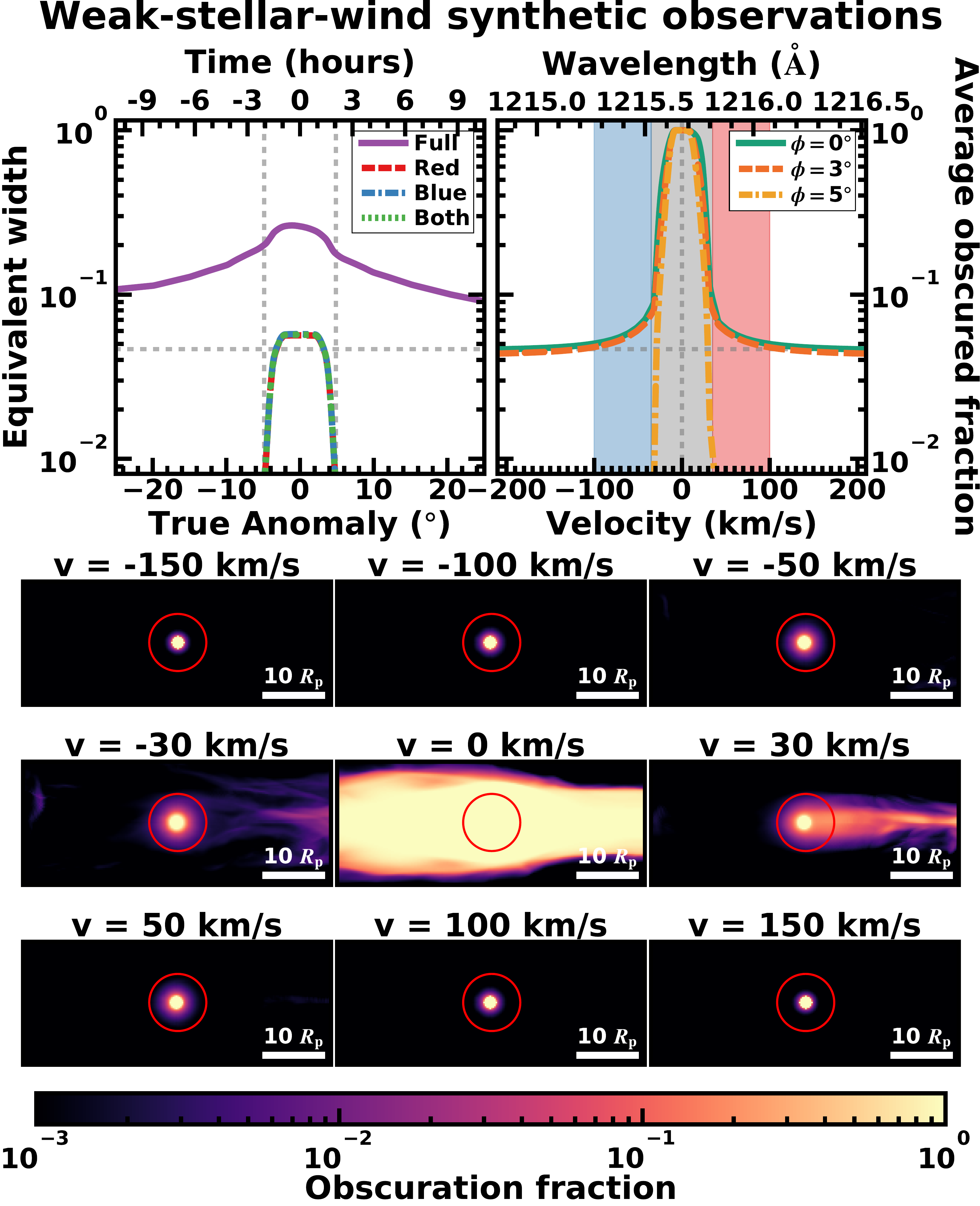}
\else
\includemedia[width=\columnwidth,
playbutton=none,
addresource=movie15.mp4,
flashvars={source=movie15.mp4&loop=true}
]{\includegraphics[width=\columnwidth]{Figure15.pdf}}{VPlayer.swf}
\fi
\caption{\textit{Upper Left}: The transit (duration shown in vertical dashed lines) for the weak stellar (\hyperref[fig:fullSnap]{Figure \ref{fig:fullSnap}(a)}) wind shows about a \SI{1}{\percent} enhancement at observable frequencies (Red, Blue, Both) over the planetary disc (dashed horizontal line). The Full line is for observations without geocoronal confusion and interstellar extinction, telling us that most of the obscuration occurs in this domain. \textit{Upper Right}: The spectrum is symmetric about line-center (vertical dashed line) with some natural broadening enhancement over the planetary disc (horizontal dashed line) for the red and blue regions during transit (\SI{0}{\degree} and \SI{3}{\degree}). Out of transit (\SI{5}{\degree}) the only appreciable absorption occurs from Doppler broadening in the geocoronal confusion domain, which is unobservable. \textit{Lower Panels}: Snapshots at direct transit of the $(50\times20) * R_{\textrm{p}}$ spatial obscuration at various spectral frequencies denoted by their equivalent Doppler velocities. The stellar disc is denoted by the red circle and the white bar provides the length scale at the plane of the planet. All line of sight projected velocities are below \SI{50}{km.s^{-1}}, with the fastest perhaps detectable feature being the dayside arm at positive velocities (redshifted absorption) occurring before direct transit. Available online as an animation, showing transit in the obscuration maps for the Red, Blue and Both frequency domains, as well as a sweep of the spectrum at $\phi = \SI{0}{\degree}$.}
\end{figure}

With a weak stellar wind, observations are symmetric both in the spectra and transit light curves, as seen in \autoref{fig:wwind}. The observable obscuration comes primarily from the disc of the planet which occults \SI{4.6}{\percent} of the stellar disc during its transit between $\phi = \SI{\pm4.9}{\degree}$, or between $t = \pm 2$ hours relative to direct transit. Since velocities of the outflowing planetary wind are well within the geocoronal confusion and interstellar extinction limits, the outflow contributes a small enhancement (\SI{\sim 1}{\percent}) in the ``Red'' and ``Blue'' spectral regions due to natural broadening near mid-transit, when the absorption column is largest.

Given that significant detections have been made far outside the geocoronal confusion limits, yet our parameters show no such features, it is worth asking how much the outflow velocity would need to increase for our results to produce significant absorption in the ``Red'' and ``Blue'' wings. To explore this question we consider an alternative frequency domain from \SIrange{1215.33}{1216.02}{\angstrom} (\SIrange{-85}{85}{km.s^{-1}}), which we call our ``Low'' domain, and ignore obscuration between \SIrange{1215.59}{1215.75}{\angstrom} (\SIrange{-20}{20}{km.s^{-1}}). We note that our weak stellar wind is the only scenario which does not produce significant obscuration in our fiducial spectral ranges, so we only consider the ``Low'' spectral range here. This adjustment mimics making observations of outflows boosted by \SI{15}{km.s^{-1}} in our standard spectral domains, which we justify as follows.

Since natural broadening wings are not substantial in our weak wind spectra, the line profile for the obscuration is primarily due to Doppler broadening. Hence, where absorption is significant, $\phi_\nu \propto \exp(-u^2/b^2)$, where $u$ is the velocity and $b = \sqrt{2 k T/\mu}$ is the Doppler broadening parameter. Then integrating over our ``Low'' domain is equivalent to integrating over our standard domain in a wind of faster outflow velocities

\begin{equation}
\label{eq:DopShift}
\int_{u_a\pm s}^{u_b\pm s} \textrm{e}^{-u^2/b^2} \, \textrm{d} u = \int_{u_a}^{u_b} \textrm{e}^{-w^2/b^2} \, \textrm{d} w.
\end{equation}

\noindent Here $[u_a, u_b]$ is our fiducial range and $s$ is the shifted boost which relates our ``Low'' domain to our fiducial range.  Note that, $w = u \mp s$ with a top sign for the red range and bottom sign for blue. Thus from \cref{eq:DopShift}, we find that comparisons between our shifted ``Low'' observations and our standard observations approximately corresponds to comparing to an outflow boosted by \SI{15}{km.s^{-1}}. This is a first-order approximation as outflows with larger velocities may have different neutral fractions and overall structure.

\begin{figure}
\label{fig:fast}
\includegraphics[width=\columnwidth]{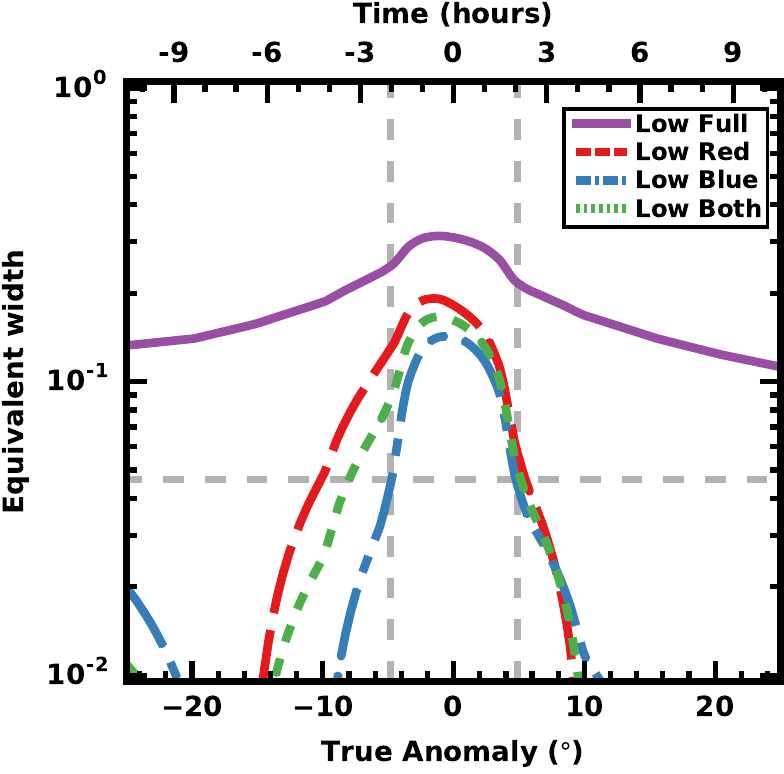}
\caption{Transit light curve of an outflow interacting with a weak stellar wind, observed in our ``Low'' wavelength window of \SIrange{1215.33}{1216.02}{\angstrom} (\SIrange{-85}{85}{km.s^{-1}}). These observations, when compared with our standard window \SIrange{1215.26}{1216.08}{\angstrom} (\SIrange{-100}{100}{km.s^{-1}}), approximate planetary outflows boosted by \SI{15}{km.s^{-1}}. With larger outflow velocities, significantly more obscuration occurs. Before transit there is significant absorption at redder wavelengths due to the leading dayside arm outflow infalling towards the star. Long before transit, blue absorption is seen due to the Coriolis force turning the dayside arm at large distances. The nightside outflow, probed after transit, is slower and produces a weaker signal. Absorption is dominated by the naturally-broadened line wings and is roughly symmetrical between the red and blue wing.}
\end{figure}

The transit measurement for the ``Low'' domain is presented in \autoref{fig:fast}. Compared to the fiducial observational range, the absorption is still symmetric at $\phi = \SI{0}{\degree}$, while prior to $\phi = \SI{0}{\degree}$ there is substantially more red absorption. Conversely, roughly three hours after transit blue is more dominant, though at a much lower amplitude than red is prior to transit. This is due to which arm of the outflow is being probed. The dayside arm (redshift) moves faster than the nightside arm moving away from the star (blueshift). Thus, a more significant absorption feature can be seen in the red wings of the spectra leading the direct transit (see the $v = \SI{30}{km.s^{-1}}$ snapshot of \autoref{fig:wwind}). Note that in the dayside arm, some gas is still moving away from the star as it has been turned by the Coriolis force (see \hyperref[fig:full_4var]{Figure \ref{fig:full_4var}(a)}). This becomes more pronounced at larger distances from the planet, and the ``Blue'' measurement eventually dominates over ``Red'' prior to $\phi = \SI{-20}{\degree}$.

\subsubsection{Intermediate-stellar-wind regime}
\label{sssec:intermed}

\begin{figure}
\label{fig:iwind}
\ifnum\movie=0
\includegraphics[width=\columnwidth]{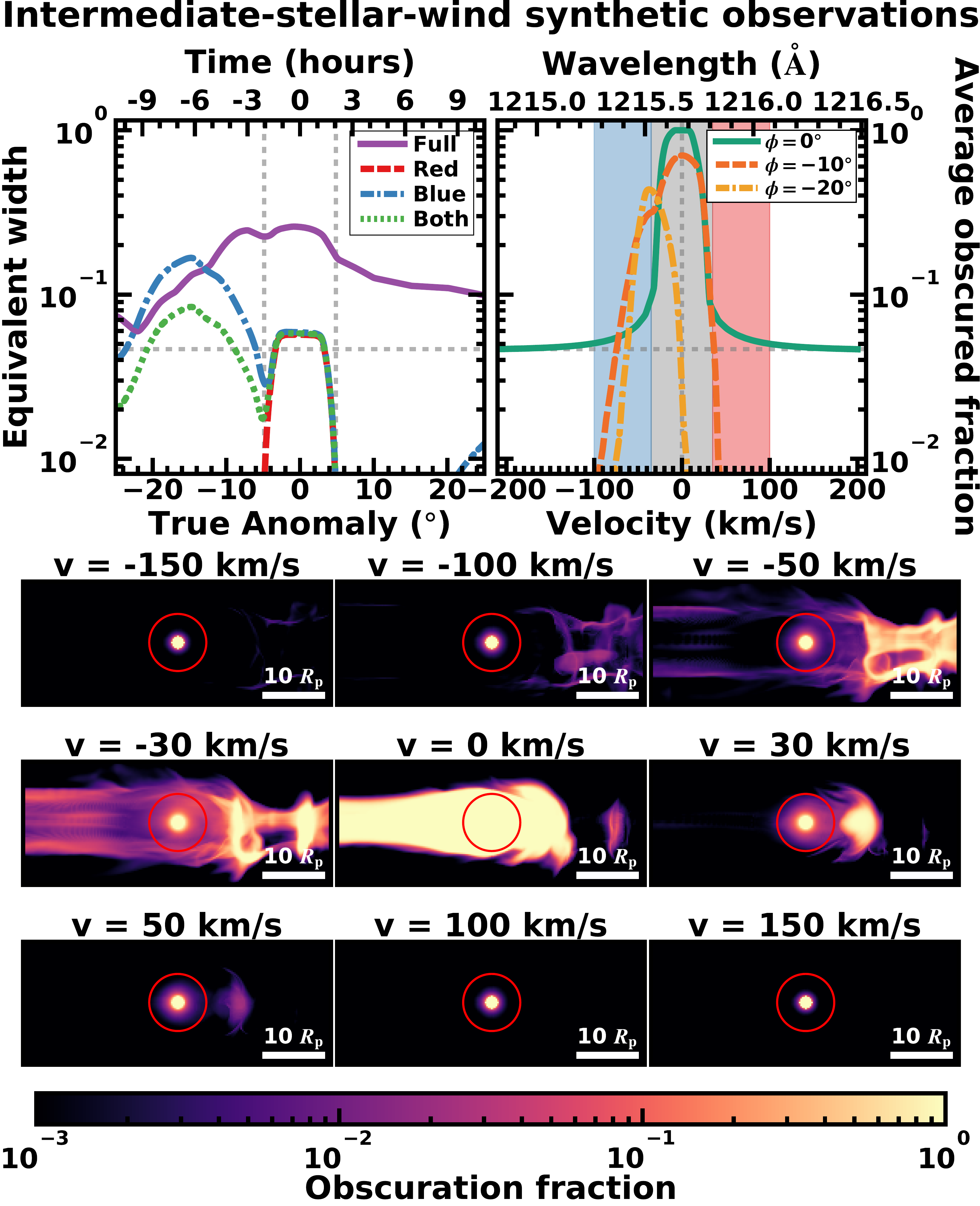}
\else
\includemedia[width=\columnwidth,
playbutton=none,
addresource=movie17.mp4,
flashvars={source=movie17.mp4&loop=true}
]{\includegraphics[width=\columnwidth]{Figure17.pdf}}{VPlayer.swf}
\fi
\caption{Panels have the same layout as \autoref{fig:wwind}, but are now for our intermediate stellar wind (specifically \hyperref[fig:full_burp]{Figure \ref{fig:full_burp}(b)}). \textit{Upper Left}: Only \SI{\sim 1}{\percent} enhancement over the planetary disc occultation occurs during transit, but prior to the transit strong blue features are visible.\textit{Upper Right}: No significant spectral asymmetry is apparent during direct transit. Prior to transit, blue features dominate. \textit{Lower Panels}: Between \SIrange{0}{50}{km.s^{-1}} one can clearly see the truncated dayside arm, while at large negative velocities we can see that further upstream the extended disrupted outflow causing significant obscuration. Available online as an animation, showing transit in the obscuration maps for the Red, Blue and Both frequency domains, as well as a sweep of the spectrum at $\phi = \SI{0}{\degree}$.}
\end{figure}

\begin{figure}
\label{fig:iwindtresolved}
\includegraphics[width=\columnwidth]{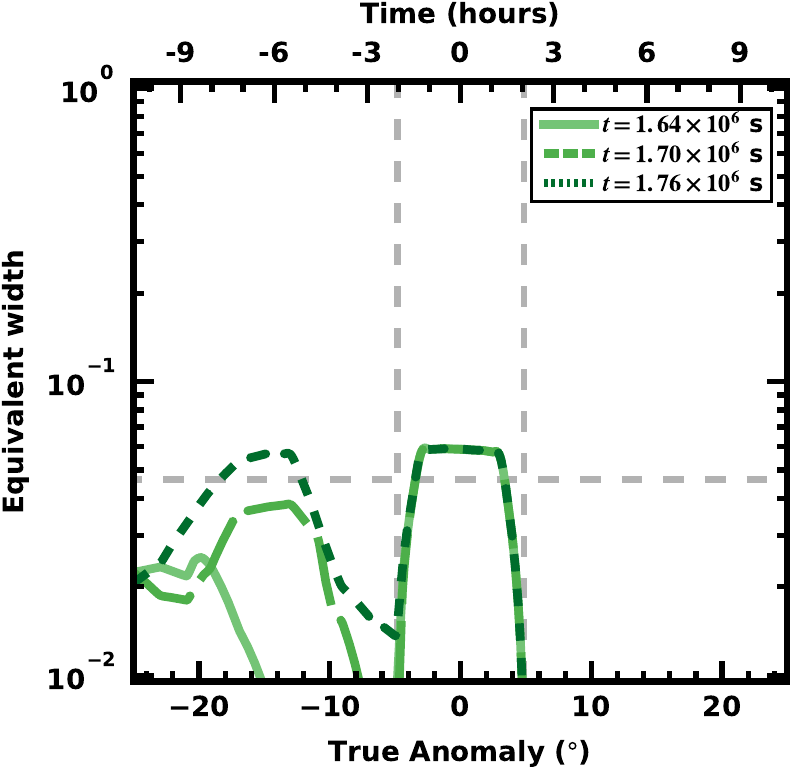}
\caption{\refemp{Three time-resolved transit observations of the unresolved \LymanA{} line (our Both domain) for the intermediate stellar wind. Each transit is labeled by the time corresponding to $\phi = \SI{-25}{\degree}$, which in time order correspond to \hyperref[fig:full_burp]{Figure \ref{fig:full_burp}(a)}, \hyperref[fig:full_burp]{\ref{fig:full_burp}(b)}, and \hyperref[fig:full_burp]{\ref{fig:full_burp}(c)}. Given the non-static disrupting dayside outflow, the pre-transit signal varies throughout the various stages of the disruption. The signal typically appears to resemble a ``double transit,'' in which a comparable signal to the transiting body appears and disappears prior to the transit of the planet. We note that this feature is strong when only the blue wing of the line is consider (\textit{c.f.} \autoref{fig:iwind}).}}
\end{figure}

For the intermediate-stellar-wind regime, we generated synthetic observations from the static frames of \hyperref[fig:full_burp]{Figures \ref{fig:full_burp}(a)}, \hyperref[fig:full_burp]{\ref{fig:full_burp}(b)}, and \hyperref[fig:full_burp]{\ref{fig:full_burp}(c)}. We found that though the transit signal quantitatively differs between snapshots, it qualitatively remains the same. Therefore, in \autoref{fig:iwind} we present the results from the \hyperref[fig:full_burp]{Figure \ref{fig:full_burp}(b)} snapshot, which showed the strongest signal, to probe the general features of the intermediate regime. As discussed in \autoref{ssec:methobs}, the disruption timescale is $\SI{\sim 50}{h}$ and a transit is only $\SI{\sim4}{h}$. \refemp{While using a static output is accurate enough for transit, if we want to create accurate transit light curves spanning $\SI{\sim22}{h}$ ($\phi = [-25^{\circ},25^{\circ}])$, we need to perform observations at times consistent with the orbital phases. Therefore, we also produce three time-resolved transit light curves displayed in \autoref{fig:iwindtresolved}, beginning at $\phi = \SI{-25}{\degree}$ for \hyperref[fig:full_burp]{Figures \ref{fig:full_burp}(a)}, \hyperref[fig:full_burp]{\ref{fig:full_burp}(b)}, and \hyperref[fig:full_burp]{\ref{fig:full_burp}(c)}.}

As in the weak stellar wind case, during the planetary transit there is not much enhancement over the planetary disc. However, unlike for the weak stellar wind, we find a significant feature that precedes the planetary transit within our standard frequency domain. As seen in the spectra of \autoref{fig:iwind}, when probing the dayside arm at $\phi = \SI{-10}{\degree}$ and $\phi = \SI{-20}{\degree}$ the features are not from the naturally-broadened wings of a Voigt profile, but from the Doppler-broadened core. Therefore, while the stellar wind may enhance the column density, the primary cause of the increased obscuration is due to negatively-accelerated material. In other words, the stellar wind is pushing the outflowing gas towards the observer.

The strong blue-shifted absorption signal is strongest, \SI{\sim 20}{\percent} for \hyperref[fig:full_burp]{Figure \ref{fig:full_burp}(b)}, but is present in all three snapshots in time. The major difference is the amplitude of the signal, which is weakest for \hyperref[fig:full_burp]{Figure \ref{fig:full_burp}(c)} at \SI{\sim5}{\percent}. Comparisons between \hyperref[fig:full_burp]{Figures \ref{fig:full_burp}(b)} and \hyperref[fig:full_burp]{\ref{fig:full_burp}(c)} show that these two phases have the most and least amount of dense gas being blown out by the stellar wind. Whether the pre-transit signal ever disappears depends on the timescale for the stellar wind to disperse the dayside material, and the timescale for another disruption to occur. For our parameters, neutral material outwardly accelerated by the stellar wind is always around, and the signal is continuously present but modulated in amplitude.

In \autoref{fig:iwindtresolved} we time-resolve the transit (in $\Delta \phi = \SI{2}{\degree}$ increments) by ray-tracing through a series of appropriate simulation snapshots. We chose three times that correspond to $\phi = \SI{-25}{\degree}$, thereby measuring the transit at various disruption stages (those being \hyperref[fig:full_burp]{Figure \ref{fig:full_burp}(a)}, \hyperref[fig:full_burp]{\ref{fig:full_burp}(b)}, and \hyperref[fig:full_burp]{\ref{fig:full_burp}(c)}). Generally, the signal seen mimics two bodies transiting the star in quick succession\textendash a double transit. The first ``object'' (which is just the outwardly accelerated dayside outflow) enters transit tens of hours prior to the true transit of the planetary body. This would be easy to disentangle with a spectrally resolved observation as it is entirely biased bluewards of line center; however, as seen in \autoref{fig:iwindtresolved}, an unresolved observation could produce such ``double transits.'' Additionally, an optical transit observation would not find any such ``Trojan body,'' indicating it was instead neutral hydrogen from the dayside outflow.

\subsubsection{Strong-stellar-wind regime}
\label{sssec:strong}

\begin{figure}
\label{fig:swind}
\ifnum\movie=0
\includegraphics[width=\columnwidth]{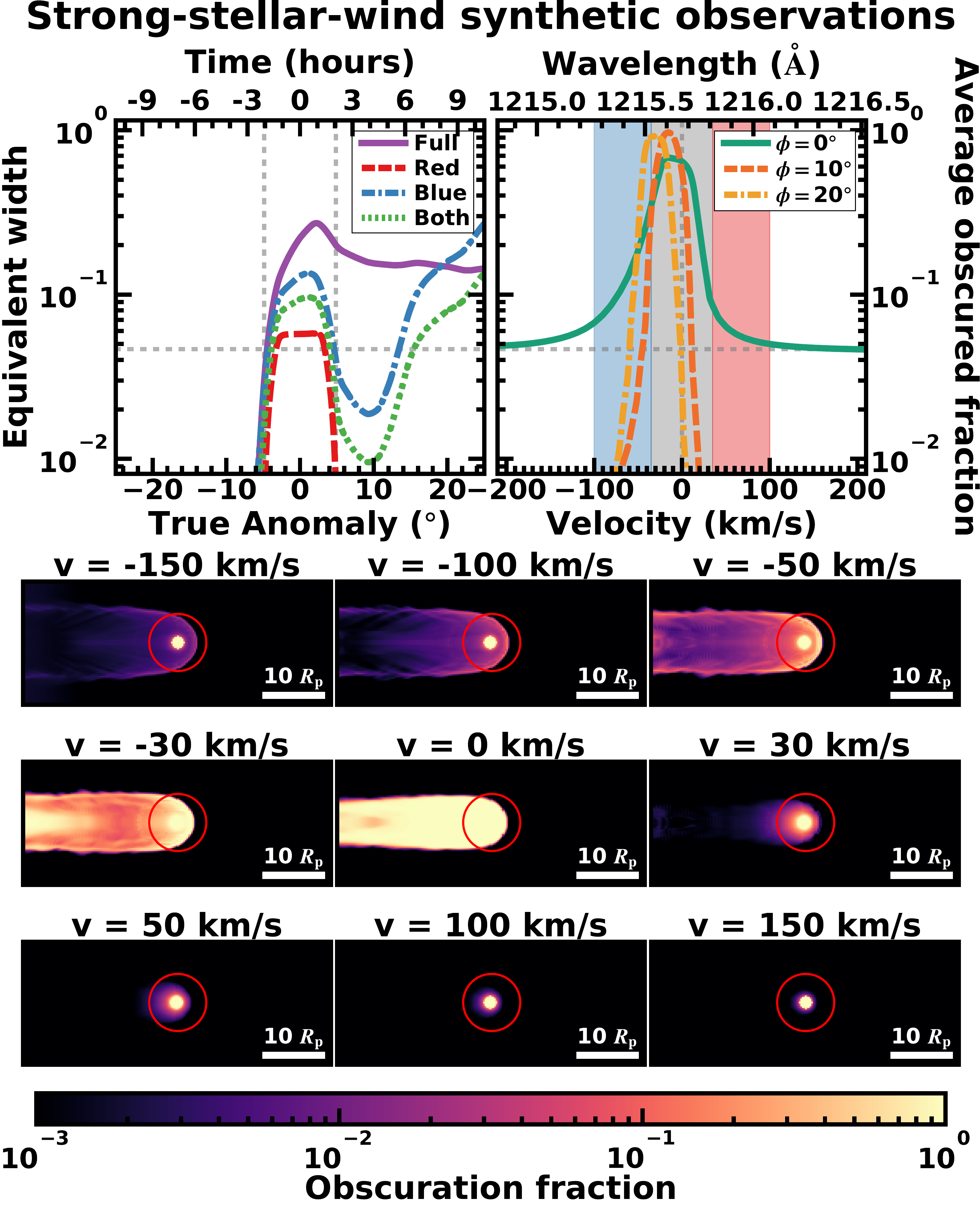}
\else
\includemedia[width=\columnwidth,
playbutton=none,
addresource=movie19.mp4,
flashvars={source=movie19.mp4&loop=true}
]{\includegraphics[width=\columnwidth]{Figure19.pdf}}{VPlayer.swf}
\fi
\caption{Panels have the same layout as \autoref{fig:wwind}, but are now for our strong stellar wind (\hyperref[fig:fullSnap]{Figure \ref{fig:fullSnap}(c)}). \textit{Upper Left}: The transit in the strong regime has significant blue enhancement over the planetary disc occultation during transit. Additionally, after the transit strong blue features are visible. \textit{Upper Right}: The asymmetry during direct transit favoring blue absorption comes from gas moving away from the star at the outer edge of the planetary outflow. Conversely, after transit there are significant blue features due to the entire outflow being accelerated outwards. \textit{Lower Panels}: The sharp confinement due to the stellar wind is easily visible in most panels. At large negative velocities the effects of the outer edge of the outflow being more strongly accelerated are visible. Available online as an animation, showing transit in the obscuration maps for the Red, Blue and Both frequency domains, as well as a sweep of the spectrum at $\phi = \SI{0}{\degree}$.}
\end{figure}

For our strongest stellar wind we observe a highly confined cometary tail-like outflow (\hyperref[fig:fullSnap]{Figures \ref{fig:fullSnap}(c)} and \hyperref[fig:fullSnap]{\ref{fig:fullSnap}(f)}). This structure leads to a transit light curve that is strongly asymmetric both temporally and in its spectra (\autoref{fig:swind}).

The blue features are due to the stellar wind accelerating the planetary outflow away from the star, so that neutral gas asymmetrically absorbs more efficiently at blueshifted velocities. Moreover, during transit substantial obscuration occurs at large velocities due to Doppler broadening skewed towards negative velocities. Shortly following the transit, blue obscuration starts increasing. This is the result of a continuously accelerated nightside arm being blown out at faster and faster velocities by the stellar wind, shifting the Doppler core to bluer frequencies. Notice that in the spectra at $\phi=0^{\circ}$, the star is not entirely obscured due to the confinement of the dayside arm, visible in the obscuration maps of \autoref{fig:swind}, and maximal obscuration is not obtained until later.

\begin{figure}
\label{fig:column}
\includegraphics[width=\columnwidth]{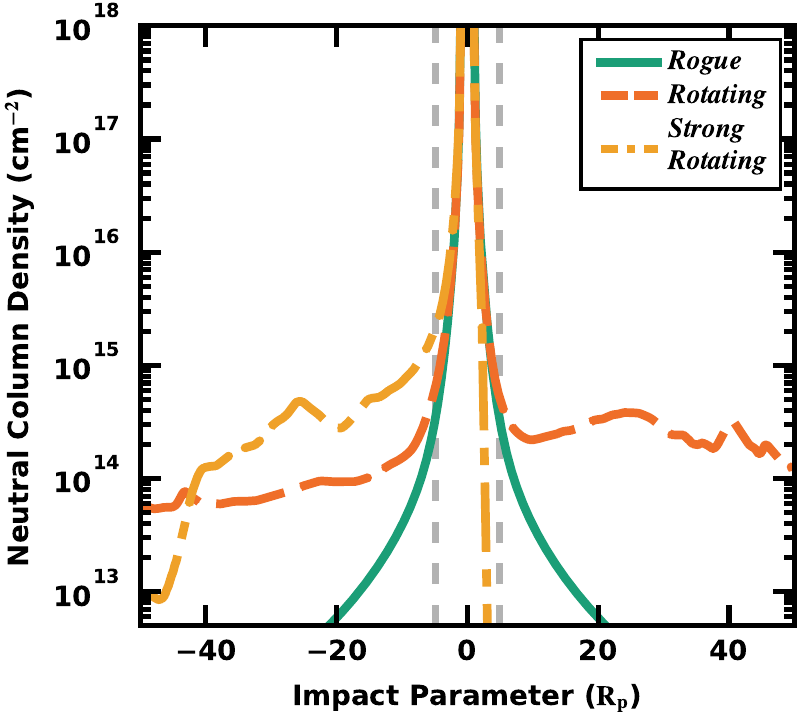}
\caption{Column densities integrated through outflows with spherical (\textit{Rogue}), toroidal (\textit{Rotating}), and constricted toroidal (\textit{Rotating} with a strong stellar wind\textemdash \textit{Strong Rotating}) geometries. For small impact parameters the column densities in all geometries match, as the outflow still behaves as if it is spherically symmetric until tidal forces or the stellar wind can significantly alter the flow into a torus. While the column density in the torus geometry should be roughly constant, notice that for the leading arm of the torus the density actually increases around $10 * R_{\textrm{p}}$. By referring to \hyperref[fig:fullSnap]{Figures \ref{fig:fullSnap}(a)}, we can see this density enhancement is due to the shock at the Coriolis length.}
\end{figure}

We now ask whether this signal is due to a density enhancement by stellar wind confinement. An enhanced column density might increase absorption in the naturally-broadened line wings. To understand the column density enhancement one obtains from stellar wind confinement we make comparisons to the \textit{Rotating} simulation with no stellar wind. This has been done previously in 2-D (\textit{c.f.} Figure 8 of Stone and Proga \citeyear{Stone09}) and is reproduced for 3-D simulations in a rotating frame in \autoref{fig:column}. We plot the column density along rays parallel to the $x$\textendash axis in the orbital plane as a function of the impact parameter measured from the center of the planet.\footnote{Here we chose ray trace as a function of impact parameters, rather than as a function of orbital phase, for direct comparison to Stone and Proga (\citeyear{Stone09}).}

To make comparisons to models not done in a proper rotating frame, we have overlain the column density for our \textit{Rogue} simulation, which has a roughly spherical outflow. In the absence of a stellar wind (\textit{Rotating}) the outflow is confined to a torus around the star due to tidal forces.  When a strong stellar wind is present (\textit{Strong Rotating}) the outflow's shape is still toroidal due to the tidal forces, but is further constricted and thus enhanced by the stellar wind.

Notice that outflow confined to a torus significantly enhances the column density over spherical outflows at large impact parameters. This can be understood by considering the area through which the planetary mass loss flux passes in each case. For a spherical outflow, this fluxing area is a sphere, which scales as $r^2$, while for the torus the fluxing area is the constant-area end caps of a cylinder. Thus, as long as the velocity does not change and the outflow has reached ionization equilibrium, the neutral column density for a torus will remain constant.

Given a fixed mass-loss rate, stellar wind constriction of the torus leads to a column density enhancement. This can be modeled by considering a cylinder of mass $M$, length $L$, and cross sectional radius $r$, radially constricted to a cylinder of length $L$ and radius $s$. While the column through the cylinder, $2r$ to $2s$, decreases by a factor of $s/r$, the density increases by a factor of $(r/s)^2$, so that column density is enhanced by $r/s$. Our strong stellar wind constricts the cross sectional radius roughly by half, seen by comparing \hyperref[fig:fullSnap]{Figures~\ref{fig:fullSnap}(d)} and \hyperref[fig:fullSnap]{\ref{fig:fullSnap}(f)}. While the mass-loss rate of a breeze is lower than the corresponding wind (\autoref{tab:massloss}), the difference is within ten percent for our parameters. Hence, relative to the weaker stellar wind cases, for our strong-stellar-wind planetary breeze, constriction increases the neutral density in the nightside arm by almost a factor of two. Another factor of two comes from redirecting the escaping gas from the dayside arm into the nightside. Thus, we estimate a grand total enhancement of a factor of four. This estimate matches what we see in the nightside arm of the constricted torus, between $\SI{-30}{R_{\textrm{p}}} \le b \le \SI{-10}{R_{\textrm{p}}}$, compared to the unconstricted torus (\autoref{fig:column}).

We conclude from \autoref{fig:column} that the stellar wind does substantially enhance the neutral column density over that from a spherical outflow. However, tidal gravity and the Coriolis force\textemdash by themselves confining the outflow to a torus\textemdash can also produce substantial enhancement in the column density at large distances from the planet. In fact, for our parameters the strong stellar wind only increases the optical depth in the naturally-broadened wings by a factor of a few over the weak-stellar-wind case. This may seem contrary to previous work that suggested stellar wind confinement could greatly enhance the transit signal, but the enhancement found in 2-D models with a stellar wind and no Coriolis force is similarly due to the geometry of constricting the flow to a torus (see Stone and Proga \citeyear{Stone09}). Therefore, while models that neglect the Coriolis force will see significant enhancement at large impact parameters due to a stellar wind, those that are done in a co-rotating frame shall not (unless the stellar wind strength is much larger compared to the tidal gravity than simulated here). We further note that our measured enhancement is modest at small impact parameters ($b \lesssim * 10 * R_{\textrm{p}}$), where most observations have so far probed. The lack of enhancement is due to the fact that the outflow will still behave roughly spherically until the outflow has been significantly altered by the tidal and Coriolis forces\textemdash becoming toroidal.

Lastly, and most importantly, an increased obscuration coming solely from naturally-broadened line wings from a low-velocity column density should be symmetric. In our simulations the predominant effect of a stellar wind is an asymmetric signal, as gas is exclusively accelerated away from the star. Therefore, symmetric observational features are likely not due to stellar wind confinement, which can be confirmed from the lack of an enhancement in the ``Red'' measurement of our strong-stellar-wind regime (\autoref{fig:swind}) relative to other stellar wind regimes. 

\subsection{Absorption outside direct transit}
\label{ssec:largeobs}

A prediction of this work is that planetary winds are capable of significant obscuration at large distances from the planet. For our parameters, the ionization timescale is $\tau_{\textrm{ion}} = (\sigma_{\textrm{HI}} F)^{-1} = \SI{3.6}{\hour}$. Generously averaging the outflow velocity to \SI{40}{km.s^{-1}}, we calculate the gas has undergone nearly seven ionization timescales by $\phi = \SI{\pm 20}{\degree}$.\footnote{Note the time relative to direct transit is not equivalent to the time it takes the gas to reach the location probed at the time relative to direct transit.} Thus, the existence of large scale signals from ionized gas may be surprising. However, note that after many ionizing timescales a gas does not become completely ionized; rather, it reaches ionization equilibrium set by the detailed balancing of ionization and recombination. Indeed for our simulations we have verified that the ionization fraction is within a few percent of equilibrium past \SI{10}{\textit{R}_{p}}.

Apparent then by our synthetic observations in \autoref{ssec:synobs}, gas in ionization equilibrium is sufficient to produce large-scale observable signals. However, these simulations were performed for a low ionizing flux analogous to the quiet Sun. We now ask if signals should persist at higher ionizing fluxes by considering each contribution to the optical depth. To order of magnitude the optical depth is $\tau_\nu = (1-X)*N_{\textrm{H}}*\sigma_{\Lya}(\nu)$, where $X$ is the ionization fraction, $N_{\textrm{H}}$ is the total column density of hydrogen, and $\sigma_{\Lya}(\nu)$ is the cross section of \LymanA{} absorption. The impact of the outflow structure on $\sigma_{\Lya}(\nu)$ is already discussed in \autoref{ssec:synobs}.

The total column density depends both on the total number density of hydrogen, $n_{\textrm{H}}$,\footnote{Note that the total number density $n = n_{\textrm{H}} + n_e \neq n_{\textrm{H}}$, where $n_{\textrm{H}} = n_{\textrm{HI}} + n_{\textrm{HII}}$ is the total number density of the hydrogen species.} and the column length, $L$. The column length scales with the Coriolis length, which depends on $F$ only through the velocity of the planetary outflow. We make the assumption that this dependence is weak and let $L$ be independent of $F$. From Murray-Clay et al. (\citeyear{Murray09}) we consider the two regimes of escape: energy-limited and radiation-limited. As more energy deposited into the atmosphere liberates more mass, in the energy-limited regime $\dot{M} \propto F$ and for the radiation-limited regime $\dot{M} \propto F^{1/2}$.\footnote{In the radiation-limited regime more energy is lost through radiative processes and escape becomes less efficient.} Since the cross section of the tidal torus only depends on $L$, which we keep constant, then $\dot{M} \propto n_{\textrm{H}}$. Therefore, in the energy-limited regime $N_{\textrm{H}} = n_{\textrm{H}} L \propto F$ and in the radiation-limited regime $N_{\textrm{H}} = n_{\textrm{H}} L \propto F^{1/2}$

With larger $F$ one might suspect lower neutral fractions $(1-X)$, as the ionization rate increases. As the outflow at large distance is in ionization equilibrium ($(1-X) \sigma_{\textrm{HI}} F = \alpha_{\textrm{B}} n_{\textrm{H}} X^2$), the ionization fraction as a function of flux

\begin{equation}
X = \frac{1}{2} \frac{\sigma_{\textrm{HI}} F}{\alpha_{\textrm{B}} n_{\textrm{H}} } \left(\sqrt{1+\frac{4\alpha_{\textrm{B}} n_{\textrm{H}} }{\sigma_{\textrm{HI}} F}} -1\right),
\end{equation}

and

\begin{equation}
\label{eqn:TaylorNeutral}
1-X = \frac{\alpha_{\textrm{B}} n_{\textrm{H}} }{\sigma_{\textrm{HI}} F} + \mathcal{O}\left(\left(\frac{\alpha_{\textrm{B}} n_{\textrm{H}} }{\sigma_{\textrm{HI}} F}\right)^2\right).
\end{equation}

\noindent Here we have Taylor expanded in $\alpha_{\textrm{B}} n_{\textrm{H}} /(\sigma_{\textrm{HI}} F)$ as we are already in the ionized regime far in the outflow such that $(1-X) \ll 1$. Then using the scalings for $n$, in the energy-limited regime, the neutral fraction does not change, $(1-X)$ is independent of $F$, while in the radiation-limited regime the neutral fraction scales as $(1-X) \propto F^{-1/2}$. So while there are more ionizing photons at higher fluxes, there is also a higher number density to facilitate recombination, preventing the neutral fraction from decreasing rapidly with increasing flux.

Taken all together $\tau_\nu \propto F$ in the energy-limited regime and $\tau_\nu$ is independent of $F$ in the radiation-limited regime. This suggests that as fluxes increase, the signal increases until the radiation-limited regime, at which point the signal is saturated. Thus in systems similar to the ones simulated, these signals may indeed be robust.

\subsection{Atmospheric escape observations}
\label{sec:realobs}

As seen in \autoref{ssec:synobs} it is possible to differentiate between our modeled stellar environments. Weak stellar winds produce the most symmetric transits and spectra. Particularly fast planetary winds can produce pre-transit absorption that is skewed red due to the dayside arm infalling towards the star. Intermediate stellar winds are marked by a quasi-static disrupted outflow, which manifests as time-varying blueshifted absorbing gas present significantly prior to transit. Lastly, strong stellar winds have strong spectral asymmetry during transit, little pre-transit obscuration, and increasing blue absorption post-transit from the strongly collimated tail being blown out by the stellar wind. Thus, to accurately distinguish between stellar regimes, measurements both in and far out of transit are required. Note that even a few hours out of transit may be too close to transit, as it takes time for the stellar environment to accelerate the outflow. Unfortunately, since \LymanA{} observations must be performed by expensive space-based telescopes, observations far out of transit are lacking.

A general theme amongst comparisons to observations is that our predicted \LymanA{} obscuration fractions are smaller than what is observed. While our planetary parameters are marginally a hot Jupiter analog, there is no one-to-one correspondence with any observed system. Yet, this is unlikely to account for the entire difference between the observations and our simulations, as previous modeling of specific systems also struggle to match observations, e.g., HD 209458 b in Murray-Clay et al. (\citeyear{Murray09}). Since outflow velocities retrieved in similar models are consistently too slow to explain the observations, it is widely thought that more exotic physics is needed (charge exchange, magnetic fields, etc.). However, we note that \cref{termv} implies that the asymptotic velocity $u_\infty$ depends on the energy spectrum of the ionizing photons. A harder spectrum produces more energy per ionization generating larger $\Delta q_\infty$ and thus larger $u_\infty$ (most easily conceptualized in the energy-limited regime, where radiative losses are minimal). Whether a more complete investigation of ionizing spectra can produce larger wind velocities merits investigation. To explore the possible consequences, we can lower the geocoronal confusion limits in post-processing to mimic faster outflows to zeroth order (see ``Low'' spectral frequency in \autoref{sssec:weak}).

To compare our models to observations we consider four different planetary systems, each of which has a distinctive signature. The first observed exoplanet undergoing escape, HD 209458 b, has a transiting decrement of \SI{15 \pm 4}{\percent} in \LymanA{} flux from \SIrange{1215.15}{1216.1}{\angstrom} (Vidal-Madjar et al. \citeyear{Vidal04}).\footnote{Two other frequency bin decrements often cited: \SI{8.9 \pm 2.1}{\percent} from \SIrange{1214.83}{1216.43}{\angstrom}, or \SI{5 \pm 2}{\percent} over the entire line \SIrange{1210}{1220}{\angstrom} (Vidal-Madjar et al. \citeyear{Vidal08}).} The spectrum was originally reported to be marginally asymmetric (Vidal-Madjar et al. \citeyear{Vidal03}), but further measurements have not detected any significant asymmetry (Ben-Jaffel \citeyear{Ben07}). Since strong stellar winds significantly skew the signal bluewards, even during direct transit, we rule out the strong stellar wind regime. Without more measurements out of transit it is hard to distinguish between a weak and intermediate stellar wind. Our ``Low'' measurements, which decreased the geocoronal confusion by \SI{15}{km.s^{-1}}, produced a roughly \SI{15}{\percent} symmetric signal during transit. This could suggest that HD 209458 b might orbit a star with a relatively weak wind, and a hard ionizing spectrum. Alternatively, all absorption from this system may result from physics, such as charge exchange, not modeled here. In particular, symmetric enhancement of the naturally-broadened line wings due to stellar wind confinement is not supported by our results.

Next, transits from HD 189733 b, with optical occultation of \SI{2.4}{\percent}, found Lyman-$\alpha$ decrements of \SI{5.0\pm1.3}{\percent} for the entire line, \SI{14.4\pm 6.6}{\percent} for the blue wing between \SIrange{-230}{-140}{km.s^{-1}}, and \SI{7.7\pm2.7}{\percent} for the red wing between \SIrange{60}{110}{km.s^{-1}} (Bourrier et al. \citeyear{Bourrier13}). We note that the observations have also seen significant variability in \LymanA{} absorption over various epochs. However, this time variability is not indicative of an intermediate stellar wind, as the variability is during direct transit and not further out in the outflow. \refemp{Rather, this implies that the mass-loss rate may be periodically modulated, possibly by a spatially variable stellar wind (Vidotto et al. \citeyear{Vidotto18}) or a temporally flaring star\textemdash consequences not probed by our steady stellar winds and ionizing flux.} We note that the numerical agreement with our strong-stellar-wind simulation for blue absorption during transit is coincidental because our modeled parameters match neither HD 189733 b nor the wavelength window of the observations. Furthermore, the asymmetric wavelength windows for these observations make evaluating the red/blue asymmetry during transit challenging. Nevertheless we note that the red-wing absorption is enhanced over the optical, a feature not seen in our models except in our ``Low'' measurements. Moreover, our blue frequency range is nearly \SI{100}{km.s^{-1}} slower than the blue range probed here, suggesting that the blue-shifted material may have been accelerated by a much stronger, and possibly faster, stellar wind than what we have simulated.

The hot Neptune GJ 436 b has the strongest and most asymmetric observations of any observed transit. The transit is within the measurement error of being undetectable in the red wings except in post-transit, with an obscuration of \SI{8.0 \pm 3.1}{\percent}. For the blue wings the obscuration is significant \SI{\sim2}{\hour} before, \SI{17.6 \pm 5.2}{\percent}, during transit, \SI{56.2 \pm 3.6}{\percent}, and \SI{\sim2}{\hour} after the transit, \SI{47.2 \pm 4.1}{\percent} (Ehrenreich et al. (\citeyear{Ehrenreich15})). It should be noted that due to lower surface gravity making the liberation of mass easier, atmospheric escape from hot Neptunes will likely be more dramatic than from hot Jupiters.\footnote{For Neptune $g_{\neptune} = \SI{\sim1100}{cm.s^{-2}}$ v.s. $g_{\jupiter} = \SI{\sim 2730}{cm.s^{-2}}$ for Jupiter, where $\neptune$ represents Neptune.} Thus, while the observations roughly correspond to velocities seen in our strong-stellar-wind regime, which shows significant blue obscuration from the cometary-like tail being blown outwards by the stellar wind, GJ 436 b's wind is able to persist at larger orbital phases on the dayside. As observations see blue absorption before transit the bow shock is not as close to the planet as in our simulation, which could be accomplished with a slightly weaker stellar wind. \refemp{Conversely, consider if the outflow disruption seen in the intermediate regime occurs at the Coriolis length. This is plausible since at the Coriolis length the velocity has deflected away from the star, leading the planetary pressure support at the stellar wind interface to be completely thermal\textemdash even if the outflow is supersonic (ram pressure dominated). Since GJ 436 b is a Neptune, its outflow velocities may be larger leading to a larger Coriolis length. Therefore, a persistent leading large-scale observational feature could be consistent with the intermediate-stellar-wind regime.}

\refemp{Bourrier et al. (\citeyear{Bourrier16}) suggest that a key component of GJ 436 b's observations can be explained by the acceleration of planetary gas by radiation pressure.  Their model and the model reported here differ in several key respects.  First, because the model in Bourrier et al. (\citeyear{Bourrier16}) is a particle simulation, it does not include forces from the pressure gradient of the gas.  As exemplified in the work reported here, pressure gradients cause gas to be launched from the planet's surface and accelerated to a non-zero bulk velocity away from the planet.  Motivated by the fact that this launch velocity is much smaller than the large velocities probed in the line wings of \LymanA{} observations, Bourrier et al. (\citeyear{Bourrier16}) use as their starting condition a population of particles at $\sim$3 planetary radii with thermal velocities but no bulk velocity.  Because this population lies within the planet's Hill radius, these particles cannot escape until they are substantially accelerated either by radiation pressure or by the stellar wind, both of which are modeled through probabilistic interactions with the gas particles, including self-shielding.}  

\refemp{These choices have subtle but important consequences. First, this leads to a highly ionized planetary outflow (relative to our simulations), as the ``static'' particles are likely to be ionized outside $\tau = 1$ while they wait to be accelerated (something they note in \S 3.3.3). In practice, what this means is that the outflow becomes ionized enough to be optically thin to radiation pressure before it is accelerated out of the planet's potential well.  In contrast, we find in our simulations that the planetary outflow is extremely optically thick out to more than 10 planetary radii, well beyond the planet's Hill sphere.  The difference comes from the outflow velocity driven by gas pressure forces\textemdash though small, this velocity is large enough to allow the planetary wind to leave the planet's Hill radius before reaching ionization equilibrium.}

\refemp{Recall that optically thin radiation ($\tau_\nu \ll 1$) will only absorb a tiny fraction of the incoming flux by definition, yet will do so throughout the entire outflow. In contrast, except at very large separations, we find that the planetary wind is predominately optically thick to the entire \LymanA{} photon band. Thus, significant optically thin radiation pressure (not modeled) is not an important accelerant in the majority of our simulation box, and certainly in the regions dominating our synthetic observations. Optically thick radiation pressure (also not modeled), in contrast, may impact the velocity structure at the boundaries of our simulated outflow, exerting a force at the wind interface rather than uniformly throughout the outflow.  As mentioned in \autoref{sec:conclusion}, optically thick radiation pressure may work in tandem with stellar wind pressure to shape the large scale structure and velocities of the planetary outflow, but it does not provide significant acceleration of the gas in the vicinity of the planet.  This contrasts with Bourrier et al. (\citeyear{Bourrier16}) who find that radiation pressure was found to be significant part of the acceleration of the outflow, a result that may not be surprising given that the model is accurate only after the particles have been significantly accelerated by the stellar wind or radiation pressure.}

\refemp{Additionally, particle models that do not model inter-particle interactions cannot generate pressure forces, which is an important part of the impact of the stellar wind and optically thick radiation pressure as the force is distributed throughout the outflow.\footnote{Our simulations show that the fluid is still collisional throughout the domain, due to Coulomb force's long range effect in the mostly ionized outflow.} The impact of both optically thin and optically thick radiation pressure far from the planet merits further investigation in hydrodynamic simulations.}

Lastly, 55 Cnc b does not have an optical transit detection, yet intriguingly a \SI{7.5\pm1.8}{\percent} decrement between \SIrange{-76.5}{0}{km.s^{-1}} was measured in the blue wings of \LymanA{} at its inferior conjunction\textemdash the location where a transit would occur if the planet was coplanar (Ehrenreich et al. \citeyear{Ehrenreich12}). No detectable obscuration in the red wings was reported. This obscuration is interpreted as coming from the transit of a fraction of an extended hydrogen atmosphere. The spectral asymmetry in the observation suggestively indicates a strong stellar wind.

\section{Conclusion and future work}
\label{sec:conclusion}

In summary, the stellar environment can play a significant role in shaping planetary outflows. Through our bottom-up approach we have examined several components of the stellar environment in 3-D: ionizing radiation, tidal gravity, the Coriolis force and a stellar wind. While spatially resolved observations of these outflows are not feasible, we demonstrated that spectrally and temporally resolved transit observations may still illuminate the overall structure of the outflow.

Alterations to the planetary outflow considered here come from orbital effects and the interaction with a stellar wind. Tidal gravity and the Coriolis force funnel the outflow into a torus with a dayside and nightside arm. The toroidal geometry enhances the column density at large distances over spherical outflows. Additionally the inclusion of non-inertial forces lead the outflow to shock on itself, justified by analysis of ballistic particle trajectory crossings. For stellar winds we find three unique regimes with respect to their effects on the planetary outflow. From our study of stellar environments and the morphology of planetary outflows we summarize the following key points

\begin{enumerate}
\item \refemp{Whether the planetary outflow is a wind or a breeze, depends in part on the stellar environment. However, if the the planetary outflow is a wind, then the mass-loss rate is insensitive to the stellar environment.}
\item Large detectable signals may be present far outside of transit.
\item These large scale signals help probe the stellar environment 
\begin{enumerate}[i.]
\item  Weak stellar winds cannot restrain the planetary outflow,  resulting in symmetric but weak transit signals.  For particularly fast planetary outflows, a pre-transit dayside arm skewed redward of line center is visible.
\item Even a spatially and temporally steady intermediate stellar wind cause periodic disruption of the growing planetary outflow, leading to large blue asymmetries preceding transit.  A double transit-like feature, from the disrupted dayside material, occurs at blue wavelengths.
\item Strong stellar winds cause cometary-like planetary outflows.  Absorption is biased bluewards during transit, and the cometary tail-like structure produces substantial blue absorption post-transit. 
\end{enumerate}
\end{enumerate}

Absent from our discussion is radiation pressure from \LymanA{}, which at the extremes behaves either in an optically thin or optically thick fashion. Recall that the total fraction of incoming flux absorbed by neutrals is the obscuration fraction, $\mathcal{O}_\nu = 1-\textrm{e}^{-\tau_\nu}$ (spatially mapped for our outflows in \autoref{ssec:synobs}). Thus, the total work done by radiation pressure on the outflow is $\Delta W_\nu \propto F_\nu \mathcal{O}_\nu$. Optically thick radiation ($\tau_\nu \gg 1 \rightarrow \mathcal{O}_\nu \approx 1$) absorbs nearly all incoming flux at the edge of the outflow, and behaves akin to a stellar wind. Optically thin radiation ($\tau_\nu \ll 1 \rightarrow \mathcal{O}_\nu \approx \tau_\nu$) on the other hand, will only absorb a tiny fraction of the incoming flux, which occurs throughout the entire outflow rather than at the edge. In our simulations, the (unmodeled) force exerted by optically thick photons would strongly dominate near the planet, and would exceed that of optically thin photons by at least a factor of two throughout the flow within our domain. We note that optically thick radiation pressure may behave similarly to stellar winds and be more appropriate to these systems. However, radiation pressure will always act radially with respect to the star. In contrast, the stellar wind has a ram-pressure headwind effect from the planet's orbit and a thermal-pressure force normal to the planetary-stellar wind contact discontinuity\textemdash neither of which are strictly radially outwards from the star.

In future work, one can use the machinery presented here to analyze various stellar spectra, stellar time variability (particularly important for M-stars), planetary and stellar magnetic fields, and the entrainment of heavier elements. Discussed in \autoref{apnss:polybern}, heat conduction should be included to allow for more realistic stellar winds, i.e., coronal winds, and because it plays an important role on the planetary nightside. As previously mentioned, magnetic fields will play a dominant roll in shaping the outflow if planetary fields exceed \SI{1}{G}. As most of the outflow is ionized, the planetary or stellar magnetic field may prevent certain features in the large-scale outflow from existing. Therefore, we emphasize again that our results will reflect systems with weak magnetic fields, and future works wishing to survey wider possibilities should implement them.

\acknowledgments
This material is based upon work supported by the National Science Foundation under Grant No. AST-
1411536. Further thanks to Jim Stone, James Owen, and Jonathan Carroll-Nellenback for insightful numerical discussions. We would also like to acknowledge the yt-project (Turk et al. \citeyear{Turk11}), which made possible the majority of our visualizations.

\bibliography{\jobname} 
\bibliographystyle{aasjournal}

\clearpage
\appendix

\section{Bernoulli Constant}
\label{apn:bern}
\subsection{Overview of Bernoulli's Constant}
\label{apnss:reviewbern}

The formalism of Bernoulli's constant is useful since it gives analytic solutions to the atmospheric escape problem, potentially negating the need for numerical simulations. One widely cited example of a quasi-analytic solution is Watson et al. (\citeyear{Watson81}), who used conduction and a delta function for ionization heating to simulate hydrodynamic winds of our solar system's early terrestrial planets.\footnote{They used numerics to iteratively solve the steady state system of equations (information contained in the Bernoulli constant plus further assumptions), as opposed to integrating the fluid equations forward to a steady state.} One method of retrieving the Bernoulli's constant is by integrating the energy equation along streamlines. \Cref{Econt} can be rewritten in steady state as

\begin{equation}
\label{eq:streamE}
\vec{u}\cdot \vec{\nabla} \left( \frac{E+P}{\rho} + \phi \right) = \frac{\mathcal{G} - \mathcal{L}}{\rho}.
\end{equation}

\noindent Take for now the following as the definition for $\Delta q$

\begin{equation}
\label{eq:Qdef}
\vec{u} \cdot \vec{\nabla} \left( \Delta q \right) \equiv \frac{\mathcal{G} - \mathcal{L}}{\rho}.
\end{equation}

\noindent Then by substituting \cref{eq:Qdef} into \cref{eq:streamE} and integrating along streamlines one retrieves the constant

\begin{equation}
\label{eq:bern}
b = \frac{1}{2} u^2 + h + \phi - \Delta q.
\end{equation}

\noindent Here we have made use of the specific enthalpy, $h = \epsilon + P/\rho$, where $\epsilon$ is the specific internal energy. \Cref{eq:bern} is the specific Bernoulli constant, which is physically motivated by conservation of energy yet is not the total specific energy. Rather $b$ is the total energy minus the total heat added to the fluid. When external heat is added to the fluid, it becomes either kinetic, internal or potential energy, so for $b$ to remain constant the added energy must be subtracted off the total energy.

At first glance it may seem challenging to invert \cref{eq:Qdef}; however, recall that the convective derivative, or directional derivative, $\nabla_{u} = \vec{u} \cdot \vec{\nabla}$, differentiates a field along streamlines of the flow. This can be thought of as a 1-D problem, e.g., by using the Frenet-Serret frame, so that $\Delta q$ can be solved as 

\begin{equation}
\label{eq:Qheat}
\Delta q = \int_{C(s)} \frac{\mathcal{G}-\mathcal{L}}{\rho * u_s} \, \textrm{d} s.
\end{equation}

\noindent Here $C(s)$ is the streamline parameterized by $s$. Note that by definition in these coordinates $u_s = | \vec{u}|$.

This prescription for Bernoulli's constant can be expanded upon to include a wide range of physics. Typically one describes the means of energy gain as either heat gained by or work done on the fluid. While conservative forces should be included in $\phi$, non-conservative forces, i.e., ones with path-dependent work, require a path-dependent term, $\Delta W$, measuring the accumulated work done along the path. We do not consider such forces in this work, but note that radiation pressure would be such a term.

Another common example of heat flux through a fluid is conduction. One can quickly see that if conduction is redistributing heat that

\begin{equation}
\label{eq:Qcond}
\Delta q = \int_{C(s)} \frac{\vec{\nabla} \cdot \left( \kappa * \vec{\nabla} T \right)}{\rho * u_s} * \textrm{d} s.
\end{equation}

\noindent The classic example of heat conduction in atmospheric escape, is Parker's spherically symmetric stellar wind model (\citeyear{Parker58}), also see Chamberlain (\citeyear{Chamberlain61}). Recall that for spherical symmetry the constant mass-loss rate can be expressed as \mbox{$\dot{M} = \rho * u * r^2$}. Then \cref{eq:Qcond} can be solved so that

\begin{equation}
\label{BernConductHeat}
b = \frac{1}{2}u^2 + h + \phi - \frac{\kappa}{\rho u} \frac{\partial T}{\partial r}.
\end{equation}

\noindent This is equivalent to Eq. 12 of Chamberlain (\citeyear{Chamberlain61}). In this special case, an integral accounting for the accumulated heat is not needed and only local quantities matter. Since in a spherically symmetric system heat conduction can act only along streamlines, it cannot alter the total energy along a streamline except at the boundaries. This however is not generally true when the assumption of spherical symmetry is relaxed, as heat flux from conduction may now flow perpendicular to streamlines\textemdash making $\Delta q$ path-dependent. 

\subsection{Reversible flows}
\label{apnss:revbern}

A perhaps more intuitive derivation of \cref{eq:Qheat} for $\Delta q$ results from considering how the energy equation is constructed from Euler's equation, \cref{momcont}, and the fundamental thermodynamic relationship for enthalpy. We use the powerful simplification that the flows are well approximated as reversible, so that a flow experiencing only reversible pressure-volume work has a specific enthalpy

\begin{equation}
\label{eq:RevFundThermRelEnthalpy}
\textrm{d} h = \dbar q + \frac{\textrm{d} P}{\rho}.
\end{equation}

\noindent Here $\, \dbar q$ is the differential heat flow into the system, e.g., a fluid element. Recall that $\, \dbar q$ is a form of energy transfer and must be represented by an inexact differential, i.e., no path-independent integrated quantity, say $q$, is well defined. However, if a path, $C$, parameterized by $s$, is specified we can define the accumulated heat along said path

\begin{equation}
\label{eq:NatQdef}
\Delta q (s) \equiv \int_{C(s)} \dbar q.
\end{equation}

For a fluid element the obvious path would be its streamline. Consider a fluid in a steady state, $\partial_t \left(\Delta q\right)= 0$, then it is clear that the infinitesimal heat added at $s$, $\, \dbar q (s)$, is the convective derivative of $\Delta q$, i.e.,

\begin{equation}
\label{HeatFlowConvect}
\dbar q (s) = \left[\vec{u} \cdot \vec{\nabla} \left(\Delta q\right)\right]\Big|_{s}.
\end{equation}

This is also apparent from \cref{eq:NatQdef}. Therefore considering \cref{eq:RevFundThermRelEnthalpy} as material derivatives

\begin{equation}
\label{eq:fundMD}
\vec{u} \cdot \frac{\vec{\nabla} P}{\rho} = \vec{u} \cdot \vec{\nabla} h - \vec{u} \cdot \vec{\nabla} (\Delta q).
\end{equation}

\noindent The term on the left-hand side of \cref{eq:fundMD} appears when you project Euler's equation into streamlines, a close analog to the energy equation. Integrating that equation along the streamline likewise retrieves the Bernoulli constant for a compressible flow with external heating, \cref{eq:bern}, which we have now shown is completely general as long as $P*\textrm{d}V$ work done by the flow is reversible, i.e., the only entropy production comes from heating.

We have thus arrived at a more natural definition for $\Delta q$ in \cref{eq:NatQdef}. Typically $\, \dbar q$ is understood in terms of a heating rate, which we would integrate with respect to time to get the total heat added. Referencing \cref{eq:Qheat}, we note that $\textrm{d}s/u_s$ is the time interval the fluid element spends near a given point and the rest of the integrand is the specific heating rate.

\subsection{Polytropic flows}
\label{apnss:polybern}

The Bernoulli constant provides a useful frame for understanding the setup of our stellar wind boundary condition, which we model as a polytropic outflow. It is often common to neglect detailed treatment of energy deposition in the stellar wind. This assumption allows us the relationship

\begin{equation}
\label{eq:polytropic}
P = P_0 \left(\frac{\rho}{\rho_0}\right)^\Gamma.
\end{equation}

\noindent In order to retrieve Bernoulli's constant from the Euler equation, as discussed in \autoref{apnss:revbern}, we can reduce the pressure force per unit mass to a scalar gradient via the polytropic relationship, for $\Gamma \neq 1$

\begin{equation}
\frac{\vec{\nabla}P}{\rho} = \frac{\Gamma}{\Gamma-1} \vec{\nabla} \left(\frac{P}{\rho}\right).
\end{equation}

\noindent So then

\begin{equation}
\label{bpolyE}
b_{\textrm{poly}} = \frac{1}{2} u^2 + \frac{\Gamma}{\Gamma-1} \frac{P}{\rho} + \phi.
\end{equation}

For an ideal gas

\begin{equation}
\label{eq:idealEnthalpy}
h = \frac{\gamma}{\gamma -1} \frac{P}{\rho},
\end{equation}

\noindent and

\begin{equation}
b_{\textrm{poly}} = b_{\textrm{isen}} + \frac{\gamma - \Gamma}{(\gamma-1)(\Gamma-1)} \frac{P}{\rho},
\end{equation}

\noindent where $b_{\textrm{isen}}$ is the isentropic Bernoulli constant (\cref{eq:bern} with $\Delta q = 0$). Note that $b_{\textrm{isen}}$ is typically thought of as the total energy.

Therefore, when $\Gamma = \gamma$, there is no bound atmosphere ($b_{\textrm{isen}} < 0$) that can become unbound ($b_{\textrm{isen}} \ge 0$), and ``adiabatic escape'' is a misnomer. A correlated point is that there exist energy (or momentum transfer) requirements at the sonic point for the outflow to go transonic. In other words, it is not only enough that energy be deposited in the fluid, but there must also be non-zero energy deposition below some threshold at the sonic point for the solution to be transonic (too large of an energy input at the sonic point reduces the outward pressure force, lessening the accelerating of the outflow). The justification of this constraint requires careful analysis of the momentum equation and is given in Lamers and Cassinelli (\citeyear{Lamers99}) \S\,4.1.4 for spherical winds.\footnote{We note that Eq. (4.31) in the reference is missing a minus sign and should read $-\partial_r \log(q r^{-1/2}) > (5/8) \partial_r M^2 > 0$ when evaluated at $r_c$, to be in agreement with both Eq. (4.29) and the discussion following.}

As discussed in \autoref{apnss:reviewbern}, the Bernoulli constant can be used to analytically solve atmospheric escape. However, the application has its limitations, namely the constant knows nothing of the ionization structure that directly corresponds to $\Delta q$. This makes outflows that are launched by ionization heating, or for which you wish to make synthetic observations, not accessible by the Bernoulli constant alone. However, for our stellar wind model we do not care about the ionization structure, and use the Bernoulli constant to derive its solution. We model the stellar wind as a spherically-symmetric, isentropic wind (polytropic outflow with $\Gamma = \gamma$). To solve we begin by writing the steady state mass continuity equation as

\begin{equation}
\label{eq:Amcont}
\rho u r^2 = \rho_0 u_0 r_0^2.
\end{equation}

This, combined with the polytropic relationship, \cref{eq:polytropic}, allows us to write the sound speed in terms of the velocity and radius

\begin{equation}
\label{eq:ATprof}
c_s(r) = \sqrt{ \frac{\gamma P_0}{\rho_0} \left( \frac{u_0 r_0^2}{u r^2}\right)^{\gamma -1} }.
\end{equation}

\noindent Combined with \cref{bpolyE} we have an equation for $u$ in terms of only constants and $r$

\begin{equation}
\label{analvel}
\frac{1}{2} u^2 + \frac{\gamma}{\gamma -1} \frac{P_0}{\rho_0} \left(\frac{u_0 r_0^2}{u r^2}\right)^{\gamma -1} + (\phi - b_{\textrm{poly}}) = 0.
\end{equation}

Thus given a mass-loss rate, $\dot{M} = \rho u r^2$, and the temperature of the outflow at $r_0$, we can analytically solve for the velocity structure of the outflow. This is precisely what we do for our stellar wind, by using Brent's method to solve for $u$ at all locations within the domain. We then take the velocity and use \cref{eq:Amcont,eq:ATprof} to solve for the density and temperature structure of the stellar wind.

\refemp{Note that for future work one should consider instead using what is sometimes called a coronal wind (Lamers \& Cassinelli \citeyear{Lamers99}, Ch.\! 5). We have already derived the solution for the coronal wind in \cref{BernConductHeat}. Notice that for coronal winds $b = b_{\textrm{isen}} - (\kappa * \partial_r T)/(\rho * u)$. Thus, if one has large heat fluxes at the wind base (large $-\kappa * \partial_r T$ and/or small $\rho *u$) it is possible to have bound atmosphere ($b_{\textrm{isen}} < 0$) near the wind base, but be unbound ($b_{\textrm{isen}} \ge 0$) further out where heat fluxes have distributed the heat ($-\kappa * \partial_r T \approx 0$). Importantly, this can be accomplished in simulations with $\gamma = 5/3$, meaning that with conduction, transonic stellar winds can be included in simulations that include adiabatically cooled planetary outflows.}

\section{Ambient medium setup and optimization}
\label{apn:ambient}

\subsection{Setup}
\label{apnss:ambsetup}

\begin{figure}
\label{fig:Rs}
\includegraphics[width=\columnwidth]{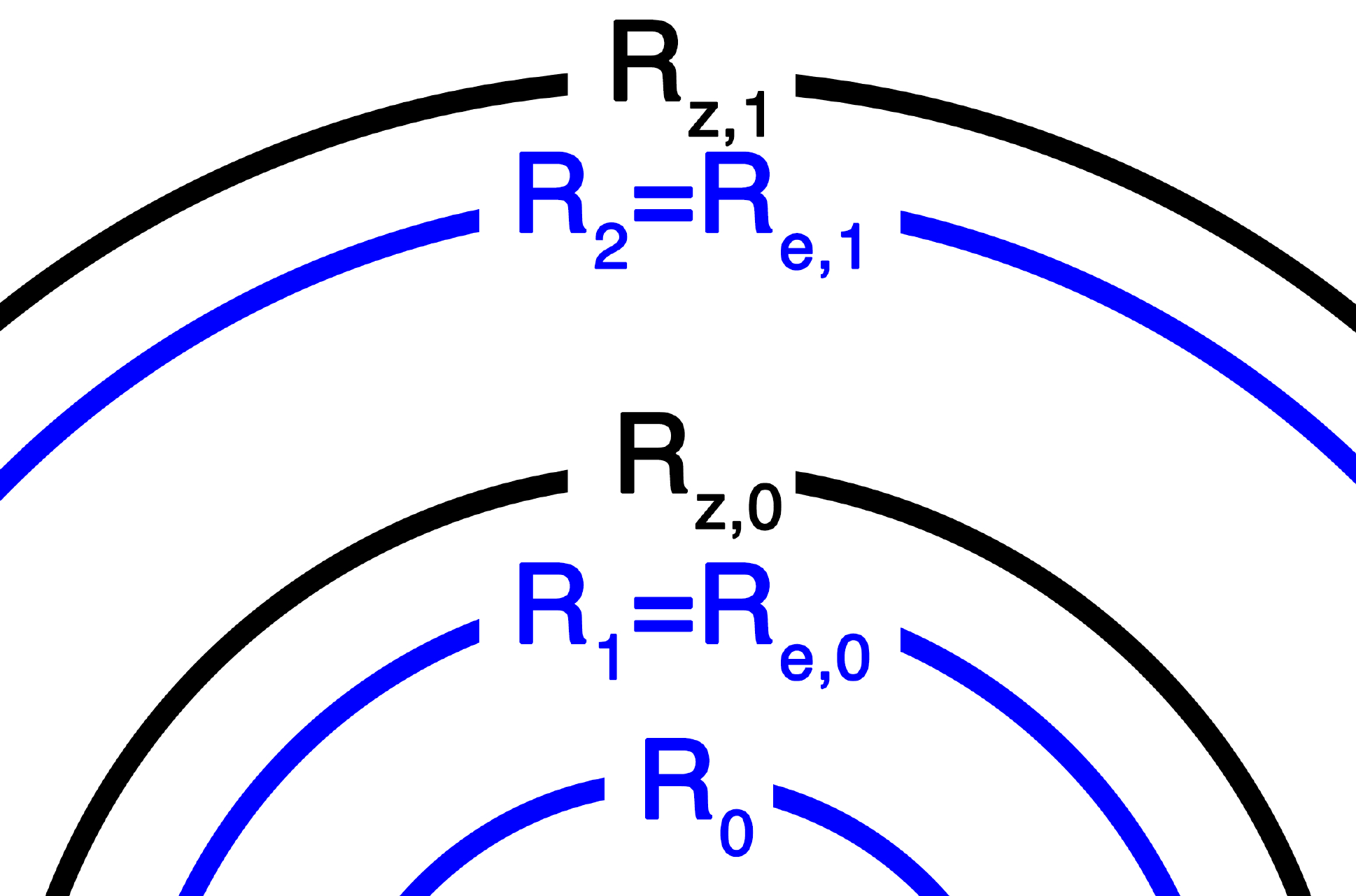}
\caption{Illustration of radial scales for the nested isentropic atmospheres that comprise our initial conditions for simulations that lack a stellar wind. Blue denotes our reference height for each atmosphere and black is the bounding surface of that atmosphere where the values go to zero. Not to scale.}
\end{figure}

When no stellar wind is present, our ambient medium takes the form of two nested isentropic atmospheres, which prevents the background gas from collapsing due to the planet's gravity. Our simulations thus contain up to three nested isentropic atmospheres, pressure-matched at the boundaries where they meet. We enumerate our isentropic atmospheres as follows: ``$0$'' corresponds to the planetary atmosphere, ``$1$'' to the primary ambient medium and ``$2$'' to the secondary ambient medium. Each isentropic atmosphere has a reference radius $R_{0}$, $R_{1}$, or $R_{2}$, a zero radius $R_{\textrm{z}}$, where the atmosphere formally reaches zero density, and a numerical edge, $R_{\textrm{e}}$, where the atmosphere is truncated in our simulation.  For example, $R_{\textrm{e},0}$ refers to the truncation radius for the planetary atmosphere.  As the atmospheres are nested and pressure matched, we take the edge of an inner atmosphere as the reference of the next atmosphere, e.g., $R_{1} = R_{\textrm{e},0}$. These radial scales are illustrated in \autoref{fig:Rs}.

As mentioned in \autoref{sssec:patm}, an isentropic bound hydrostatic atmosphere has a bounding surface at which the density and pressure go to zero.  To avoid the difficulty of handling this numerically, our atmospheres are cut off at a numerical edge such that $R_{\textrm{e}} < R_{\textrm{z}}$. It is desirable to chose $R_{\textrm{e}}$ well away from the $\tau =1$ surface for photoionization, This prevents unnecessary evolution of $\tau =1$, although the final solution should be insensitive to any such choices. More importantly, the closer $R_{\textrm{e}}$ is to $R_{\textrm{z}}$, the lower the pressure and density of the pressure-matched ambient medium can be. We find that setting the $R_{\textrm{e}} = R_{\textrm{z}} - \updelta x/2$ produces numerical stability, where $\updelta x$ is the size of a computational cell at the locally highest level of static mesh refinement. This choice is successful because as part of the hydrodynamic update we must extrapolate the state variables from cell centers to cell edges. Truncating at $R_{\textrm{z}} - \updelta x/2$ ensures that a well-defined solution exists at the cell edge to which we are extrapolating. Note that our stellar wind is also technically an isentropic atmosphere; however, it is unbounded so that it does not have a zero radius, $R_{\textrm{z}}$. Therefore, we only consider a reference radius, $r_{\star,0}$, at which to prescribe boundary conditions.

\subsection{Optimization}
\label{apnss:ambsetup}

The numerical difficulties in simulating our ambient medium are twofold. First, for numerical stability, areas with large gradients should be well resolved. To efficiently achieve this we need to avoid regions of the isentropic ambient medium that have small scale heights, e.g., near their zero radius. The second difficulty comes from pressure matching a low- and high-density region, since the speed of sound in the low density region will be larger. Lowering the ambient density then in turn, leads to exceedingly expensive computations as we must satisfy the Courant condition.

When trying to optimize our simulations, the limitations placed by these two constraints turn out to be at odds with one another. Since we must prioritize accuracy over efficiency, we foremost avoid large gradients in the flow, and we will treat the sound speed as a secondary optimization to consider only if we can ensure small gradients. Why these constraints are at odds with one another is intuitively understood by considering the atmosphere's temperature. A hotter atmosphere has a larger sound speed, which by the Courant condition requires higher temporal resolution. Conversely a colder atmosphere has smaller scale heights that requires higher spatial resolution to resolve. As the density is related to the temperature by the equation of state, we will now examine the bounds these constraints place on our ambient density.

As already stated, the first difficulty (large gradients) is averted by avoiding the edge of the ambient atmosphere. From \cref{amb1} we know $R_{\textrm{z},1}$ occurs where $\phi(R_{\textrm{z},1}) = h_1 + \phi_1$. To avoid a ``snowplow'' phase as our planetary wind launches, we wish to reduce the reference density of the ambient medium, $\rho_1$, relative to that of the edge of the planetary atmosphere, $\rho(R_{\textrm{e},0})$. Then, as the ambient medium and planetary atmosphere are pressure matched ($P_{\textrm{1}} = P(R_{\textrm{e,0}})$), $h_1 > h(R_{\textrm{e},0})$ and thus $R_{\textrm{z},1} > R_{\textrm{z},0}$ (see \cref{fn:azvspz} for detailed proof). However, rather than setting $\rho_1$ and solving for $R_{\textrm{z},1}$, we instead solve for $\rho_1$ after choosing $R_{\textrm{z},1}$. Operationally we chose the $R_{\textrm{z},1}$ to either be outside of our domain or at some special radius, as discussed below.

To bound $\rho_1$, consider a potential such that $\phi(r) \geq \phi_1$ for $r > R_1$. This happens for a point mass or, in the full rotating reference frame, inside the Roche lobe. To ensure that the density does not go to zero inside $R_{\textrm{e},1}$, we require $\phi(R_{\textrm{e},1}) > \phi(R_{\textrm{z},1})$, or

\begin{equation}
\label{conup}
\rho_1 \leq \frac{\gamma P_1}{(\gamma-1) \left(\phi (R_{\textrm{e},1}) - \phi_1\right)}.
\end{equation}

For the lower bound, it turns out that inside our isentropic planetary atmosphere the largest sound speed is found in the inner masked region. Therefore, we seek an ambient medium that at most has this speed of sound to add no further computational expense. Using the Courant condition to keep the adiabatic sound speed below that of the masked region, the ambient gas density should satisfy

\begin{equation}
\label{conlo}
\rho_1 \geq \rho_{\textrm{mask}} \frac{P_1}{P_{\textrm{mask}}}.
\end{equation}

Here the ``mask'' subscript is the value inside $R_{\textrm{mask}}$. Combining \cref{conup,conlo} provides upper and lower constraints on our ambient density 

\begin{equation}
\rho_{\textrm{mask}} \frac{P_1}{P_{\textrm{mask}}} \leq \rho_1 \leq \frac{\gamma P_1}{(\gamma-1) \left(\phi (R_{\textrm{z},1}) - \phi_1\right)}.
\end{equation}

Ideally, one should chose the lowest density possible to avoid causing the wind to enter a ``snowplow'' phase when expanding. Yet, there is no guarantee that these constraints are consistent, and as mentioned we must favor satisfying \cref{conup}. If these constraints turn out to be inconsistent then there is additional computational expense in our ambient medium over the planetary atmosphere. However, since we are using structured static mesh refinement we can further exploit the larger cell sizes in the lower resolution regions to perhaps make inconsistent bounds consistent, or at the very least ``less'' inconsistent.

Let the cell size ratio between the coarsest and finest resolution be $f$. By the Courant condition on the coarser mesh we can have a factor of $f$ larger sound speed compared to the fine resolution mesh, which translates into a factor of $f^2$ lower density. Therefore, if we extend our first ambient medium layer out only to the coarse refinement, e.g., if it is given by a cube of side $2 * R_{\textrm{refine}}$ then $\sqrt{3} * R_{\textrm{refine}}$, we can use a factor of $f^2$ smaller density here while satisfying the Courant conditions with the same time step. Then from there a second ambient medium layer would extend out to infinity, $R_{\textrm{z},2} \rightarrow \infty$, so that $\phi(R_{\textrm{z},2} ) = 0$. Then the constraints on our secondary ambient medium layer become

\begin{equation}
\frac{1}{f^2} \left(\rho_{\textrm{mask}} \frac{P_2}{P_{\textrm{mask}}} \right) \leq \rho_2 \leq \frac{\gamma P_2}{(\gamma-1) \left(- \phi_2\right)}.
\end{equation}

Here the subscript ``$2$'' denotes the second ambient layer's reference point, which is at $\sqrt{3} * R_{\textrm{refine}} = R_{\textrm{z},1}$. There is still no guarantee that these conditions are consistent. However, by exploiting the refinement factor and using multiple ambient layers one can optimize their initial setup relative to a single ambient layer. This is only appropriate when lower resolution does not lead to numerical instability or loss of numerical convergence. We therefore need the scale height of our wind to be larger than our isentropic hydrostatic atmosphere, which is what we find.

\section{Scale heights within the atmosphere}
\label{apn:sheight}

We will now review four distinct definitions of a scale height and explore their uses. For simplicity we will work with a generic variable $X$ in a spherically symmetric atmosphere. Often the first scale height one encounters, and perhaps the best defined, is the isothermal scale height, $H_{\textrm{iso}} = k_{\textrm{B}} T/(\mu g)$. It is derived from the plane-parallel, isothermal atmosphere, for which the thermodynamic variables are of the form $X(z) = X_0 \exp(-(z-z_0)/H_{\textrm{iso}})$. Thus, for each $H_{\textrm{iso}}$ away from $z_0$, the thermodynamic variables have an e\textendash folding in value.

The second way in which we will define a scale height is an often-used order of magnitude definition that can be thought of as the natural extension of isothermal scale height

\begin{equation}
\label{Fdef}
H' = \left( - \frac{\textrm{d} \log(X)}{\textrm{d} r}\right)^{-1}.
\end{equation}

\noindent Solving this differential equation for constant $H'$ retrieves the isothermal atmosphere. However, this definition is often used even when the atmosphere is not an exponential, i.e., $H'$ is not constant. For example, we have used this definition for our length scale in our Knudsen number calculation, $(L = ({\nabla} \log P)^{-1})$. That is because it roughly encapsulates the scale on which the variables change the order of themselves ($H' \sim (X/\Delta X) * \Delta r$). When one integrates \cref{Fdef}, it counts the number of e\textendash foldings between two locations,

\begin{equation}
\label{nheight}
N_{\textrm{e}}(a,b) = -\int_a^b H'^{-1} \textrm{d} r = \log\left(\frac{X(b)}{X(a)}\right),
\end{equation}

\noindent but what does $H'(r)$ itself mean in an atmosphere that is not an isothermal plane-parallel atmosphere? One will find that in general it is not the distance from $r$ at which variables have e\textendash folded. Instead it describes the e\textendash folding length in the analogous isothermal plane\textendash parallel atmosphere that has the same conditions at $r$ as our atmosphere. If we want the scale height, $H(r)$, over which the atmosphere folds by a factor of $s$, we need to solve

\begin{equation}
X(r + H(r)) = s X(r).
\end{equation}

This is usually what we mean when we talk about a scale height in our isentropic atmospheres, since it is useful to determine the actual folding within a numerical cell to quantify a condition for numerical stability. Even for a spherically symmetric isothermal atmosphere the form is not simple; however, both the isothermal and polytropic spherically symmetric cases have closed forms. For polytropic atmospheres all thermodynamic variables are a power, $n$, of an underlaying function $f(r)$, i.e., $X(r) = X_0 f(r)^n$, e.g., $n = 1/(\gamma-1)$ for density, $n=\gamma/(\gamma-1)$ for pressure, and $n=1$ for temperature. Then for the corresponding variable to the index $n$, the scale height $H(r)$ has a $s$-folding after

\begin{equation}
\label{actheight}
H_n(r, s) = \frac{R_{\textrm{z}} - r }{r + R_{\textrm{z}} (s^{-n^{-1}} - 1)^{-1}} \, r,
\end{equation}

\noindent where $R_{\textrm{z}}$ is the edge of your atmosphere, which for a hydrostatic isentropic atmosphere in a point mass potential is given by

\begin{equation}
R_{\textrm{z}} = \left( \frac{1}{R_0} - \frac{\gamma c_{\textrm{s},0}^2}{(\gamma-1)GM_{\textrm{p}}} \right)^{-1}.
\end{equation}

The last way which we define a scale height only applies for atmospheres that tend to zero in the limit of infinity. It is defined as

\begin{equation}
\int_r^\infty X(r') \textrm{d} r' = X(r) \widetilde{H}(r).
\end{equation}

\noindent Clearly this scale height is undefined in atmospheres that do not vanish at infinity, such as the spherically symmetric isothermal atmospheres.\footnote{Note that a plane-parallel isothermal atmosphere may extend infinity far out, but it tends to zero in the limit, so it is well defined there.} For a hydrostatic polytropic atmosphere, this scale height has a closed form in terms of incomplete beta functions

\begin{equation}
\label{Gheight}
\widetilde{H}(r) = R_{\textrm{z}} \left( \frac{R_{\textrm{z}}}{r} - 1\right)^{-3/2} \left(-\frac{3\pi}{2} - \beta \left(\frac{r}{R_{\textrm{z}}},1-n, 1 + n\right) \right).
\end{equation}

This last scale height, $\widetilde{H}$, is useful since its definition parallels that of optical depth

\begin{equation}
\label{Gtau}
\tau(r) = \int_{r}^\infty \sigma \, n_{\textrm{HI}}(r') \, \textrm{d} r' = \sigma \, n_{\textrm{HI}}(r) \, \widetilde{H}(r).
\end{equation}

\noindent Therefore, rather than doing a numerical integral to calculate the optical depth in our isentropic atmospheres, we only need to calculate the incomplete beta function. We use this calculation when picking an initial number density at $R_{\textrm{p}}$ (\cref{eq:planetn}). This is can be justified by considering where $\tau = 1$ occurs in steady state.


From \cref{Gtau}, one can always pick a $n_{\textrm{HI}}(R_{\textrm{p}})$ such that $\tau = 1$ at $R_{\textrm{p}}$ in steady state. However, the issue is knowing a priori the steady-state $\widetilde{H}$ of the outflow. For the energy-limited regime, where the outflow is not strongly ionized and neutrals are replenished by advection rather than recombination, we find that the optical depth one surface does not change substantially from its initial location. We calculate the number density required to place $\tau =1$ at $R_{\textrm{p}}$ using \cref{Gheight} and \cref{Gtau}.

However, in the radiation-limited regime the assumption that the ionization front does not move breaks down. We therefore no longer suggest using \cref{Gtau}, and instead appeal to the fact that $\tau = 1$ will occur where recombination balances ionization ($(1-X) \sigma_{\textrm{HI}} F = \alpha_{\textrm{B}} n_{\textrm{H}} X^2$).\footnote{This is not true in the energy-limited as advection primarily balances ionization.} Solving for $n_{\textrm{H}}$ gives

\begin{equation}
n_{\textrm{H}} = \frac{\sigma_{\textrm{HI}}*F_0}{\alpha_{\textrm{B}}} \frac{1-X}{X^2}.
\end{equation}

\noindent For the ionization fraction a small value, perhaps $X \sim 0.1$ is appropriate, as in the radiation-limited regime the ionization structure will be extremely sharp. The ideal value of $X$ merits future investigation.

Note that, we do not provide a first-order principles method of knowing a priori if you are in the energy-limited or radiation-limited regime, but it suffices to say that from previous studies we knew our parameters would firmly be in the energy-limited regime (Tripathi et al. \citeyear{Tripathi15}).

We conclude by noting, perhaps unsurprisingly, all three scale heights: $H'$, $H$, and $\widetilde{H}$ reduce to $H_{\textrm{iso}}$ for the plane-parallel isothermal atmospheres.

\end{document}